\documentclass{elsarticle}
\usepackage{subfigure}
\usepackage{graphicx}
\usepackage{booktabs}
\usepackage{algorithm}
\usepackage[noend]{algorithmic}
\newcommand{\OUTPUT}{\STATE \textbf{output~}}
\newcommand{\RET}{\STATE \textbf{return~}}

\newcommand{\no}[1]{}

\newcommand{\esch}{\mathtt{escherichia}}
\newcommand{\influ}{\mathtt{influenza}}
\newcommand{\para}{\mathtt{para}}
\newcommand{\fitags}{\mathtt{fiwikitags}}
\newcommand{\dna}{\mathtt{DNA1}}

\newcommand{\dnaaa}{\mathtt{DNA001}}
\newcommand{\dnaaaa}{\mathtt{DNA0001}}

\newcommand{\M}{\mathtt{RP}}
\newcommand{\rsa}{\mathtt{rsa}}
\newcommand{\rank}{\mathtt{rank}}
\newcommand{\access}{\mathtt{access}}
\newcommand{\select}{\mathtt{select}}
\newcommand{\extract}{\mathtt{extract}}
\newcommand{\gcc}{\mathtt{GCC}}
\newcommand{\gccc}{\mathtt{GCC.C}}
\newcommand{\gccn}{\mathtt{GCC.N}}
\newcommand{\rrr}{\mathtt{RRR}}
\newcommand{\cm}{\mathtt{CM}}
\newcommand{\deltae}{\mathtt{DELTA}}
\newcommand{\wt}{\mathtt{WT}}
\newcommand{\wm}{\mathtt{WM}}
\newcommand{\mwt}{\mathtt{MWT}}
\newcommand{\mwm}{\mathtt{MWM}}
\newcommand{\mwth}{\mathtt{MWTH}}
\newcommand{\wth}{\mathtt{WTH}}
\newcommand{\wmh}{\mathtt{WMH}}
\newcommand{\ap}{\mathtt{AP}}
\newcommand{\aprep}{\mathtt{AP.RP}}
\newcommand{\gol}{\mathtt{GMR}}
\newcommand{\cut}{\mathtt{cut}}
\newcommand{\cuto}{\mathtt{cut_o}}

\newcommand{\wtrp}{\mathtt{WTRP}}
\newcommand{\wmrp}{\mathtt{WMRP}}
\newcommand{\rpbmp}{\mathtt{RPB}}

\newcommand{\sxsi}{\mathtt{SXSI}}
\newcommand{\gct}{\mathtt{GCT}}
\newcommand{\libcds}{{\sc Libcds}}
\newcommand{\countp}{\mathtt{count}}
\newcommand{\locate}{\mathtt{locate}}

\begin{document}

\begin{frontmatter}

\title{Grammar Compressed Sequences \\ with Rank/Select Support
	\footnote{An early partial version of this paper appeared in
	 	{\em Proc. SPIRE 2014} \cite{NOspire14}.
Funded in part by  European Union's Horizon 2020 research and innovation
programme under the Marie Sk{\l}odowska-Curie grant agreement No 690941
(project BIRDS),
Fondecyt Grant 1-140796, Chile,
CDTI EXP 000645663/ITC-20133062 (CDTI, MEC, and AGI),
Xunta de Galicia (PGE and FEDER) ref.\ GRC2013/053, and by
MICINN (PGE and FEDER) refs.\ TIN2009-14560-C03-02, TIN2010-21246-C02-01,
TIN2013-46238-C4-3-R and TIN2013-47090-C3-3-P
and AP2010-6038 (FPU Program).}
}

\author{Alberto Ord\'o\~nez$^\star$  ~~~~~~
        Gonzalo Navarro$^\dag$ ~~~~~~
	Nieves R. Brisaboa$^\star$   \\ \ \\
{\small $^\star$ Database Laboratory, Universidade da Coru\~na, Spain} \\
{\small $^\dag$ Department of Computer Science, University of Chile, Chile}}


\begin{abstract}

Sequence representations supporting not only direct access to their symbols,
but also rank/select operations, are a fundamental building block in many
compressed data structures. Several recent applications need to represent 
highly repetitive sequences, and classical statistical compression proves 
ineffective. We introduce, instead, grammar-based representations for 
repetitive sequences, which use up to 6\% of the space needed by
statistically compressed representations, and support direct access and 
rank/select operations within tens of microseconds. We demonstrate the impact 
of our structures in text indexing applications.

\end{abstract}
\end{frontmatter}

\section{Introduction}

Given a sequence $S[1,n]$ over an alphabet $\Sigma=[1,\sigma]$, an intensively
studied problem in recent years has been how to represent $S$
space-efficiently while supporting these three operations:
\begin{itemize}
\item $\access(S,i)$, which returns $S[i]$, with $1 \le i \le n$.
\item $\rank_b(S,i)$, which returns number of occurrences of $b\in\Sigma$ in $S[1,i]$, with $0 \le i \le n$.
\item $\select_b(S,i)$, which returns the position of the $i$-th occurrence of $b\in\Sigma$ in $S$, with $0\leq i \leq \rank_b(S,n)$ and $\select_b(S,0)=0$.
\end{itemize} 

The data structures supporting these three operations will be called 
$\rsa$ structures (for $\rank$, $\select$, $\access$).
Their popularity owes to the wide number of applications in which they are particularly useful. For instance, we can simulate and improve the functionalities of {\em inverted indices} \cite{baeza1999modern,witten1999managing} by concatenating the posting lists and representing the resulting sequence with an $\rsa$ structure 
\cite{BN13,AGMOS12,AGCGMO12}. We can also build full-text {\em self-indices}
like the FM-Index \cite{FM05,FMMN07} on an $\rsa$-capable representation of the Burrows-Wheeler Transform \cite{BWT} of the text. 
Several other applications of $\rsa$ structures have been studied, for example
document listing in sequence collections \cite{NAVacmcs14}, XML/XPath 
systems \cite{SXSI}, positional inverted indices \cite{AGO10}, graphs
\cite{CNtweb10}, binary relations \cite{BCN13}, tries and labeled trees \cite{FLMM09}. 

In many applications, keeping the data in main memory is essential for high
performance. Therefore, one aims at using little space for an $\rsa$ structure.
The best known such sequence representations \cite{GGV03,CNOis14,BHMR11,GOR10,BCGNNalgor13,BNesa12} use 
{\em statistical} compression, which exploits the frequencies of the symbols
in $S$. The smallest ones achieve $nH_k(S)+o(n\log\sigma)$ bits for any
$k=o(\log_\sigma n)$. The measure $H_k(S)$ is the minimum bit-per-symbol rate
achieved by a statistical compressor based on the frequencies of each symbol 
conditioned to the $k$ symbols preceding it. Statistically-compressed 
representations can, on a RAM machine with word size $w$, answer $\access$ 
in $O(1)$ time and $\select$ in any time in $\omega(1)$, or vice versa, and 
$\rank$ in time $O(\log\log_w\sigma)$. These times
match lower bounds \cite{BNesa12}.

Although statistical compression is appropriate in many contexts, it is
unsuitable in various other domains. This is the case of an increasing number
of applications that deal with highly repetitive sequences: software
repositories, versioned document collections, genome datasets of individuals
of the same species, and so on, which contain many near-copies of the same
source code, document, or genome \cite{Nav12}. In this scenario, statistical
compression does not take proper advantage of the repetitiveness \cite{KN12}:
for $k=0$, the entropy does not change if we concatenate many copies of the 
same sequence, and for $k>0$ the situation is similar, as in most cases the 
near-copies are much farther apart than $k=o(\log_\sigma n)$ positions.

Instead, grammar \cite{KY00,CLLPPSS05} and Lempel-Ziv \cite{LZ76,ZL77}
compressors are very efficient to represent repetitive sequences, and thus
could be excellent candidates for applications that require $\rsa$
functionality on them. However, even supporting $\access$ is difficult on
those formats. The fastest schemes take $O(\log n)$ time, using either
$O(g\log n)$ bits of space on a grammar of size $g$ \cite{BSODA11}, or more 
than $O(z\log n)$ bits on a Lempel-Ziv parsing of $z$ phrases \cite{GGKNP14}. 
This time is essentially optimal \cite{VY13}. Therefore, supporting $\access$ 
is intrinsically harder than with statistically compressed sequence 
representations. 

The support for $\rank$ and $\select$ is even more rare on repetitive sequences.
Only for bitmaps (i.e., bit sequences) compressed with balanced grammars (whose grammar tree is
of height $O(\log n)$), the $O(g\log n)$ bits and $O(\log n)$ time obtained 
for $\access$ on grammar-compressed strings is extended to all $\rsa$ queries 
\cite{NPVjea13}. However, for larger $\sigma$, the space becomes
$O(g\sigma\log n)$ bits and the time raises to $O(\log\sigma\log n)$.

In this paper we propose two new solutions for $\rsa$ queries over grammar
compressed sequences, and compare them with various alternatives on a number
of real-life repetitive sequences. Our first structure, tailored to sequences over small
alphabets, extends and improves the current representation of bitmaps
\cite{NPVjea13}. On a balanced grammar of size $g$, it obtains $O(\log n)$
time for all the $\rsa$ operations with $O(g\sigma\log n)$ bits of space,
using in practice similar space while being much faster than previous work
\cite{NPVjea13}. We dub this solution $\gcc$ (Grammar Compression with 
Counters). It can be used, for example, on sequences of XML tags or DNA.

Our second structure
combines $\gcc$ with alphabet partitioning \cite{BCGNNalgor13} and is aimed at 
sequences with larger alphabets. Alphabet partitioning splits the sequence $S$ 
into subsequences over smaller alphabets. If these alphabets are small enough, 
we apply $\gcc$ on them. On the subsequences with larger alphabets, we 
use representations similar to previous work \cite{NPVjea13}. The resulting
time/space guarantees are as in previous work \cite{NPVjea13}, but the scheme 
is much faster in practice while using about the same space. 
Recent work \cite{BPTesa15} (see next section) shows that time complexities
of $\gcc$ are essentially optimal.

While up to an order of magnitude faster than the alternative
grammar-compressed representation, our solutions are still an order of 
magnitude slower than statistically compressed representations, but they are 
also an order of magnitude smaller on repetitive sequences. We also evaluate 
our data structures on two applications: full-text self-indices and XML 
collections.

This paper is organized as follows: Section~\ref{sec:related} describes the basic concepts and previous work; Section~\ref{sec:gcc} explains our $\rsa$ data structures for small alphabets; Section~\ref{sec:large} presents our solution for $\rsa$ on large alphabets; Section~\ref{sec:exp} experimentally evaluates our proposals; Section~\ref{sec:app} explores their performance in several applications; and finally Section~\ref{sec:concl} gives conclusions and future research lines. 

\section{Basic Concepts and Related Work}\label{sec:related}

\subsection{Statistical compression measures} \label{sec:empentropy}

Given a sequence $S[1,n]$ over $\Sigma=[1,\sigma]$, let $0\le p_i\le 1$ be the
relative frequency of symbol $i$ in $S$. The {\em zero-order 
empirical entropy} of $S$ is defined as\footnote{We use $\lg$ to denote the
logarithm in base 2.} $$H_0(S) = \sum_{i=1}^{\sigma} p_i \lg \frac{1}{p_i},$$
and it is a lower bound of the bit-per-symbol rate achievable by a compressor that
encodes $i$ only considering its frequency $p_i$. A richer model considers the
frequency of each symbol within the context of $k$ symbols preceding it. This
leads to the {\em $k$-order empirical entropy} measure, 
$$H_k(S)=\sum_{C \in \Sigma^k}\frac{|S_C|}{n}\,H_0(S_C),$$ where $S_C$ is the string formed by collecting the symbols that follow each occurrence of the context $C$ in $S$. It holds $H_k(S) \le H_{k-1}(S) \le H_0(S) \le \lg
\sigma$ for any $k \ge 1$.

\subsection{Grammar compression} \label{sec:grammar}

Grammar-compressing a sequence $S$ means finding a context-free grammar that generates (only) $S$. Finding the smallest such grammar is NP-complete \cite{CLLPPSS05}, but heuristics like RePair \cite{LM00} run in linear time and find very
good grammars.

% //MENTION ALTERNATIVE GRAMMAR COMPRESSORES
% //PSEUDO-CODE FOR REPAIR
% //DETAILS OF REPAIR IMPLEMENTATION

\begin{figure}[t]
 \centering
 \includegraphics[width=1.0\textwidth]{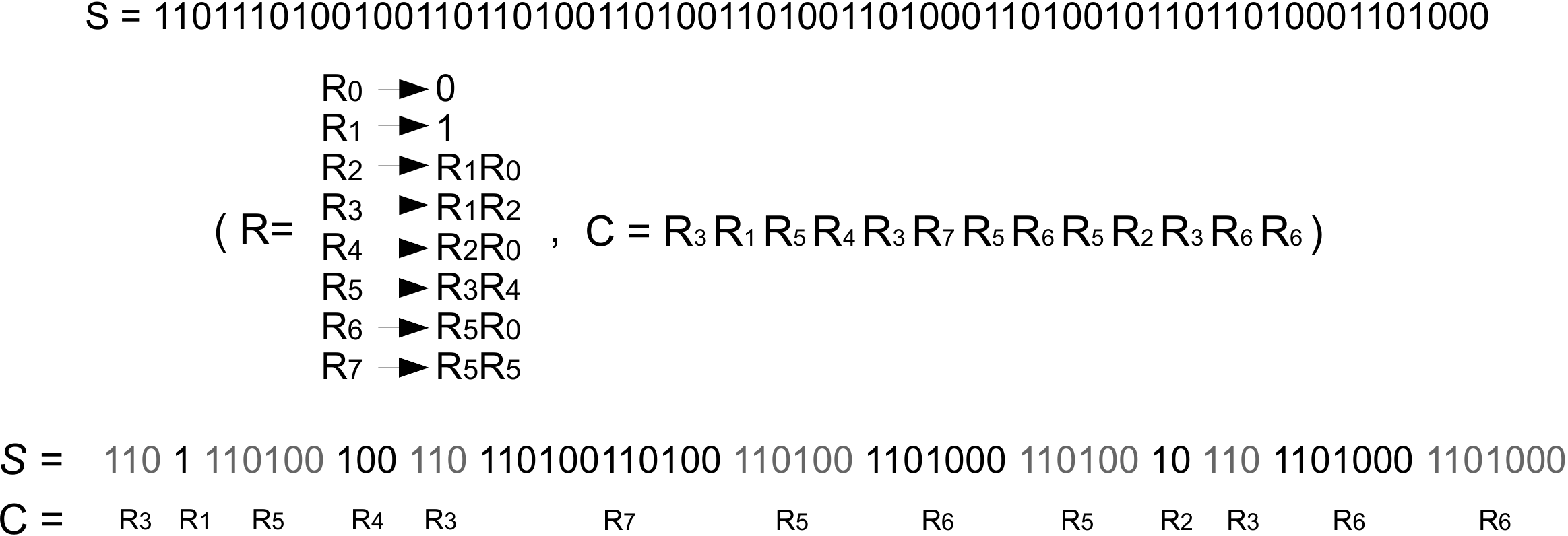}
  \caption{The data structures $(R,C)$ are the result of executing the RePair algorithm on the input sequence $S$ with $\sigma=2$.}
 \label{fig:repair}
 \end{figure}

RePair finds the most frequent pair of symbols $ab$ in $S$, adds a rule
$X\rightarrow ab$ to a dictionary $R$, and replaces each occurrence of $ab$ in
$S$ by $X$.\footnote{Note that, if $a=b$, we can only replace every other
occurrece of $aa$ in a sequence of $a$s.}
This process is repeated ($X$ can be involved in future pairs)
until the most frequent pair appears only once. The result is a pair $(R,C)$, where the dictionary $R$ contains $r=|R|$ rules and $C$, of length $c=|C|$, is the final result of $S$ after all the replacements are done. 
Note that $C$ is drawn from the alphabet of terminals and nonterminals.
For simplicity we assume that the first $\sigma$ rules generate the
$\sigma$ terminal symbols, so that $r$ counts terminals plus nonterminals. 
Thus, the total output size of $(R,C)$ is
$(2(r-\sigma)+c)\lg r$ bits. Figure~\ref{fig:repair} shows an example of 
applying RePair on a binary input $S$. 

By using the technique of Tabei et al.~\cite{TTS13}, it is possible to 
represent the dictionary in $r\lg r+O(r)$ bits, reducing the total space to 
$(r+c)\lg r +O(r)$ bits. However, our experiments in the conference version
\cite{NOspire14} show that the resulting access method is much slower,
so in this paper we use a plain representation of the rules.

Finally, it is possible to force the grammar to be {\em balanced}, that is, 
with the grammar tree being of height $O(\log n)$ \cite{Sak05}. We use instead
a simple heuristic that modifies RePair so that the newly created pairs are
added at the end of the list of the pairs with the same frequency. This is
sufficient to make the grammars balanced in all the cases we tested.

\subsection{Variable-length encoding of integers} \label{sec:vbyte}

In several cases one must encode a sequence of numbers, most of which are
small. A variable-length integer encoding aims to use fewer bits when encoding
a smaller number. For example, $\gamma$-codes \cite{witten1999managing} encode
a number $x>0$ using $2\lg x$ bits, by writing its length $|x|$ in unary
followed by $x$ itself in binary (devoided of its highest 1). For larger
numbers, $\delta$-codes \cite{witten1999managing} encode $|x|$ using
$\gamma$-codes instead of unary codes, and thus require 
$\lg x + O(\lg\lg x)$ bits to encode $x$.

For even larger numbers, the so-called Variable Byte \cite{vbyte} (VByte)
representation is interesting, as it offers
fast decoding by accessing byte-aligned data. The idea is to split 
each integer into 7-bit chunks and encode each chunk in a byte. The highest
bit of the byte is used to indicate whether the number continues in the next
byte or not. Then encoding $x$ requires at most $(8/7)\lg x + 7$ bits.

\subsection{Statistically compressed bitmaps} \label{sec:bitmaps}

Several classical solutions represent a binary sequence $B[1,n]$ with $\rsa$
support. Clark and Munro \cite{CPhd98,Mun96} ($\cm$) use $o(n)$ bits on top of
$B$ and answer the $\rsa$ queries in $O(1)$ time. 

Raman et al.~\cite{RRR02} ($\rrr$) also support the operations in $O(1)$ time, but they compress $B$ statistically, to $nH_0(B)+o(n)$ bits. This solution is well suited for scenarios where the distribution of $0$/$1$ is skewed. However, it is not adequate to exploit repetitiveness in the bitmaps.

If the bitmaps are very sparse, the $o(n)$-bits term of the previous solution
may be dominant. In this case, it is better to encode the differences between
consecutive positions of the $1$s with an encoding that favors small numbers,
like $\delta$-codes, and add absolute pointers to regularly sampled positions.
This encoding uses $nH_0(B) + o(nH_0(B))$ bits and handles $\rsa$
operations in $O(\log n)$ time. This folklore idea, which we call $\deltae$, has
been used repeatedly; see e.g. \cite{KN12}.

\subsection{Grammar-compressed bitmaps} \label{sec:rpb}

The only bitmap representation we are aware of that exploits repetitiveness in
the bitmaps is due to Navarro et al.~\cite{NPVjea13} ($\rpbmp$). They
RePair-compress $B$ with a balanced grammar and enhance the output $(R,C)$
with extra information to answer $\rsa$ queries. For each rule $X\in R$, let $exp(X)$ be the string of terminals $X$ expands to. Then they store two numbers per nonterminal $X$: 
\begin{itemize}
	\item $\ell(X) = |exp(X)|$,
	\item $z(X) = \rank_0(exp(X),\ell(X))$ (the number of $0$s contained in $exp(X)$).
\end{itemize}

Note that these values can be recursively computed: If $X \rightarrow YZ$,
then  $exp(X) = exp(Y)exp(Z)$; $\ell(X)=\ell(Y)+\ell(Z)$, with $\ell(0)=\ell(1)=1$; and $z(X)=z(Y)+z(Z)$, with $z(0)=1$ and $z(1)=0$.

To save space, they store $\ell(\cdot)$ and $z(\cdot)$ only for a subset of nonterminals, and compute the others recursively by partially expanding the nonterminal. Given a parameter $\delta$, they guarantee that, to compute any $\ell(X)$ or $z(X)$, we have to expand at most $2\delta$ rules. The sampled rules are marked in a bitmap $B_d[1,r]$ and the sampled values are stored in two vectors, $S_{\ell}$ and $S_{z}$, of length $\rank_1(B_d,r)$. To obtain $\ell(X)$ we check whether $B_d[X]=1$. If so, then $\ell(X)=S_{\ell}[\rank_1(B_d,X)]$. Otherwise $\ell(X)$ is obtained recursively as $\ell(Y)+\ell(Z)$. The process for $z(X)$ is analogous. 

Finally, every $s$th position of $B$ is sampled, for a parameter $s$. Note that
$B = exp(C[1])\,exp(C[2])\ldots exp(C[c])$, where the position where each
$exp(C[p])$ starts in $B$ is $L(p) = 1+ \sum_{k=1}^{p-1}\ell(C[k])$. Then, the
sampling array $S_n[0,n/s]$ stores a tuple $(p,o,rnk)$ at $S_n[k]$, where $exp(C[p])$ contains $B[k\cdot s]$, that is, $p=\max\{j,L(j)\leq k\cdot s\}$. The other components are $o= k\cdot s-L(p)$, that is, the offset of $B[k \cdot s]$ within $exp(C[p])$; and 
$rnk=\rank_0(B,L(p)-1)$ is the number of $0$s before $exp(C[p])$ starts.
We also set $S_n[0]=(0,0,0)$.

To answer $\rank_0(B,i)$, let $S_n[\lfloor i/s\rfloor]=(p,o,rnk)$ and set $l=s \cdot \lfloor i/s\rfloor -o$. Then  we move forward from $C[p]$, updating $l=l+\ell(C[p])$, $rnk=rnk+z(C[p])$, and $p=p+1$, as long as $l+\ell(C[p]) \leq i$. When $l \le i < l+\ell(C[p])$, we have reached the rule $C[p]=X\rightarrow YZ$ whose expansion contains $B[i]$. Then, we recursively traverse $X$ as follows. If $l+\ell(Y)> i$, we recursively traverse $Y$. 
Otherwise we update $l=l+\ell(Y)$ and $rnk=rnk+z(Y)$, and recursively traverse $Z$. This is repeated until $l= i$ and we reach a terminal symbol in the 
grammar. Then we return $rnk$. Obviously, we can also compute
$\rank_1(B,i)=i-\rank_0(B,i)$. Supporting $access(B,i)$ is completely equivalent, but instead of maintaining $rnk$ we just return the terminal symbol we reach when $l=i$. 

To answer $\select_0(B,j)$, we binary search $S_n$ to find $S_n[i]=(p,o,rnk)$ and $S_n[i+1]=(p',o',rnk')$ such that $rnk<j\leq rnk'$. Then we proceed as for $\rank_0$, but updating $l$ and $rnk$ as long as $rnk+z(C[p]) \le j$, and then traversing by going left (to $Y$) when $rnk+z(Y)>j$, and going right (to $Z$) otherwise. At the end, we return $l$. The process for $\select_1(B,j)$ is analogous (note that $X$ contains $\ell(X)-z(X)$ 1s).

On a balanced grammar, a rule is traversed in $O(\log n)$ time. The time to
iterate over $C$ between samples is $O(s)$. Therefore, if we set
$s=\Theta(\log n)$, the total time for $\rsa$ queries is $O(s+\log n)=O(\log
n)$ and the total space is $O(r\log n + (n/s) \log n) + c\lg r = O((r+c)\log
n+n)$ bits.\footnote{We can obtain $O((r+c)\log n)$ bits and the same time by
sampling $C$ instead of $B$, as we show later.} The time is multiplied by $\delta$ if we use sampling to avoid storing all the information for all the rules.

\subsection{Wavelet trees} \label{sec:wt}

The wavelet tree \cite{GGV03,Navjda13} ($\wt$) is a complete balanced binary
tree that represents a sequence $S[1,n]$ over alphabet $\Sigma=[1,\sigma]$.
Assume we assign a plain encoding of $\lceil \lg \sigma \rceil$ bits to the
symbols. Let us call $S[i]\langle j\rangle$ the $j$th most significant bit of
the code associated with $S[i]$. The $\wt$ construction proceeds as follows:
At the root node it splits the alphabet $\Sigma$ into two halves, $\Sigma_1$
and $\Sigma_2$. A symbol belongs to $\Sigma_1$ iff $S[i]\langle 1
\rangle=0$, and to $\Sigma_2$ otherwise. We store that
information in a bitmap $B[1,n]$ associated with the node, being $B[i]=0$ iff
$S[i]\in\Sigma_1$ and $1$ otherwise. The left child of the root will then
represent the subsequence of $S$ containing symbols in $\Sigma_1$, while the
right node will do the same with $\Sigma_2$. The process is then recursively
repeated in both children until the alphabet of the current node is unary. The height the $\wt$ is $\lceil \lg \sigma \rceil$. 

The only information we need to store from a $\wt$ are the bitmaps stored in the internal tree nodes and the tree pointers. The total space for the sequences is $n\lceil \lg \sigma \rceil$ bits, while for tree pointers we use $O(\sigma \log n)$ bits. Thus, the total space becomes $n\lg \sigma + O(n + \sigma \log n)$ bits. 

Although we will focus on the binary case, we can generalize the concept of
$\wt$ to the multi-ary case: Instead of recursively dividing the alphabet into
two halves, we can split it into $2^b$ disjoint sets. This is known as
Multi-ary $\wt$ or $\mwt$. Now the internal $\mwt$ nodes store sequences drawn
over alphabet $[1,2^b]$ instead of bitmaps, and the height is reduced to 
$\lceil (\log \sigma)/b \rceil$.

% ,instead of a binary $\wt$, we can build an r-ary tree (Multi-ary $\wt$ or
% $\mwt$). In this case, at each node we divide the alphabet into $r$ disjoint sets to obtain a $r$-ary tree (). Instead of bitmaps, nodes will hold sequences with alphabet $r$ and the tree will have a maximum height of $\lceil \log_{r} \sigma \rceil$.

% The only information we need to store from a $\wt$ or $\mwt$ is the inner bitmaps or sequences stored in the tree nodes and the tree pointers. The total space for inner sequences is $n\lceil \log \sigma \rceil$ bits while for tree pointers is $O(\sigma \log_{r} n)$ bits. Thus, the total space becomes $n\log \sigma + O(n + \sigma \log_{r} n)$ bits. 

Algorithm~\ref{alg:wt} shows how $\rsa$ queries on $S$ are built on $\rsa$
queries on the bitmaps or sequences of the $\mwt$ of $S$.
A key aspect in $\wt$'s performance is how we represent those bitmaps or
sequences. In the binary case (Section~\ref{sec:bitmaps}), if we use $\cm$ for
bitmaps, the total space is $n\lg\sigma + o(n\log\sigma) +O(\sigma\log n)$
bits and $\rsa$ times are $O(\log \sigma)$. By using $\rrr$, the time
complexity is retained (although its times are higher in practice) but the
space shrinks to $nH_0(S)+o(n\log\sigma)+O(\sigma \log n)$ bits. Zero-order
compression is also obtained by using a Huffman \cite{Huffman:1952} encoding
for the symbols and giving the $\wt$ the shape of the Huffman tree: using
$\cm$ for the bitmaps results in $n(H_0(S)+1)(1+o(1))+O(\sigma \log n)$ bits,
whereas using $\rrr$ for the bitmaps the space becomes
$nH_0(S)(1+o(1))+O(\sigma \log n)$ bits \cite{BN13}. This solution is called 
Huffman-shaped $\wt$ ($\wth$). The main advantage of using a $\wth$ is that,
if queries follow the same statistical distribution of symbols, then the
average query time for any $\rsa$ query becomes $O(1+H_0(S))$ instead of
$O(\log \sigma)$ \cite{BN13}. A Huffman-shaped multi-ary wavelet tree will
be called $\mwth$. For any $2^b = o(\log n/\log\log n)$, the $\mwth$ retains 
the same space complexities of a $\wth$, whereas the worst-case and average 
query time are divided by $b$ \cite{BN13}.

\begin{figure}[t!]
 \centering
 \includegraphics[width=0.6\textwidth]{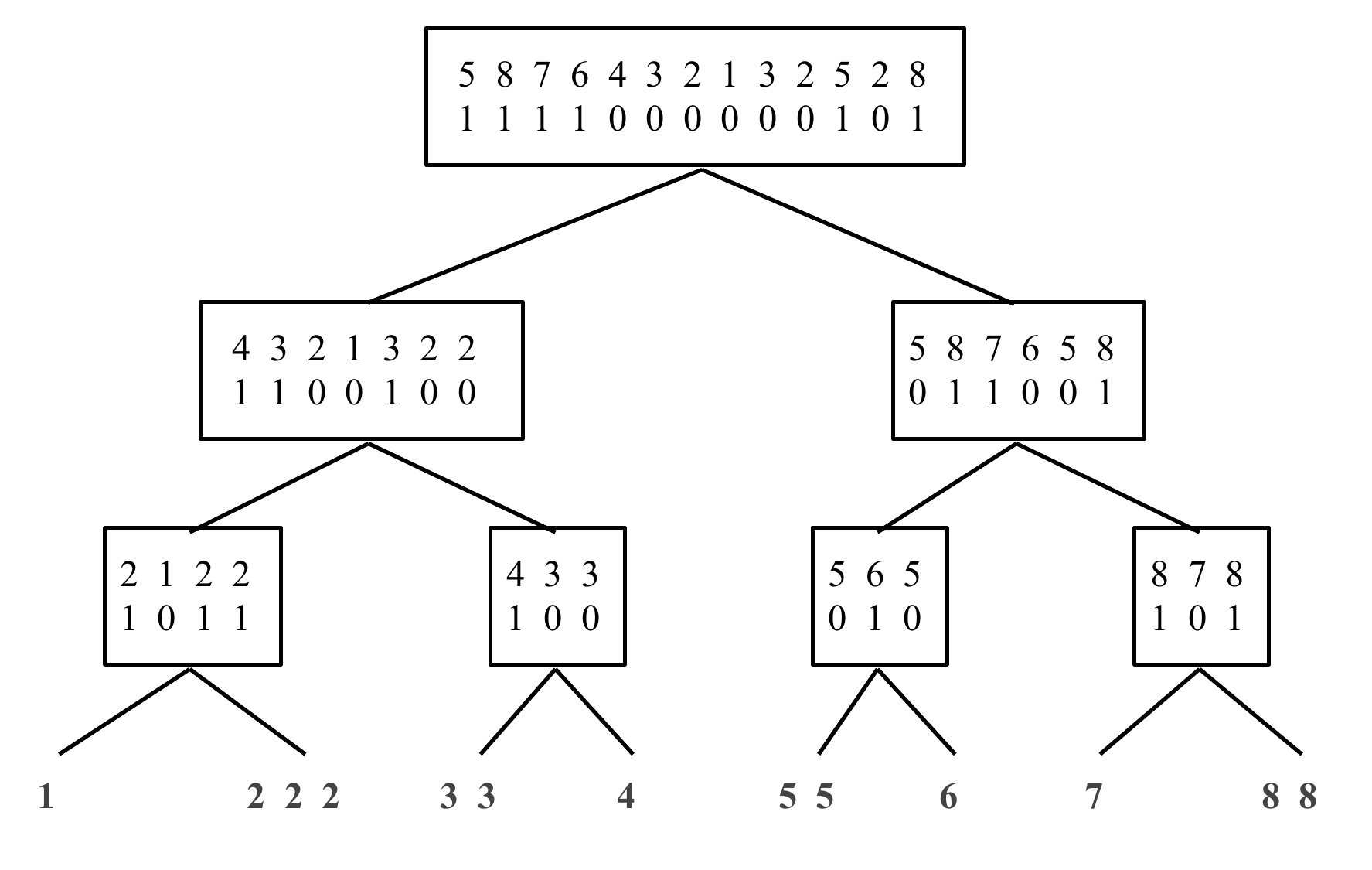}
 \includegraphics[width=0.48\textwidth]{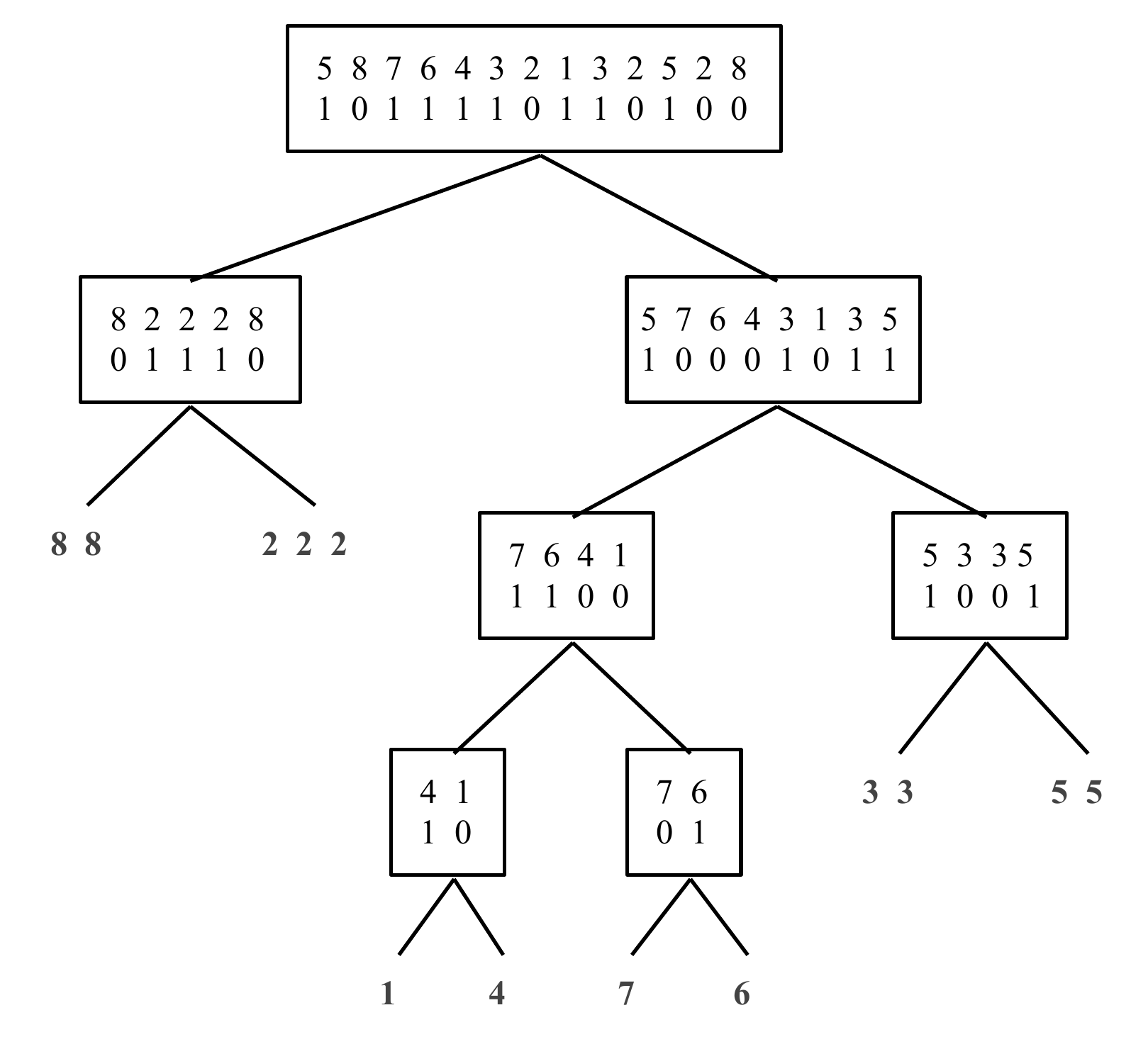} 
\begin{minipage}{0.48\textwidth}
  \vspace*{-6cm}
 \includegraphics[width=\textwidth]{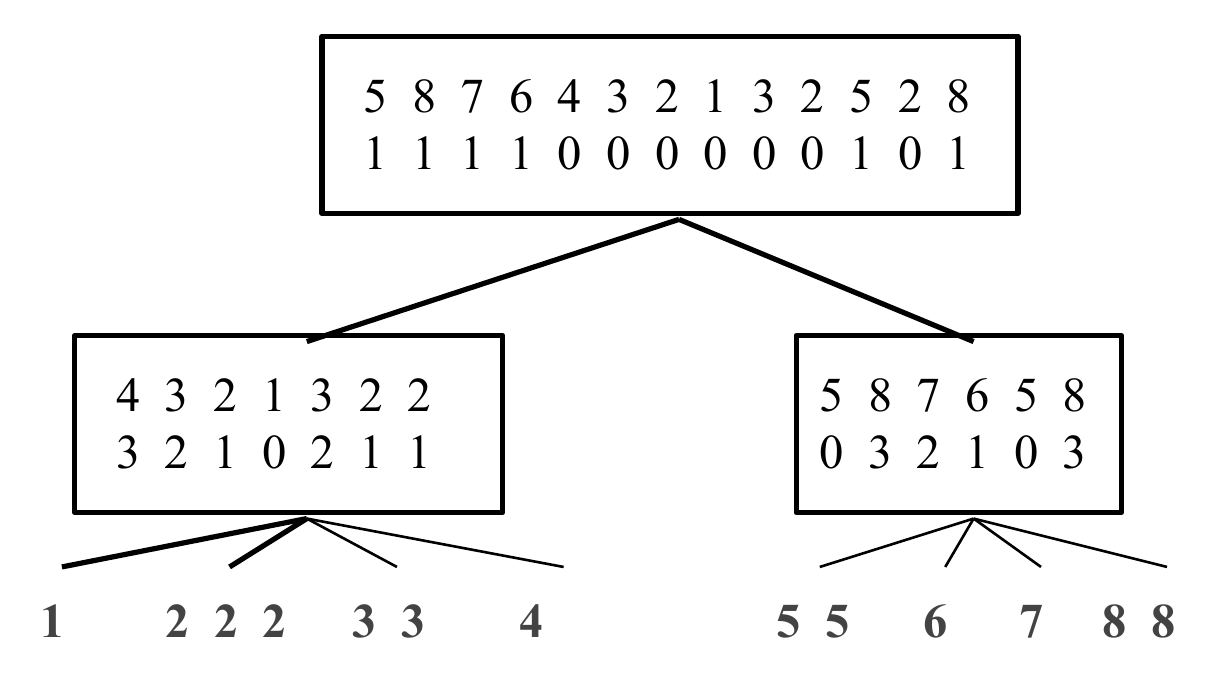}
\end{minipage}
  \caption{Wavelet tree representations of sequence
$S = 5876432132528$. On the top a $\wt$, on the bottom left $\wth$, and on the 
bottom right a $\mwt$ with $2^b=4$ (the first level can only have arity 2).}
 \label{fig:wt.hswt}
 \end{figure}

Figure \ref{fig:wt.hswt} exemplifies all these wavelet tree variants.

\begin{algorithm}[t]
\caption{Standard $\mwt$ algorithms on a sequence $S$. The sequence associated
with node $v$ is $S_v$ and its $i$th child is $v_i$. For $\access(S,i)$ we
return $\mathbf{acc}(root,i,0)$, where $root$ is the $\mwt$ root;
$\rank_a(S,i)$ returns $\mathbf{rnk}(root,a,i,\lceil (\log \sigma)/b\rceil)$;
and $\select_a(S,j)$ returns $\mathbf{sel}(root,a,j,\lceil(\log \sigma)/b
\rceil)$. Function $\mathit{leaf}(v)$ returns whether node $v$ is a
leaf, and $chunk(a,b,\ell)=(a \gg (\ell-1)b) ~\&~ ((1 \ll b)-1)$ takes the $\ell$th chunk of $b$ most significant bits from $a$.}
\label{alg:wt}
\small
\begin{tabular}{ccc}

\begin{minipage}{0.29\textwidth}

$\mathbf{acc}(v,i,c)$
\begin{algorithmic}
\vspace{-0.4cm}
\IF{$\mathit{leaf}(v)$}
   \RET $c$
\ENDIF
\STATE $c\leftarrow (c \ll b) ~|~ S_v[i]$
% \IF{$B_v[i]=0$}
   \STATE $i \leftarrow \rank_{S_v[i]}(S_v,i)$
   \RET $\mathbf{acc}(v_{S_v[i]},i,c)$
% \ELSE
%    \STATE $i \leftarrow \rank_1(B_v,i)$
%    \RET $\mathbf{acc}(v_r,i,c|1)$
% \ENDIF
\end{algorithmic}
\end{minipage}

&

\begin{minipage}{0.33\textwidth}
$\mathbf{rnk}(v,a,i,\ell)$
\begin{algorithmic}
\vspace{-0.4cm}
\IF{$\mathit{leaf}(v)$}
   \RET $i$
\ENDIF
% \IF{$a\langle \ell \rangle=0$}
   \STATE $c \leftarrow chunk(a,b,\ell)$
   \STATE $i \leftarrow \rank_c(S_v,i)$
   \RET $\mathbf{rnk}(v_c,a,i,\ell-1)$
% \ELSE
%    \STATE $i \leftarrow \rank_1(B_v,i)$
%    \RET $\mathbf{rnk}(v_r,a,i,\ell+1)$
% \ENDIF
\end{algorithmic}
\end{minipage}

&

\begin{minipage}{0.33\textwidth}
$\mathbf{sel}(v,a,j,\ell)$
\vspace{-0.4cm}
\begin{algorithmic}
\IF{$\mathit{leaf}(v)$}
   \RET $j$
\ENDIF
% \IF{$a\langle \ell \rangle=0$}
   \STATE $c \leftarrow chunk(a,b,\ell)$
   \STATE $j \leftarrow \mathbf{sel}(v_c,a,j,\ell-1)$
   \RET $\select_c(S_v,j)$
% \ELSE
%    \STATE $j \leftarrow \mathbf{sel}(v_r,a,j,\ell+1)$
%    \RET $\select_1(B_v,j)$
% \ENDIF
\end{algorithmic}
\end{minipage}
\end{tabular}
\end{algorithm}

%  \begin{figure}[t!]
%  \centering
% \begin{minipage}{0.49\textwidth}
% \vspace*{-4cm}
% \begin{center}
%  \includegraphics[width=\textwidth]{wt.pdf}
%  \includegraphics[width=0.7\textwidth]{mwt.pdf}
% \end{center}
% \end{minipage}
%  \includegraphics[width=0.49\textwidth]{hwt.pdf}
%  \includegraphics[width=0.49\textwidth]{lwt.pdf}
%  \includegraphics[scale=0.727]{wm.pdf}
%   \caption{Wavelet tree/matrix representations of sequence
% $S = 5876432132528$. On the top left a $\wt$, below it a $\mwt$ with $2^b=4$ in
% its first level,
% on the top right a $\wth$, on the bottom left a levelwise $\wt$, and on the
% bottom right a $\wm$.}
%  \label{fig:wt.hswt}
%  \end{figure}

 \begin{figure}[t!]
 \centering
 \includegraphics[width=0.49\textwidth]{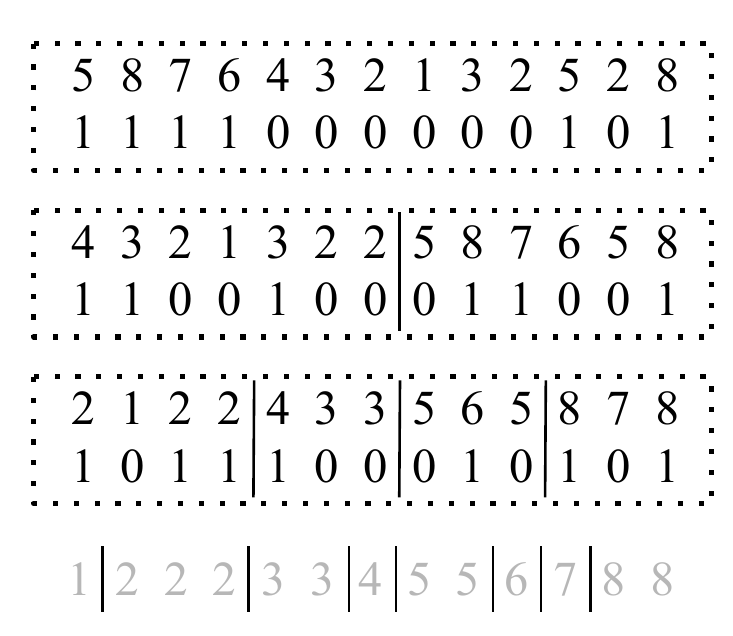}
 \includegraphics[scale=0.727]{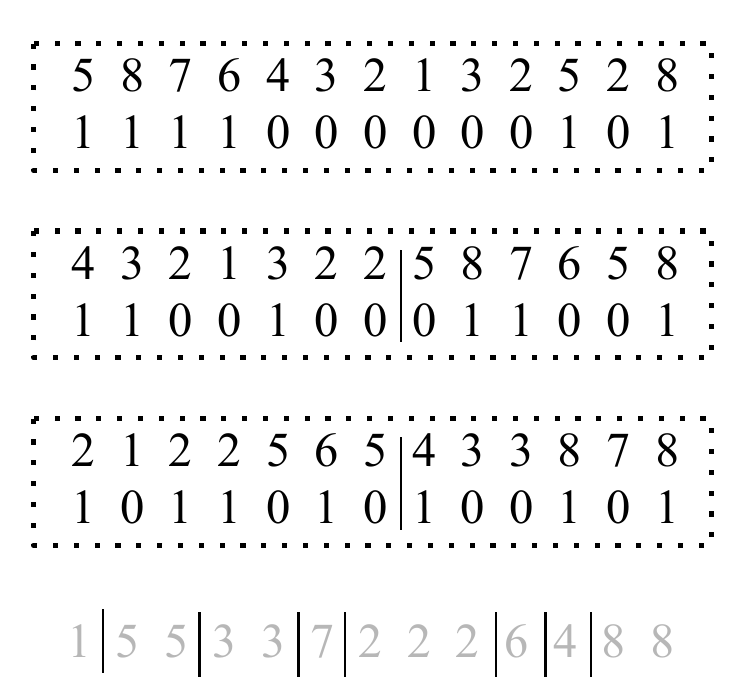}
  \caption{Wavelet tree/matrix representations of sequence
$S = 5876432132528$. On the left a levelwise $\wt$, and on the right a $\wm$.}
 \label{fig:lwt.wm}
 \end{figure}

\subsection{Wavelet matrix}\label{sec:wm}

If $\sigma$ is close to $n$, the $O(\sigma \log n)$ bits to store the tree pointers in a $\wt$ will become dominant. To skip this term, the {\em levelwise} $\wt$ \cite{MNlatin06} concatenates all the bitmaps at the same depth and simulates the tree pointers with $\rsa$ operations. This variant obtains the same space of the $\wt$ or $\mwt$ but without the $O(\sigma \log n)$ term. The time performance is asymptotically the same, but it is slower in practice because pointers are simulated. More recently, the {\em wavelet matrix} ($\wm$) \cite{CNOis14} was proposed, which speeds up the levelwise $\wt$ by reshuffling the bits at each level in a different way so that the tree pointers can be simulated with fewer $\rsa$ operations. 
Assume we start with $S_l=S$ at level $l=1$; then the wavelet matrix is built
as follows:

\begin{enumerate}
\item Build a single bitmap $B_l[1,n]$ where $B_l[i]= S_l[i]\langle l\rangle$;
\item Compute $z_l=\rank_0(B_l,n)$; 
\item Build sequence $S_{l+1}$ such that, for $k \le z_l$, $S_{l+1}[k] =S_l[\select_0(B_l,k)]$, and for $k > z_l$, $S_{l+1}[k] = S_l[\select_1(B_l,k-z_l)]$;
\item Repeat the process until $l=\lceil \log \sigma \rceil$. 
\end{enumerate}

This reshuffling of the bits of $S[i]\langle j\rangle$, akin to radix sorting 
the symbols of $S$, uses $n\lceil\lg\sigma\rceil$ bits in total (plus 
$\lg n\lg\sigma$ for the values $z_l$). Therefore, 
the total space of the $\wm$ is $n \lg \sigma+o(n\log\sigma)$. 
Figure~\ref{fig:lwt.wm} exemplifies the levelwise $\wt$ and the $\wm$.
As in the case of the $\wt$, this space can be further reduced to
$nH_0(S)+o(n\log \sigma)$ if we use $\rrr$ (Section~\ref{sec:bitmaps}) to
compress the bitmaps, or to $n(H_0(S)+1)(1+o(1))+O(\sigma\log n)$ by using
plain bitmaps ($\cm$) and giving Huffman shape to the $\wm$ \cite{CNOis14}
(Section~\ref{sec:wt}). The latter is called a $\wmh$.
We can also convert a $\mwt$ into a multi-ary $\wm$ ($\mwm$) by increasing the 
number of counters $z_l$ at each level: if $2^b$ is the arity, we need $2^b-1$
counters $z_l$ per level. 

Algorithm~\ref{alg:wm} shows how the algorithms are implemented on a $\wm$.
Although better than the levelwise $\wt$, it still requires more operations on
the bitmaps than the $\wt$.

\begin{algorithm}[t]
\caption{Standard $\wm$ algorithms on a sequence $S$. The bitmap at level $l$
is denoted by $B_{l}$ and $z_{l}=\rank_0(B_{l},n)$. For $\access(S,i)$ we
return $\mathbf{acc}(1,i,0)$; $\rank_a(S,i)$ returns $\mathbf{rnk}(1,a,i,0)$;
and $\select_a(S,j)$ returns $\mathbf{sel}(1,a,j,0)$.}
\label{alg:wm}
\small
\begin{tabular}{ccc}

$\!\!\!$\begin{minipage}{0.34\textwidth}

$\mathbf{acc}(l,i,c)$
\begin{algorithmic}
%\vspace{-0.4cm}
	\IF{$l = \lceil \lg \sigma \rceil$}
		\RET $c$
	\ENDIF
	\STATE $c\leftarrow (c \ll 1) ~|~ B_{l}[i]$

   \STATE $i\!\leftarrow\!\rank_{B_{l}\![i]}\!(B_{l}{,}i){+}z_l{\cdot}B_l[i]$
   \RET $\mathbf{acc}(l+1,i,c)$ 
\end{algorithmic}
\end{minipage}

&

$\!\!\!\!\!\!\!\!\!\!\!\!\!\!$\begin{minipage}{0.33\textwidth}
$\mathbf{rnk}(l,a,i,p)$
\begin{algorithmic}
\vspace{-0.4cm}
	\IF{$l = \lceil \lg \sigma \rceil$}
		\RET $i-p$
	\ENDIF
	\STATE $i\!\leftarrow\!\rank_{a\langle l \rangle}\!(B_{l}{,}i){+}z_l{\cdot}%
a\!\langle l \rangle$
	\STATE $p\!\leftarrow\!\rank_{a\!\langle l \rangle}\!(B_{l}{,}p){+}z_l{\cdot}%
a\langle l \rangle$
  \RET $\mathbf{rnk}(l+1,a,i,p)$
% \ELSE
\end{algorithmic}
\end{minipage}

&

$\!\!\!\!\!\!\!\!\!\!\!\!\!\!$\begin{minipage}{0.33\textwidth}
$\mathbf{sel}(l,a,j,p)$
%\vspace{-0.4cm}
\begin{algorithmic}
	\IF{$l = \lceil \lg \sigma \rceil$}
		\RET $p+j$
	\ENDIF
	\STATE $p\!\leftarrow\!\rank_{a\!\langle l
\rangle}(B_{l}{,}p){+}z_l{\cdot}a\langle l \rangle$
	\STATE $j\leftarrow \mathbf{sel}(l+1,a,j,p)$\vspace*{-4mm}
	\RET $\select_{a\!\langle l \rangle}\!(B_{l}{,}j{-}z_l{\cdot}a\langle l \rangle)$
\end{algorithmic}
\end{minipage}
\end{tabular}
\end{algorithm}

\subsection{Alphabet partitioning}\label{sec:ap}

\begin{figure}[t]
 \centering
 \includegraphics[width=0.7\textwidth]{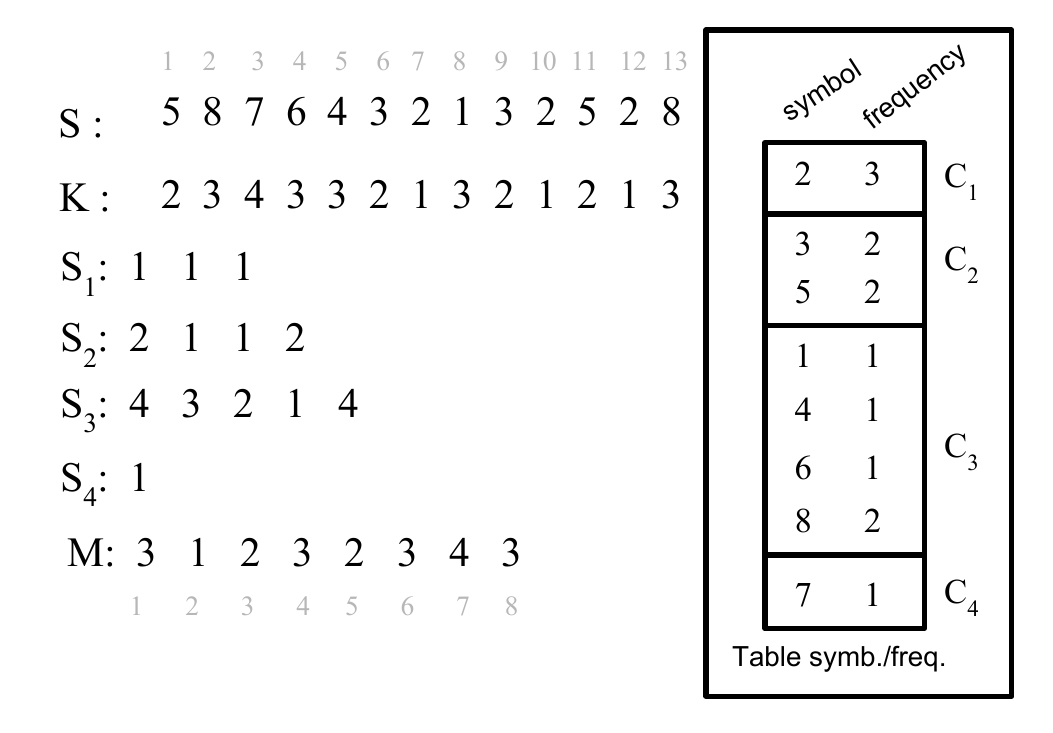}
  \caption{Alphabet partitioning example.}
 \label{fig:ap}
 \end{figure}

%introduction
An alternative solution for $\rsa$ queries over large alphabets is {\em Alphabet Partitioning} ($\ap$) \cite{BCGNNalgor13}, which obtains $nH_0(S)+o(n(H_0(S)+1))$ bits and supports $\rsa$ operations in $O(\log\log\sigma)$ time.
%construction
The main idea is to partition $\Sigma$ into several subalphabets $\Sigma_j$,
and $S$ into the corresponding subsequences $S_j$, each defined over
$\Sigma_j$ (see Figure~\ref{fig:ap}). The practical variant sorts the $\sigma$
symbols by decreasing frequency and then splits that sequence into disjoint
subsets, or subalphabets, of increasingly exponential size, so that $\Sigma_j$
contains the $2^{j-1}$th to the $(2^j-1)$th most frequent symbols. The
information on the partitioning is kept in a sequence $M$, where $M[i]=j$ iff
$i\in \Sigma_j$. A new string $K[1,n]$ indicates the subalphabet each symbol
of $S$ belongs to: $K[i] = M[S[i]]$. Analogously to wavelet trees, the
sequences $S_j$ are defined as $S_j[i] = \rank_j(M,S[\select_j(K,i)])$. Note that the
number of subalphabets is at most $\lfloor \lg \sigma\rfloor +1$, and this is
the alphabet size of $M$ and $K$. Therefore, a binary $\wt$ representation of
$M$ and $K$ handles $\rsa$ operations in time $O(\log\log\sigma)$. Further,
the symbols in each $\Sigma_j$ are of roughly the same frequency, thus a fast
compact (but not compressed) representation of $S_j$ ($\gol$ \cite{GMR06})
yields $O(\log\log\sigma)$ time and retains the statistical compression of $S$
\cite{BCGNNalgor13}.

%operations
Algorithm~\ref{alg:ap} shows how the $\rsa$ operations on $S$ translate into 
$\rsa$ operations on $M$, $K$, and on some subsequence $S_j$, thus obtaining 
$O(\log\log\sigma)$ times. In practice, the sequences $S_j$ with the smallest
alphabets are better integrated directly into the $\wt$ of $K$.

\begin{algorithm}[t]
\caption{Alphabet partition algorithms for $\access$, $\rank$, and $\select$}
\label{alg:ap}
\small
\begin{tabular}{ccc}
\begin{minipage}{0.30\textwidth}
	$\access(S,i)$
		\vspace{-0.4cm}
		\begin{algorithmic}
		\STATE $j \leftarrow K[i]$
		\STATE $v \leftarrow S_j[\rank_j(K,i)]$
		\RET  $\select_j(M,v)$ \\ \ 
	\end{algorithmic}
\end{minipage}

&

\begin{minipage}{0.33\textwidth}
	$\rank_a(S,i)$
	%\vspace{-0.1cm}
	\begin{algorithmic}
		\STATE $j \leftarrow M[a]$
		\STATE $v \leftarrow \rank_j(M,a)$
		\STATE $r \leftarrow \rank_j(K,i)$
		\RET   $\rank_v(S_j,r)$
	\end{algorithmic}
\end{minipage}	

&
\begin{minipage}{0.33\textwidth}
	$\select_a(S,i)$
	\vspace{-0.4cm}
	\begin{algorithmic}
		\STATE $j \leftarrow M[a]$
		\STATE $v \leftarrow \rank_j(M,a)$
		\STATE $s \leftarrow \select_v(S_j,i)$
		\RET $\select_j(K,s)$
	\end{algorithmic}
\end{minipage}
\end{tabular}
\end{algorithm}

There are other representations that improve upon this solution in theory, but
are unlikely to do better in practice. For example, it is possible to retain
similar time complexities while reducing the space to $nH_k(S)+o(n\log\sigma)$ 
bits, for any $k=o(\log_\sigma n)$ \cite{BHMR11,GOR10}. It is also possible,
within zero-order entropy space, to
support $\access$ and $\select$ in $O(1)$ and any $\omega(1)$ time, or vice versa,
and $\rank$ in time $O(\log\log_w\sigma)$, on a RAM machine with word 
size $w$, which matches lower bounds \cite{BNesa12}.

% \vspace*{-2mm}
\subsection{RePair compressed $\wt$} \label{sec:wtrp}

As far as we know, what we will call $\wtrp$ \cite{NPVjea13} is the only solution to support $\rsa$ on grammar-compressed sequences. The structure is a levelwise $\wt$ where 
each bitmap $B_l$ is compressed with $\rpbmp$ (Section~\ref{sec:rpb}). The rationale is that the repetitiveness of $S$ is reflected in the bitmaps of the $\wt$.

However, since the $\wt$ construction splits the alphabet at each level, those repetitions are cut into shorter ones at each new level, and become blurred after some depth.
Therefore, the bitmaps of the first few $\wt$ levels are likely to be compressible with RePair, while the remaining ones are not. The authors \cite{NPVjea13} use at each level $l$ the technique to represent $B_l$ that yields the least space, $\rpbmp$, $\rrr$, or $\cm$ (Sections~\ref{sec:bitmaps} and \ref{sec:rpb}). In case of a highly compressible sequence, the space can be drastically reduced, but the search performance degrades by one or more orders of magnitude compared to using $\cm$ or $\rrr$: If all the levels use $\rpbmp$, the $\rsa$ time becomes $O(\log \sigma \log n)$. 

On the other hand, as repetitiveness is destroyed at deeper levels, the total
space is far from that of a plain RePair compression of $S$. A worst-case
analysis, albeit pessimistic, can be made as follows: Each node stores a
subsequence of $S$, whose alphabet is mapped onto a binary one (or of size $r$
in an $r$-ary wavelet tree). We could then take the same grammar that
compresses $S$ for each node, remove all the terminal symbols not represented
in that node, and map the others onto $\{0,1\}$ or $[1,r]$. This is not the
best grammar for that node, but it is correct and at most of the same size $g$
of the original one. Therefore, each node can be grammar-compressed to at most
$O(g\log n)$ bits, and summed over all the wavelet tree nodes, this yields
$O(g\sigma\log n)$. Therefore, the size grows at most linearly with $\sigma$.

\subsection{Other grammar-compressed $\rsa$ solutions}

Let a grammar compressor produce a grammar of size $g$ with $r$ nonterminals 
for $S[1,n]$. Thus $S$ can be represented in $g\lg (r+\sigma)$ bits. Bille et 
al.~\cite{BSODA11} show how to represent $S$ using $O(g\log n)$ bits so that 
$\access(S,i)$ is answered in $O(\log n)$ time. This time is essentially optimal
\cite{VY13}: any structure using $g^{O(1)}\log n$ bits requires 
$\Omega(\log^{1-\epsilon} n/\log g)$ time for $\access$, for any $\epsilon>0$.
If $S$ is not very compressible and $g = \Omega(n^\alpha)$ for some constant 
$\alpha$, then the time is $\Omega(\log n / \log\log n)$ for any structure 
using $O(n\,\mathrm{polylog}\,n)$ bits. 

As said, we are not aware of any previous $\rsa$ structure building on grammar
compression apart from $\wtrp$ \cite{NPVjea13}, which
handles queries in $O(\log\sigma\log n)$ time and uses $O(g\sigma\log n)$ bits. 
Our simplest variant, $\gcc$, obtains $O(\log n)$ time for the three $\rsa$ 
operations within $O(g\sigma\log n)$ bits. For larger alphabets, we can 
increase the time to $O(\log\sigma\log n)$ and keep the worst-case space in
$O(g\sigma\log n)$ bits (yet in practice the solution takes less space and 
time than $\wtrp$, and less space than $\gcc$). 
Alternatively, we can retain
the $O(\log n)$ time but lose the space guarantee.

After the publication of the conference version of our article
\cite{NOspire14}, Belazzougui et al.~\cite{BPTesa15} gave more theoretical 
support to our results. They obtained our same $O(\log n)$ time for $\rsa$
operations with $O(g \sigma \log n)$ bits on arbitrary grammars of size $g$
(not only balanced ones). They also show how to obtain $O(\log n / \log\log n)$
time using $O(g\sigma\log(n/g)\log^{1+\epsilon} n)$ bits, for any constant
$\epsilon>0$. Most importantly, they prove that it is unlikely that these times
for $\rank$ and $\select$ can be significantly improved, since long-standing
reachability problems on graphs would then be improved as well. This shows that 
the time complexity of their (and our) solutions are essentially the best one 
can expect.

Lempel-Ziv \cite{LZ76,ZL77} compression is able to outperform grammar 
compression 
\cite{KY00,CLLPPSS05}, because the number of phrases it generates is never 
larger than the size $g$ of the best possible grammar. However, its support 
for $\rsa$ queries is even more difficult. Let $z$ be the number of phrases 
into which a 
Lempel-Ziv parser factors $S$. Then a Lempel-Ziv compressor can represent $S$ 
in $z(\lg n + \lg\sigma)$ bits. We are not aware of any scheme supporting 
$O(\log n)$ time access within $O(z\log n)$ bits. Gagie et al.~\cite{GGKNP14}
do achieve this time, but they use $O(z\log n \log(n/z))$ bits, which is
superlinear in the compressed size of $S$. A more recent work \cite{BGGKOPT15} 
supports $\access$ and $\rank$ in time $O(\log (n/z))$ and $\select$ in time
$O(\log(n/z)\log\log n)$. The lower bound \cite{VY13} also
holds for this compression, replacing $g$ by $z$.

\subsection{Directly Addressable Codes}\label{chap:pconcetps:stat:dacs}

A Directly Addressable Code~\cite{BLN12} (DAC) is a variable-length encoding
for integers that supports direct access operations ($\access$) efficiently,
but not $\rank$ and $\select$. Assume we have to encode a sequence
$X=x_1\dots x_n$ of integers and are given a chunk size $b$. Then we divide
each $x_i=X[i]$ into $j=\lceil (\lfloor \lg x_i \rfloor+1) /b\rceil$ chunks,
from least to most significant. At the most significant position of each chunk 
we will prepend a bit $0$ if that chunk is the last one, and a $1$ otherwise. 
Therefore, the number $x_i$ is encoded as
$$b_{1,i}a_{1,i}b_{2,i}a_{2,i}\dots b_{k,i}a_{k,i},$$
where $b_{j,i}$ is the bit prepended to the chunk $a_{j,i}=x_i\langle
jb,(j-1)b+1\rangle$.
Note the similarity with the VByte codes of
Section~\ref{sec:vbyte}. 

\begin{figure}[]
\centering
\includegraphics[scale=0.8]{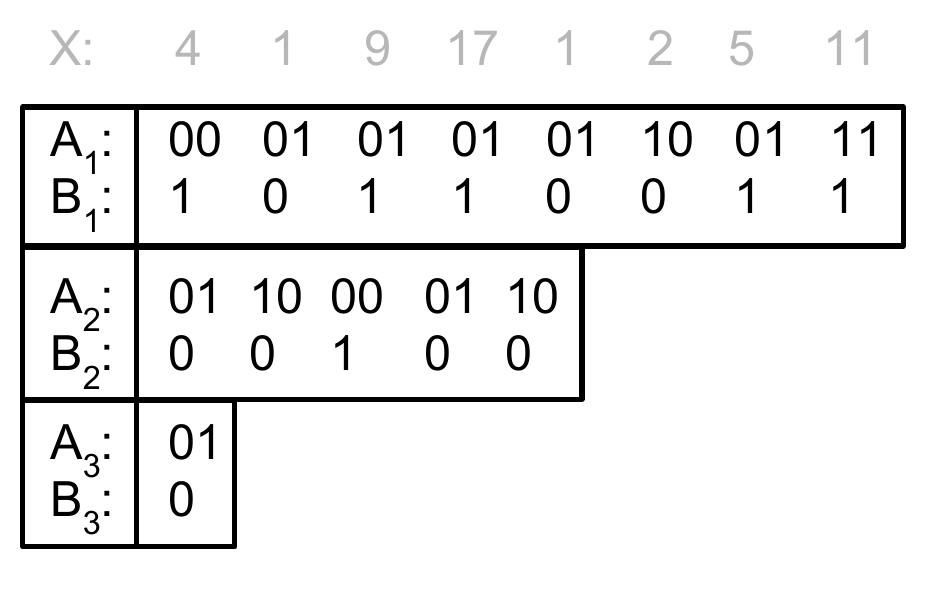}
\caption[Example of a DAC]{Example of a DAC for the sequence $X=4,1,9,17,1,2,5,11$ and $b=2$.}
\label{fig.dac}
\end{figure}

Instead of concatenating the encoding of $x_{i+1}$
after that of $x_i$, however, we build a multi-layer data structure. At each
layer $l \ge 1$, we concatenate the $l$th chunks of all the numbers that have 
one, and do the same with the bits prepended to each chunk. For instance, for 
layer $l=1$ we obtain a binary sequence $B_1$ and a sequence $A_1$ as follows: 
$$B_1 = b_{1,1}b_{1,2}\dots b_{1,n},$$
$$A_1 = a_{1,1}a_{1,2}\dots a_{1,n}.$$
The next layer is then built by concatenating the second chunk of each number
that has one, and the process is repeated for $M$ layers, where
$M=\lceil  (\lfloor\lg (\max x_i) \rfloor+1) /b\rceil$. Figure~\ref{fig.dac} 
shows an example DAC over a sequence $X$ using $b=2$.  

To provide efficient
direct access, we preprocess each
sequence of prepended bits ($B_i$) to support $\rank$ and $\access$
queries in $O(1)$ time. Thus, we access $X[i]$ as follows.
We start by setting $i_1=i$, and reading $A_1[i_1]=a_{1,i_1}$. We set
$res=A_1[i_1]$ and if $B_1[i_1]=0$ we are done because this chunk is the last
of $x_i$. If, instead, $B_1[i_1]=1$, $x_i$ continues in the next layer. To
compute the position of the next chunk in the next layer we set
$i_2=\rank_1(B_1,i)$. In the second layer we concatenate $A_2[i_2]$ with the
current result: $res= A_2[i_2]A_1[i_1]$ and then check $B_2[i_2]$, repeating
the process until we get a $B_k[i_k]=0$. Then the time to extract an
an element of $X$ when represented with a DAC is worst-case $O(M)$.
 
It is possible to define a different $b$ value for each level, and to choose
them so as to optimize the total space used, even with a restriction on $M$
\cite{BLN12}.

\section{Efficient $\rsa$ for Sequences on Small Alphabets}\label{sec:gcc}

Our first proposal, dubbed $\gcc$ \textit{(Grammar Compression with Counters)}
is aimed at handling $\rsa$ queries on grammar-compressed sequences with small
alphabets. We first generalize the existing solution for bitmaps ($\rpbmp$,
Section~\ref{sec:bitmaps}), to sequences with $\sigma>2$. We also introduce
several enhancements regarding how we store the additional information to
handle $\rsa$ queries. 
%TABEI , and a new technique \cite{TTS13} to further compress the dictionary of rules $R$. 
Finally, we propose two different sampling approaches that yield different space-time tradeoffs, both in theory and in practice. 

Let $(R,C)$ be the result of a balanced RePair grammar compression of $S$. We store $S_{\ell}[X]=\ell(X)$ for each grammar rule $X\in R$. In addition, we store an array of counters $S_a[X]$ for each symbol $a \in \Sigma$: $S_a[X] = \rank_a(exp(X),\ell(X))$ is the number of occurrences of $a$ in $exp(X)$.

The input sequence $S$ is also sampled according to the new scenario: each element $(p,o,rnk)$ of $S_n[1,n/s]$ is now replaced by $(p,o,lrnk[1,\sigma])$, where $lrnk[a]=\rank_a(S,L(p)-1)$ for all $a\in\Sigma$, $s$ being the sampling period.

The extra space incurred by $\sigma$ can be reduced by using the same
$\delta$-sampling of $\rpbmp$, which increases the time by a factor $\delta$.
In this case we also use the bitmap $B_d[1,r]$ that marks which rules store
counters. We further reduce the space by noting that many rules are short, and
therefore the values in $S_\ell$ and $S_a$ are usually small. We represent
them using direct access codes (DACs, recall
Section~\ref{chap:pconcetps:stat:dacs}), which store variable-length numbers while retaining direct access to them. The $o$ components of $S_n$ are also represented with DACs for the same reason.

On the other hand, the $p$ and $lrnk[1,\sigma]$ values are not small but are
increasing. We reduce their space using a two-layer strategy: we sample $S_n$
at regular intervals of length $s'$. We store $SS_n[j]=S_n[j\cdot s']$, and
then represent the values of $S_n[i]=(p,o,lrnk[1,\sigma])$ in differential
form, in array $S'_n[i] = (p',o,lrnk'[1,\sigma])$, where $p' = p - p^*$ and
$lrnk'[a] = lrnk[a]-lrnk^*[a]$, with $SS_n[\lfloor i/s'\rfloor] = (p^*,o^*,lrnk^*[1,\sigma])$. 

The total space for the $p$ and $lrnk[1,\sigma]$ components is
$O(\sigma((n/s)\log (s\cdot s') + (n / (s\cdot s')) \log n))$ bits, whereas
the $o$ components use $O((n/s)\log n)$ bits in the worst case. For example,
if we use $s' = \lg n$ and $s = \log^{O(1)}n$ (a larger value would imply an
excessively high query time), the space becomes $O(r\sigma\log n +
(n/\log^{O(1)}n)(\sigma\log\log n+\log n)) +c\lg(\sigma+r)$ bits. 

A further improvement is aimed to reduce the space on extremely repetitive sequences. In this scenario, many elements of $S_n$ may contain the same values: if a rule covers a wide range of $S$, we store the same $S_n$ values for many samples of $S$. Thus, we sample the vector $C$ instead of sampling the whole sequence $S$. Instead of $(p,o,lrnk[1,\sigma])$ we store a tuple $(i,lrnk[1,\sigma])$, where $i$ is the position where the sampled cell of $C$ starts in $S$, and $lrnk$ is computed up to $i-1$. On the other hand, the two-layer scheme cannot be 
applied, because now the samples may cover arbitrarily long ranges of $S$.

The total space with this sampling then becomes $O(r\sigma\log n + \sigma(c/s)\log n) + c\lg(\sigma+r) = O((r+c)\sigma\log n)$ bits. This removes any linear dependency on $n$ from the space formula. The size of the RePair grammar is
$g = O(r+c)$, thus the space can be written as $O(g\sigma\log n)$ bits. 

The $\rsa$ algorithms stay practically the same as for $\rpbmp$; now we use
the symbol counter of $a$ for $\rank_a$ and $\select_a$. The resulting data
structure performs $\rsa$ operations in time $O(s+\log n)$. In case $C$ is sampled instead of $S$,
there is an additional $O(\log c)$ time to binary search for the right sample.
This is still within $O(s+\log n)$. If we choose $s=O(\log n)$, then the time
is $O(\log n)$. The space is still $O(g\sigma\log n)$ if we sample $C$.

When $\sigma$ is small and the sequence is repetitive, this data structure is 
very space- and time-efficient. It outperforms $\wtrp$ \cite{NPVjea13}
(Section~\ref{sec:wtrp}) in time: $\wtrp$ takes $O(\log\sigma \log n)$ time
and our $\gcc$ uses $O(\log n)$. In terms of space, both use $O(g\sigma\log n)$
bits and perform similarly in practice. In the next section we develop a
variant for large alphabets that uses much less space in practice, even if 
the worst-case guarantee it offers is still as bad as $O(g\sigma\log n)$ bits.

\section{Efficient $\rsa$ for Sequences on Large Alphabets}\label{sec:large}

Our main idea for large alphabets is to use wavelet trees/matrices or alphabet
partitioning (Sections~\ref{sec:wt} to \ref{sec:ap}) as a mechanism to cut 
$\Sigma$ into smaller alphabets, which can then be handled with $\gcc$. This is
in the same line of $\wtrp$ (Section~\ref{sec:wtrp}), which also partitions
the alphabet. Our techniques deal better with the problem of loss of
repetitiveness when the alphabet is partitioned.

The most immediate approach is to generalize $\wtrp$ to use a $\mwt$, since now
we can use $\gcc$ on small alphabets $[1,r]$ to represent the sequences $S_v$
stored at the internal nodes of the $\mwt$. Compared to a binary $\wt$, a
$\mwt$ takes more advantage of repetitiveness before
splitting the alphabet, and reduces the time complexity from 
$O(\log\sigma \log n)$ to $O(\log_r \sigma \log n)$. The worst-case space is
still $O(g\sigma\log n)$ bits. The use of a $\wm$ requires only $\log_r
\sigma$ grammars, one per level, but still the guarantee on their total size 
is the same.

A less obvious way to use $\gcc$ is to combine it with $\ap$ (Section~\ref{sec:ap}). Note that the string $K$ is a projection of $S$, and therefore it retains
all its repetitiveness. Further, it contains a small alphabet, of size 
$\lg\sigma$, and therefore we can use $\gcc$ on it. The resulting 
representation takes at most $O(g \log\sigma \log n)$ bits.

The other important sequences are the $S_j$, which have alphabets of size
$2^{j-1}$. For the smallest $j$, this is small enough to use $\gcc$ as well.
For larger $j$, however, we must resort to other representations, like
$\wtrp$, $\gol$, or $\wt/\wm$, depending on how compressible they are.

An interesting fact of $\ap$ is that it groups symbols of
approximately the same frequency. The symbols participating in the most
repetitive parts of $S$ have a good chance of having similar frequencies and
thus of belonging to the same subalphabet $S_j$, where their repetitiveness will
be preserved. On the other hand, the larger alphabets, where $\gcc$ cannot be
applied, are likely to contain less frequent symbols, whose representation
using faster structures like $\gol$ or $\wt/\wm$ do not miss very important
opportunities to exploit repetitiveness.

Note that, if we do not use $\wtrp$ for the larger subalphabets, then the time
performance for $\rsa$ queries stays within $O(\log n)$, independently of the 
alphabet size. In exchange, we cannot bound the size of the representation in 
terms of the size of the grammar that represents $S$. Instead, if we use
$\wtrp$, our worst-case guarantees are the same as for $\wtrp$ itself, but in
practice our structure will prove to be much better, especially in time.

\subsection{$\ap$ with $\gcc$ in practice}\label{sec:ap.rep.practice}

We introduce two new parameters for the combination of $\ap$ and $\gcc$. The first parameter, $\cut$, tells that the $2^{\cut}$ most frequent symbols will be directly represented in $K$. This parameter must be set carefully to avoid 
increasing too much the alphabet of $K$, since $K$ is represented with $\gcc$.

Our second parameter is $\cuto$, which tells how many of the first $S_j$
classes are to be represented with $\gcc$. For the remaining sequences $S_j$
we consider two options: (a) if $S_j$ is not grammar-compressible, we use
$\gol$~\cite{GMR06}, which does not compress but is very fast, or
(b) if $S_j$ is still grammar-compressible, we use $\wtrp$, which is the
grammar-based variant that performed best.

\section{Experimental Results}
\label{sec:exp}

\subsection{Setup and Datasets}\label{sec:datasets}

We used an Intel(R) Xeon(R) E5620 at $2.40$GHz with $96$GB of RAM memory,
running GNU/Linux, Ubuntu 10.04, with kernel 2.6.32-33-server.x86\_64. All our
implementations use a single thread and are coded in {\tt C++}. The compiler
is \verb|g++| version $4.7$, with \verb|-O9| optimization. We implemented our
solutions on top of \libcds\ ({\tt github.com/fclaude/libcds}) and use Navarro's implementation of RePair ({\tt www.dcc.uchile.cl/gnavarro/software/repair.tgz}).

Table \ref{table:datasets} shows statistics of interest about the datasets
used and their compressibility: length ($n$), alphabet size ($\sigma$),
zero-order entropy ($H_0$), bits per symbol (bps) obtained by RePair (RP,
assuming $(2(r-\sigma)+c)\lceil\lg r\rceil$ bits, see
Section~\ref{sec:grammar}), bps obtained by p7zip (LZ, {\tt www.7-zip.org}), a Lempel-Ziv compressor, and finally $r/n$ is the number of runs of the BWT \cite{BWT} of each dataset divided by $n$ (see Section~\ref{sec:app-fm}).

We use various DNA collections from the Repetitive Corpus of {\em
Pizza\&Chili}\footnote{\tt http://pizzachili.dcc.uchile.cl/repcorpus}. On one
hand, to study precisely the effect of repetitiveness in the performance of
our $\rsa$ proposals, we generate four synthetic collections of about 100MB:
{\tt DNA 1\%}, {\tt DNA 0.1\%}, {\tt DNA 0.01\%}, and {\tt DNA 0.001\%}. Each
{\tt DNA $p$\%} text is generated starting from 1MB of real DNA text, which is
copied 100 times, and each copied base is changed to some other value with
probability $p/100$. This simulates a genome database with different
variability between the genomes. As real genomes, we used collections {\tt
para}, {\tt influenza}, and {\tt escherichia}, also obtained from {\em
Pizza\&Chili}. From the statistics of Table~\ref{table:datasets}, we see that
{\tt para} and {\tt influenza} are actually very repetitive, while {\tt
escherichia} is not that much. Collection {\tt einstein} corresponds to
Wikipedia versions of articles about Albert Einstein in German (also available at {\em Pizza\&Chili}) and is the most repetitive dataset we have. Text {\tt einstein.words} is the same collection but regarded as a sequence of words, instead of characters. Sequence {\tt fiwiki} is a prefix of a Wikipedia repository in Finnish\footnote{\tt http://www.cs.helsinki.fi/group/suds/rlcsa} tokenized as a sequence of words instead of characters. Sequence {\tt fiwikitags} corresponds to the XML tags extracted from a prefix from the same Finnish Wikipedia repository. Finally, {\tt indochina} is a subgraph of the Web graph \textit{Indochina2004} available at the WebGraph project\footnote{{\tt http://law.dsi.unimi.it}} containing $2{,}531{,}039$ nodes and $97{,}468{,}933$ edges. Each node has an adjacency list of nodes, which is stored as a sequence of integers. Each list is separated from the next with a special separator symbol. 

\begin{table}[tb]
\centering
\begin{tabular}{l@{~}|@{~~}r@{~~}r@{~~}r@{~~}r@{~~}r@{~~}r}
\textbf{dataset} &$n/10^6$&$\sigma$&$H_0$&RP&LZ&$r/n$\\  
\midrule
{\tt DNA.1}				&	99	&	5		&	2.00	&	0.819	&	0.172		&	0.094	\\
{\tt DNA.01}			&	99	&	5		&	2.00	&	0.178	&	0.042		&	0.016	\\
{\tt DNA.001}			&	99	&	5		&	2.00	&	0.075	&	0.024		&	0.007	\\
{\tt DNA.0001}			&	99	&	5		&	2.00	&	0.063	&	0.021		&	0.006	\\
{\tt para}				&	429	&	5		&	2.12	&	0.376	&	0.191		&	0.036	\\
{\tt influenza}			&	154	&	15		&	1.97	&	0.280	&	0.132		&	0.019	\\
{\tt escherichia}		&	112	&	15		&	2.00	&	1.048	&	0.524		&	0.133	\\
{\tt fiwikitags}		&	48	&	24		&	3.37	&	0.110	&	0.219		&	0.031	\\
{\tt einstein}			&	92	&	117		&	5.04	&	0.019	&	0.009		&	0.001	\\
{\tt software}			&  210  &   134 	& 	4.69 	&	0.139  	&   0.214		& 	0.009	\\
{\tt einstein.words}	&	17	&	8,046	&	9.92	&	0.076	&	0.003		&	0.001	\\
{\tt fiwiki}			&	86	&	102,423	&	11.06	&	0.235	&	0.034		&	0.008	\\
{\tt indochina}			&  100	& 2,576,118	&	15.39	&	1.906	&	0.159		&	0.076	\\
\end{tabular}

\vspace{2mm}
\caption{Statistics of the datasets. The length is measured in millions of
symbols and rounded.}
\label{table:datasets}
\end{table}

\subsection{Parameterizing the data structures}\label{sec:param}

We compare our data structures with several others. The list of structures compared, along with the parameters used, is listed next. These parameter ranges are chosen because they have been proved adequate in previous work, or because we have obtained the best space/time tradeoffs with them.
\begin{itemize}
\item $\gccn$ is our structure for small alphabets where we sample $S$ at
regular intervals. We set the sampling rate to
$s=\{2^{10},2^{11},2^{12},2^{13},2^{14}\}$, the rule sampling to
$\delta=\{0,1,2,4\}$, and the superblock sampling to $s'=\{5,8\}$.  

\item $\gccc$ is our structure for small alphabets where we sample $C$ at
regular intervals. We set the sampling rate to
$s=\{2^{6},2^{7},2^{8},2^{9},2^{10}\}$ and the rule sampling to
$\delta=\{0,1,2,4\}$. %, and the super sampling to $s'=\{5,8\}$. omitido

\item $\{\wt | \wm | \wth | \wmh \}$.$\{\cm|\rrr\}$ is a wavelet tree, a wavelet matrix, a Huffman-shaped wavelet tree or a Huffman-shaped Wavelet Matrix with bitmaps represented either with $\cm$ or $\rrr$. For $\cm$ we use the  
implementation~\cite{GGMN05} with one level of counters over the plain bitmap,
while $\rrr$ corresponds to the implementation~\cite{CN08} of the compressed 
bitmaps of Raman et al.~\cite{RRR02}. In both cases, the sampling rate for 
the counters was set to $\{32,64,128\}$. 

\item$\{\wt | \wm | \wth | \wmh \}.\M$ are the $\wt$, $\wm$, $\wth$ or $\wmh$,
with the bitmaps compressed with RePair. Therefore, $\wm.\M$ is
equivalent to $\wtrp$ \cite{NPVjea13}, but with our improved implementation
using a wavelet matrix and $\gcc$ for the bitmaps.
As in $\wtrp$, we use several bitmap representations depending on the compressibility of the bitmap: $\gcc$ varying the parameters as described above, $\rrr$ or $\cm$ with sampling set to $32$. We choose the one using the least space among these.

\item $\ap$ is a plain alphabet partitioning implementation
\cite{BCGNNalgor13}. We used parameter values
$\cut=\{2^{3},2^{4},2^{5},2^{6}\}$ and $\cuto=\{1,3,5\}$. The sequence $K$ is
represented with $\wt.\rrr$ with sampling set to $32$. The sequences $S_j$ are represented with $\gol$ using the default configuration provided in the libcds tutorial\footnote{\tt https://github.com/fclaude/libcds/blob/master/tutorial/tutorial.pdf}.

\item $\aprep.\{\wmrp|\gol\}$ is our $\ap$-based variant for large alphabets. We
use the same values $\cut$ and $\cuto$ as for $\ap$. The sequence $K$ and the first $\cuto$ sequences $S_j$ are represented with $\gcc$. The remaining sequences $S_j$ are represented either with $\wm.\M$ or with $\gol$, using their already described configurations.

\item $\mwth.\M$ is a $\mwth$ using RePair-compressed sequences in the nodes. 
As for $\aprep$, we use two different representations for the node sequences. 
The first $\cut=\{2,3,4\}$ levels are represented with $\gcc$, and the rest 
with a $\wt.\rrr$ with fixed sampling $32$. We tested arities in $\{4,8,16\}$. 
We did not try combining with the $\wm$ because it is slower (requires more
operations) and the overhead of $\sigma/2^b$ nodes is not as large as for
$\sigma$ nodes of the binary case. Also, the Huffman-shaped variants are shown
to be always superior.
\end{itemize}

Among all the data points resulting from the combination of all the parameters, in the experiments we only show those points which are space/time dominant. 

Regarding queries, those for $\access$ are positions at random in $S[1,n]$.
For $\rank$, we used a random position $p$ in $S[1,n]$ and the symbol is
$S[p]$. Finally, for $\select$, we took a random position $p$ in $S[1,n]$,
using $S[p]$ and a random rank in $[1,\rank_{S[p]}(S,n)]$. We generated $10,000$ queries of each type, reporting the average time for each operation. 

\begin{figure}[]
 \centering
 
 \includegraphics[width=0.49\textwidth]{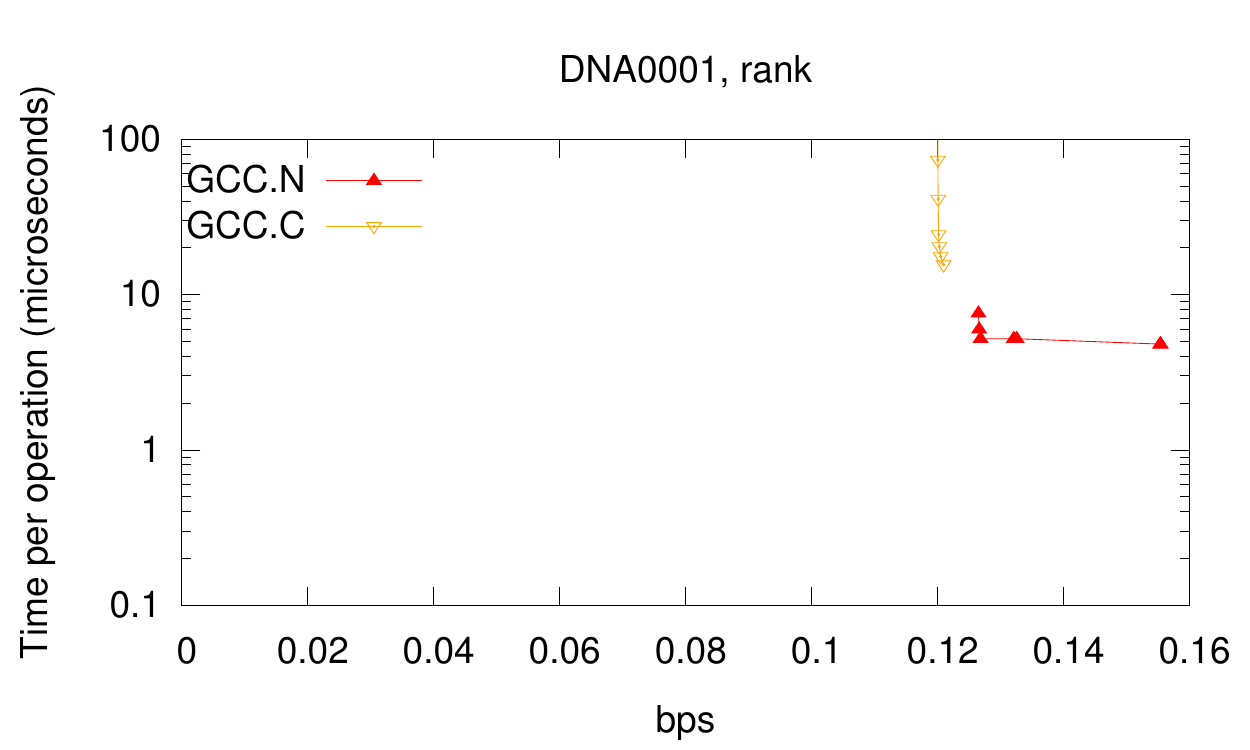}
 \includegraphics[width=0.49\textwidth]{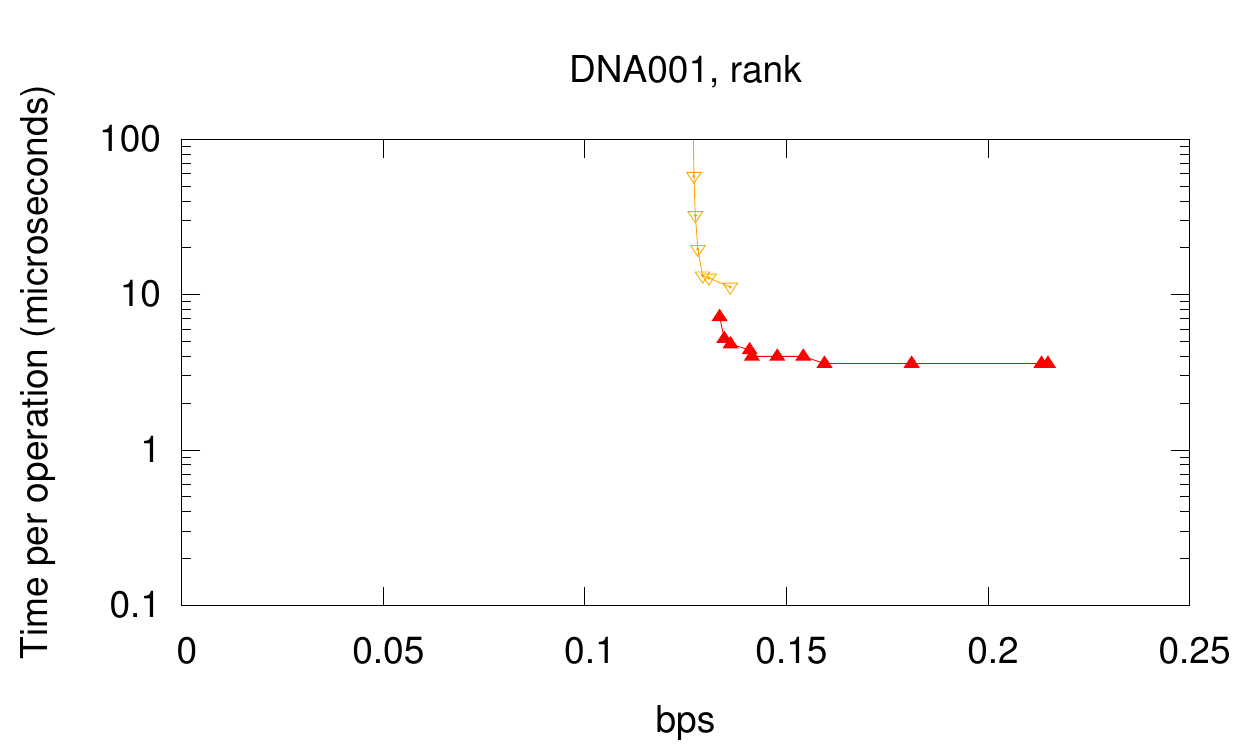}
 
 \includegraphics[width=0.49\textwidth]{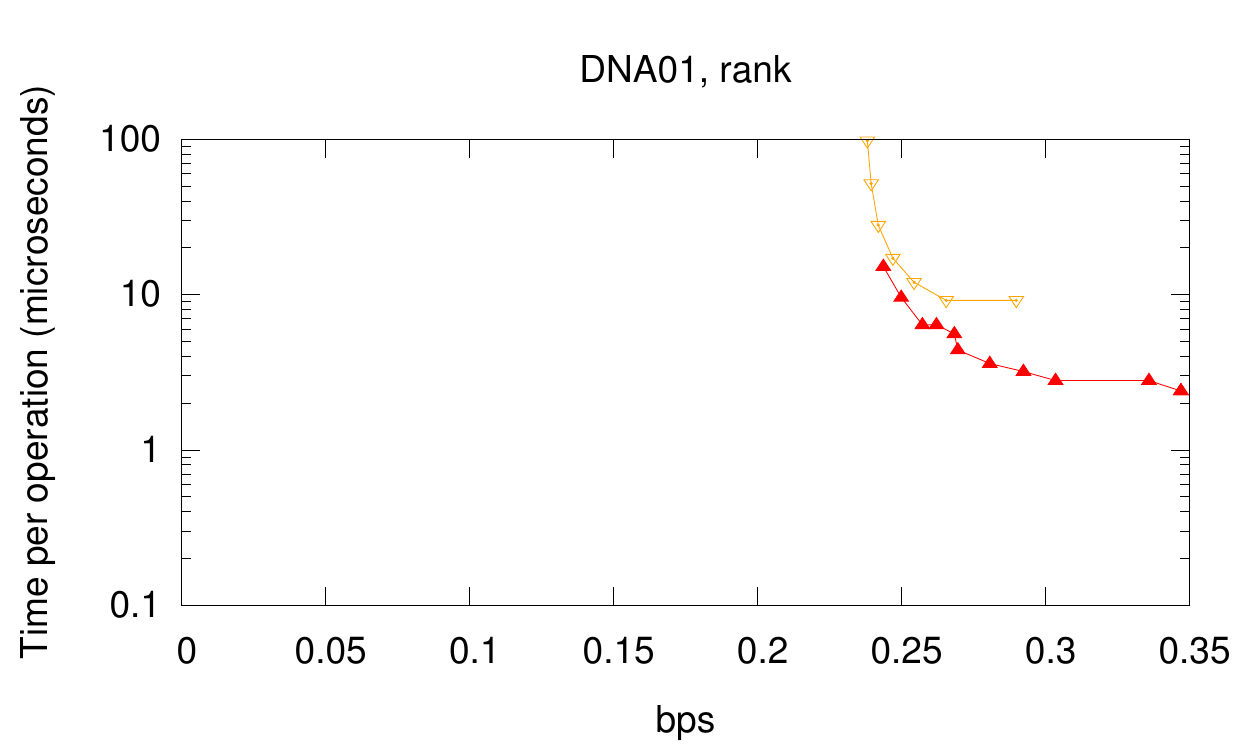}
 \includegraphics[width=0.49\textwidth]{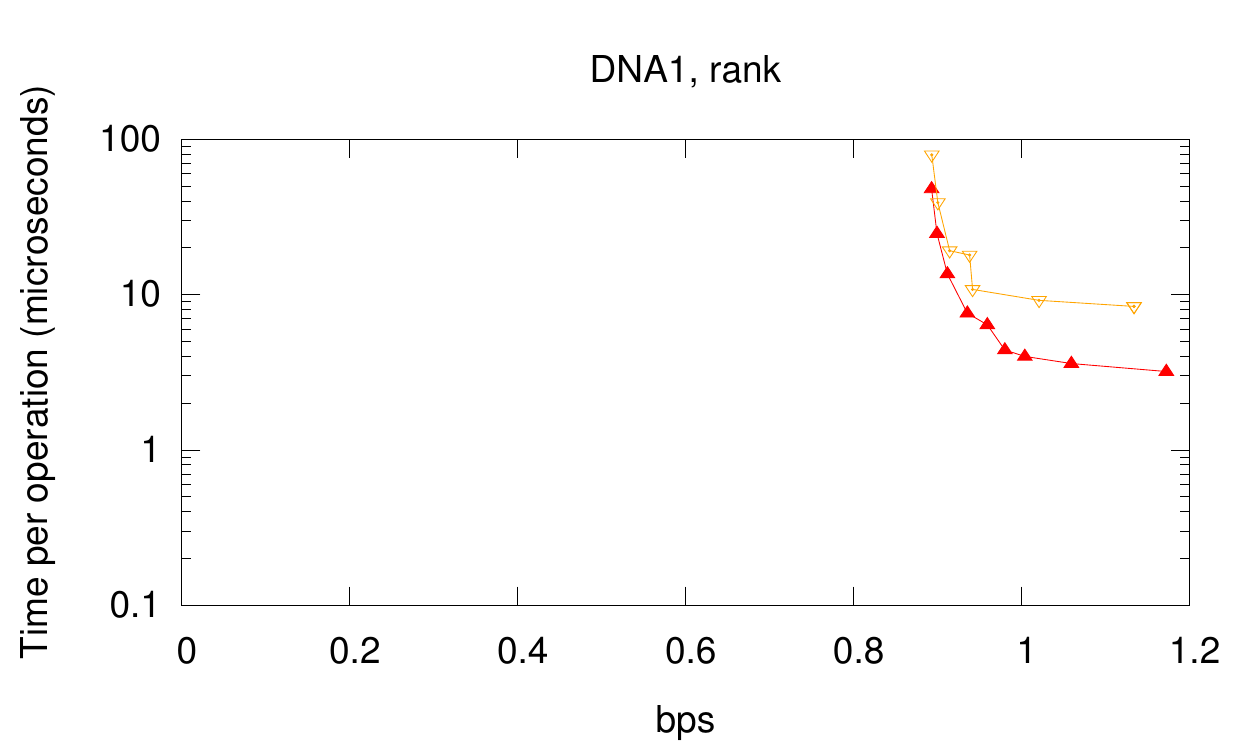}
 
 \includegraphics[width=0.49\textwidth]{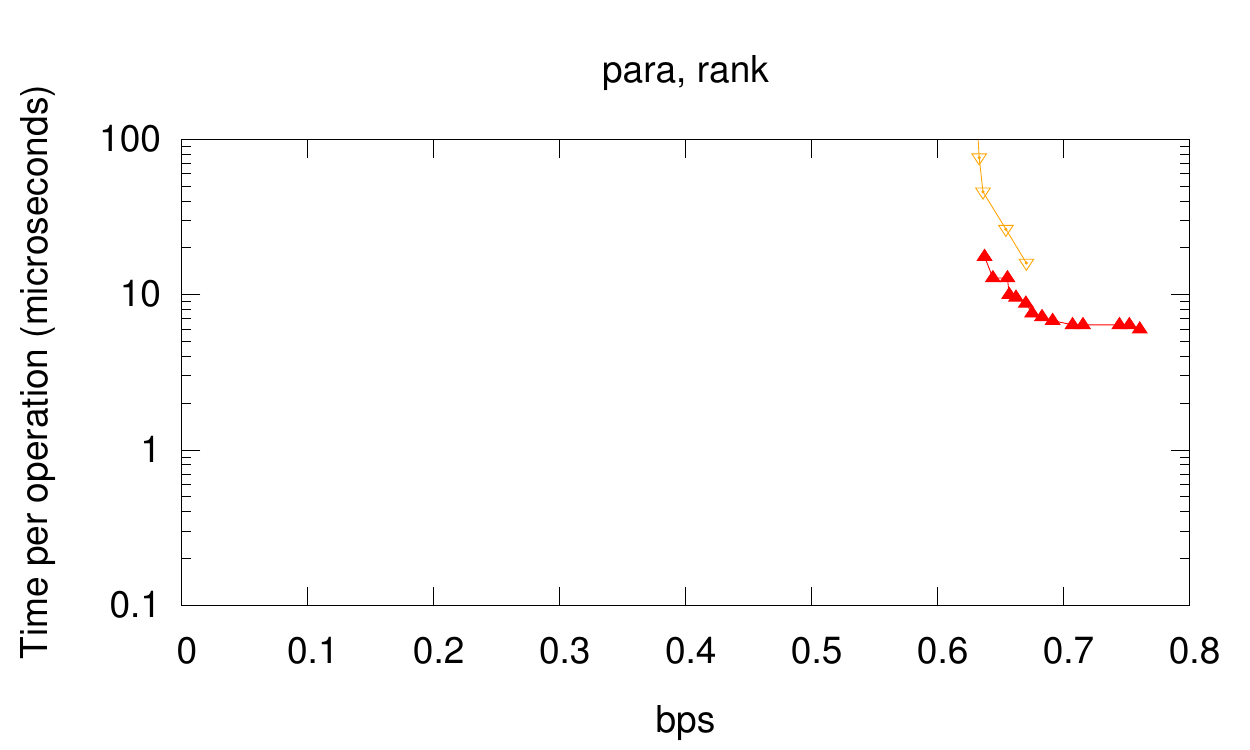}
 \includegraphics[width=0.49\textwidth]{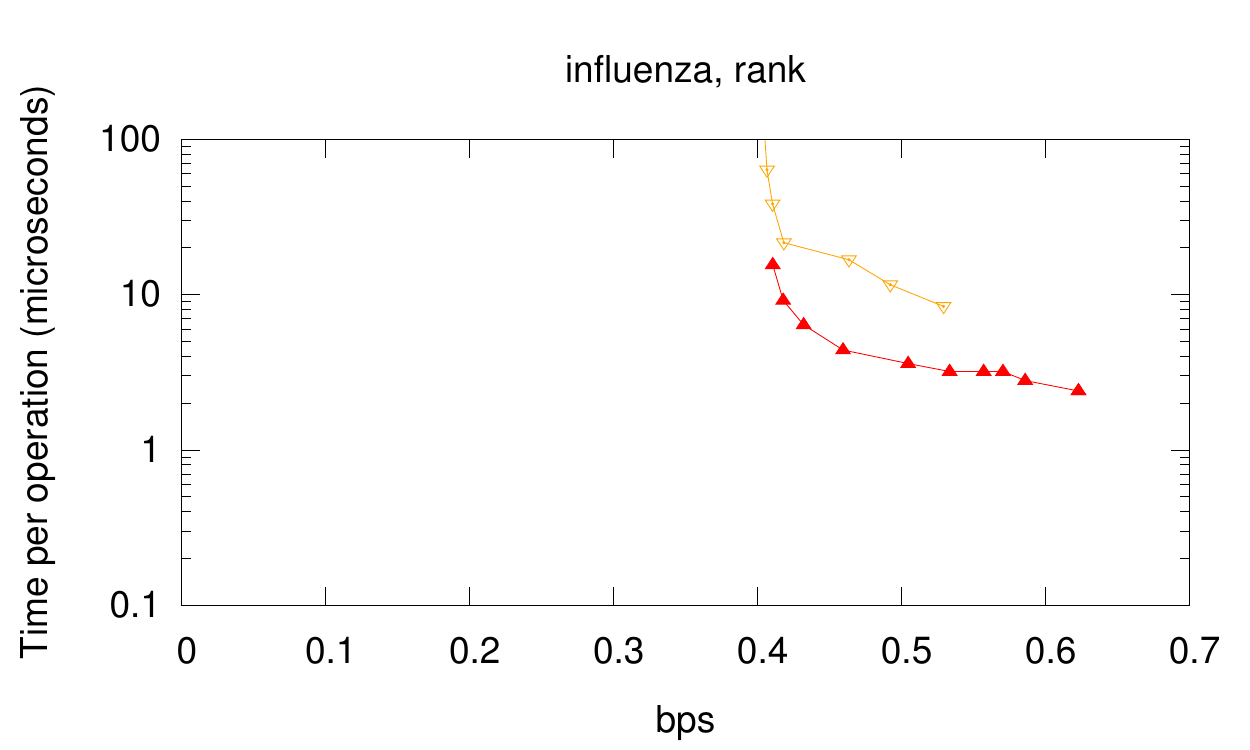}
 
 \includegraphics[width=0.49\textwidth]{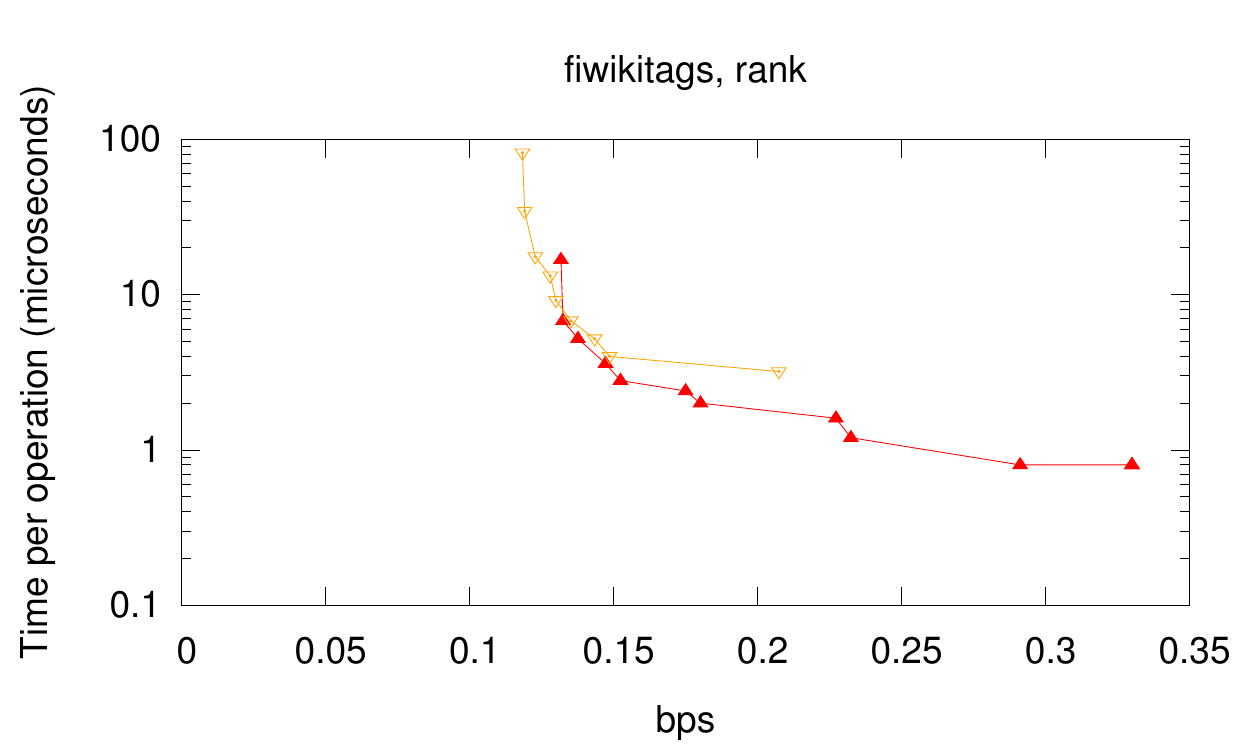}
 \includegraphics[width=0.49\textwidth]{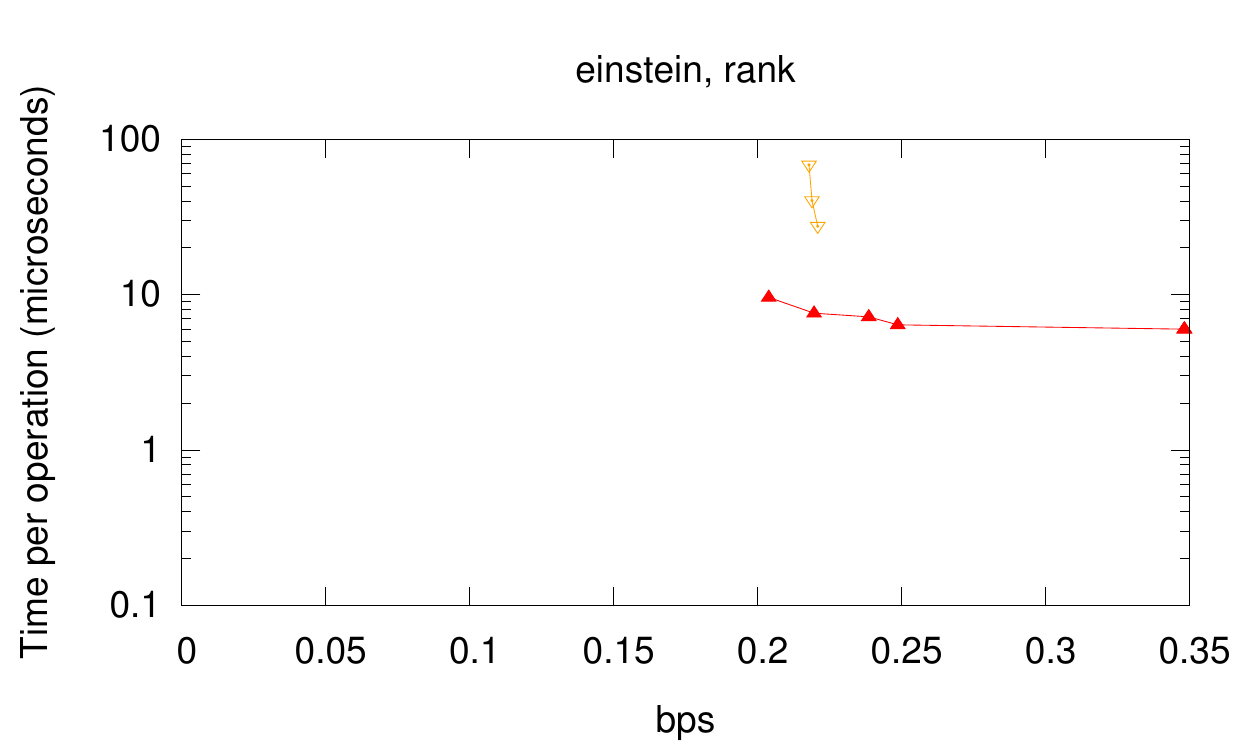}

\vspace*{-14.4cm}
 \includegraphics[width=0.49\textwidth]{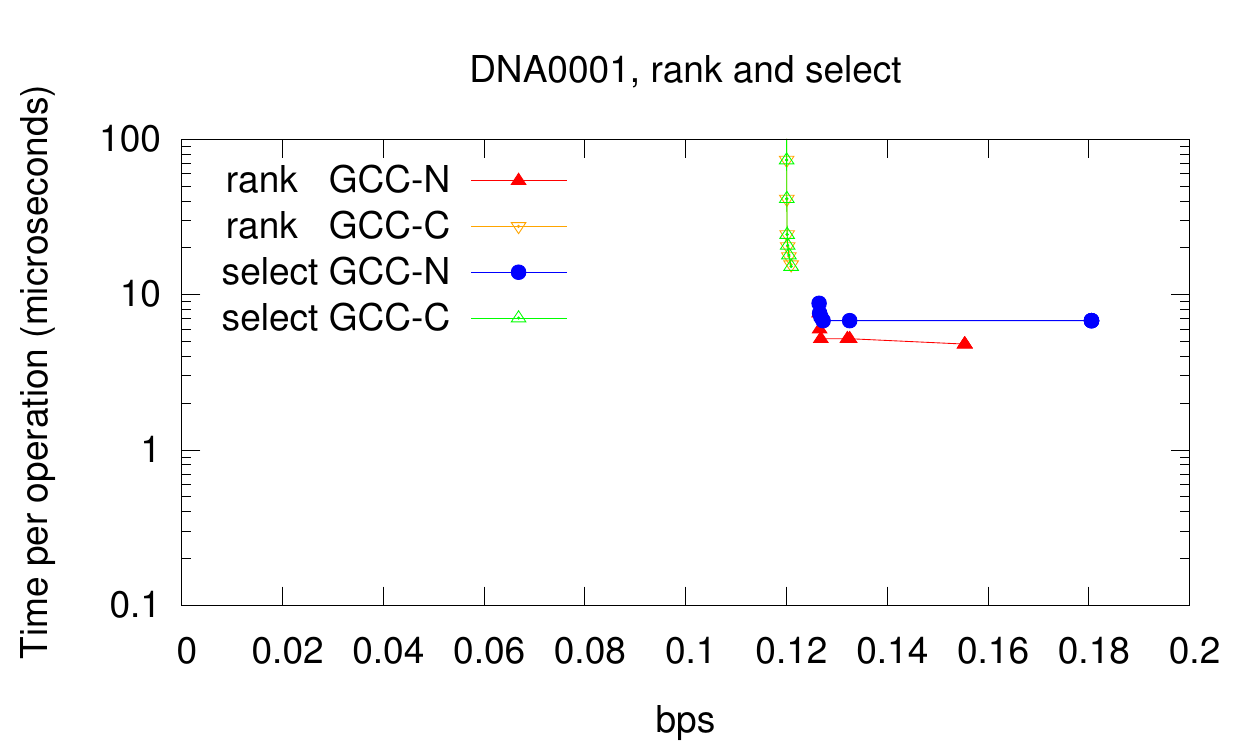}
 \includegraphics[width=0.49\textwidth]{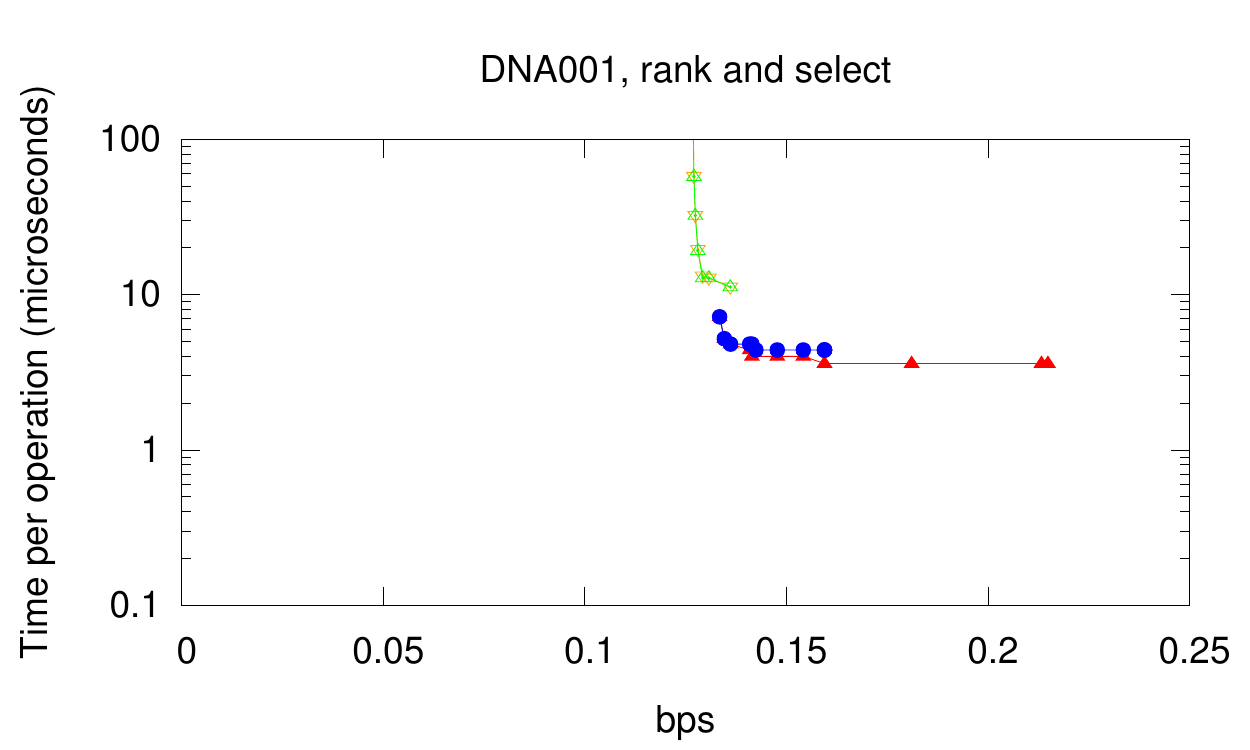}
 
 \includegraphics[width=0.49\textwidth]{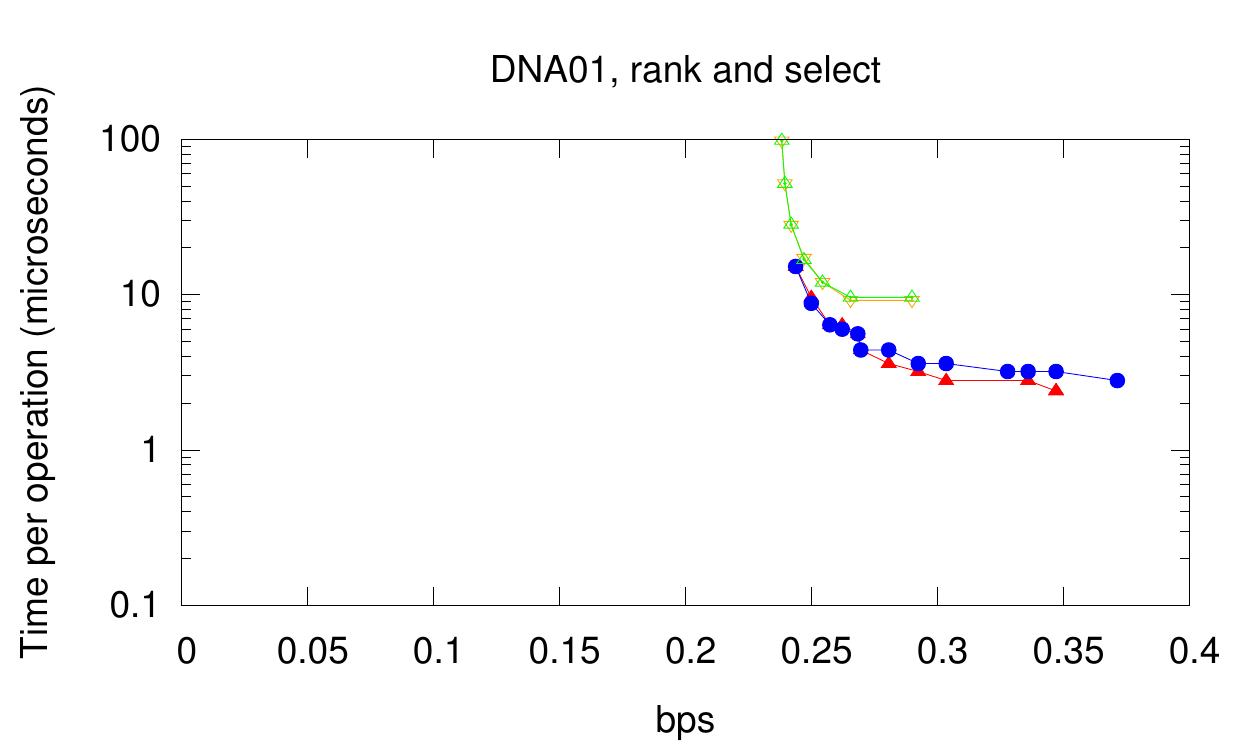}
 \includegraphics[width=0.49\textwidth]{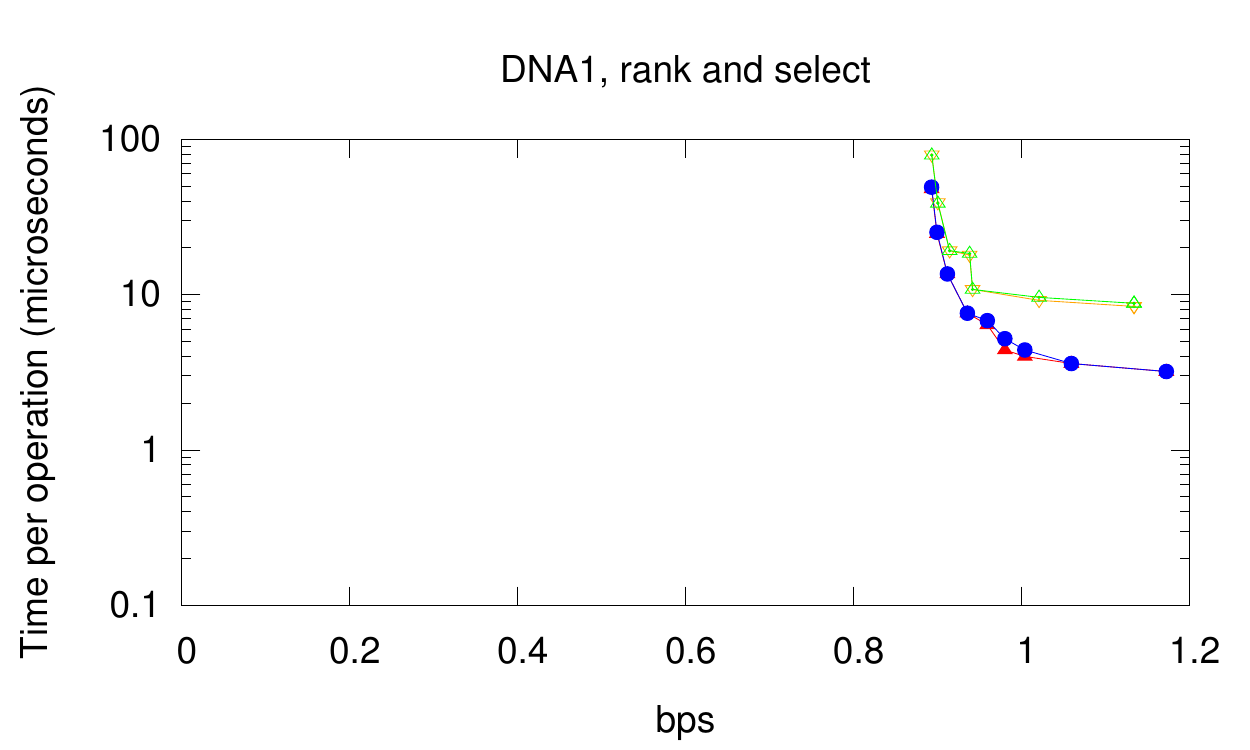}
 
 \includegraphics[width=0.49\textwidth]{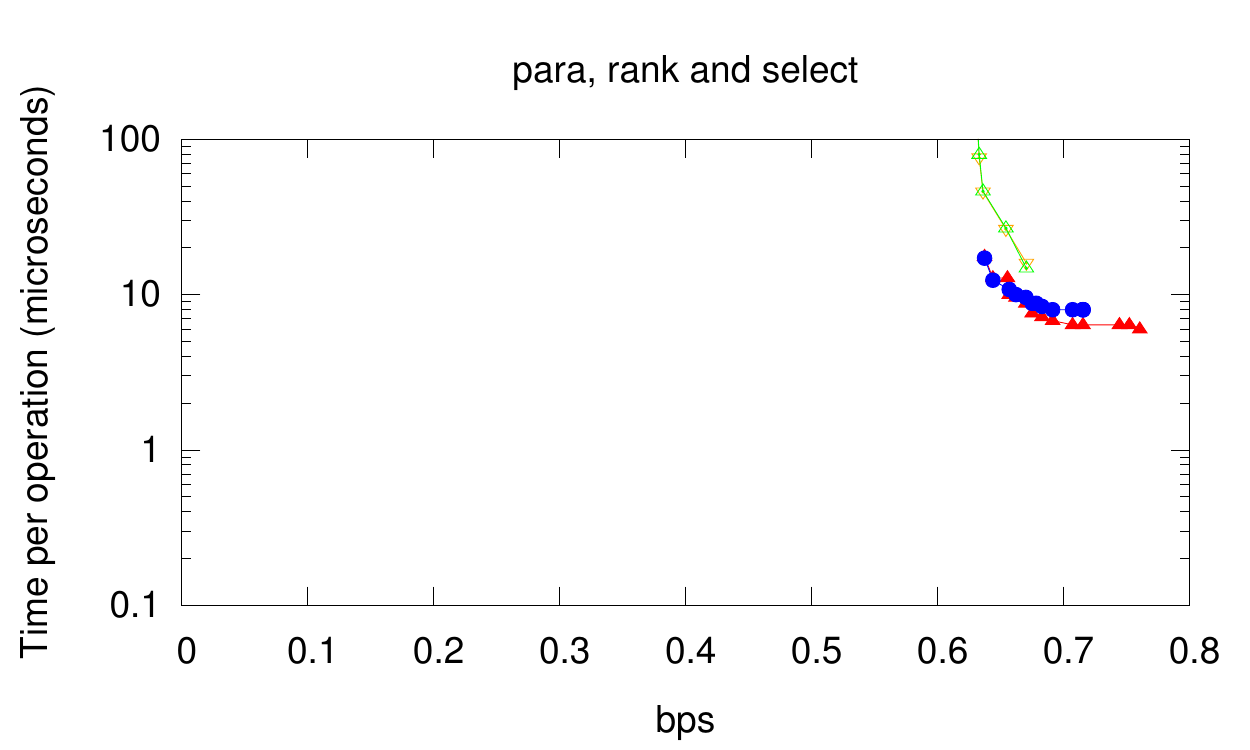}
 \includegraphics[width=0.49\textwidth]{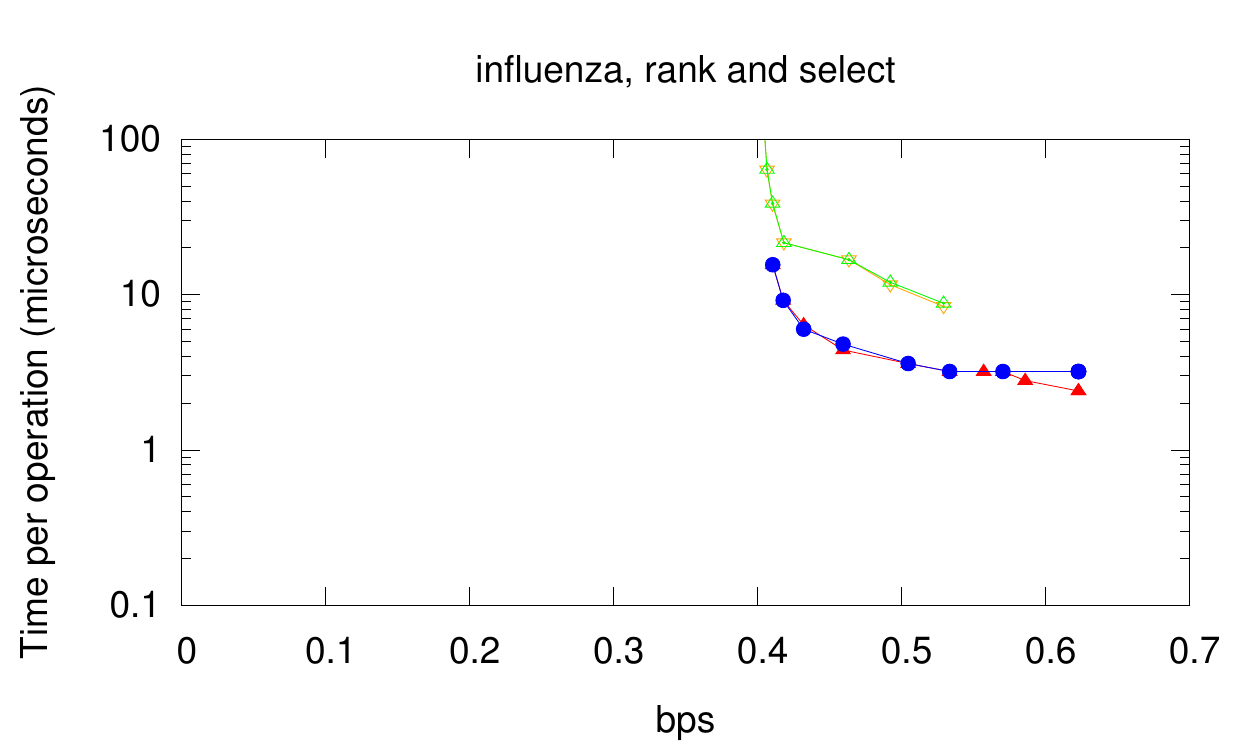}
 
 \includegraphics[width=0.49\textwidth]{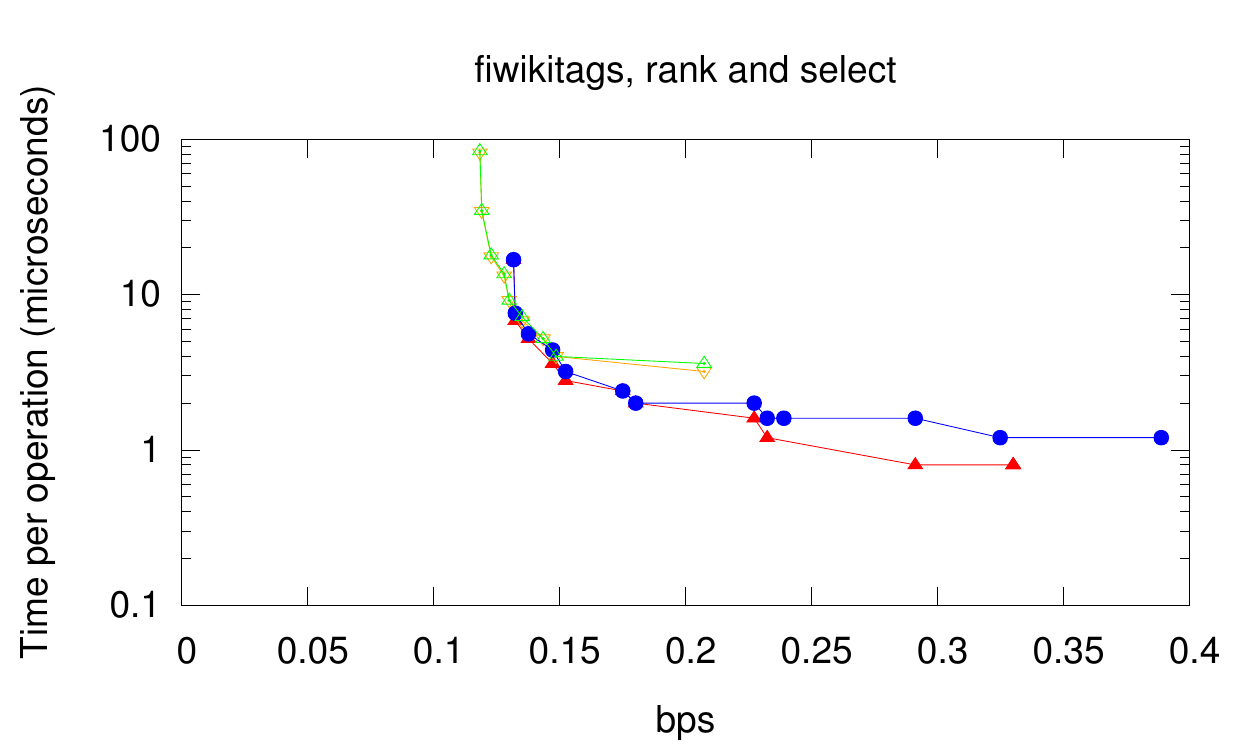}
 \includegraphics[width=0.49\textwidth]{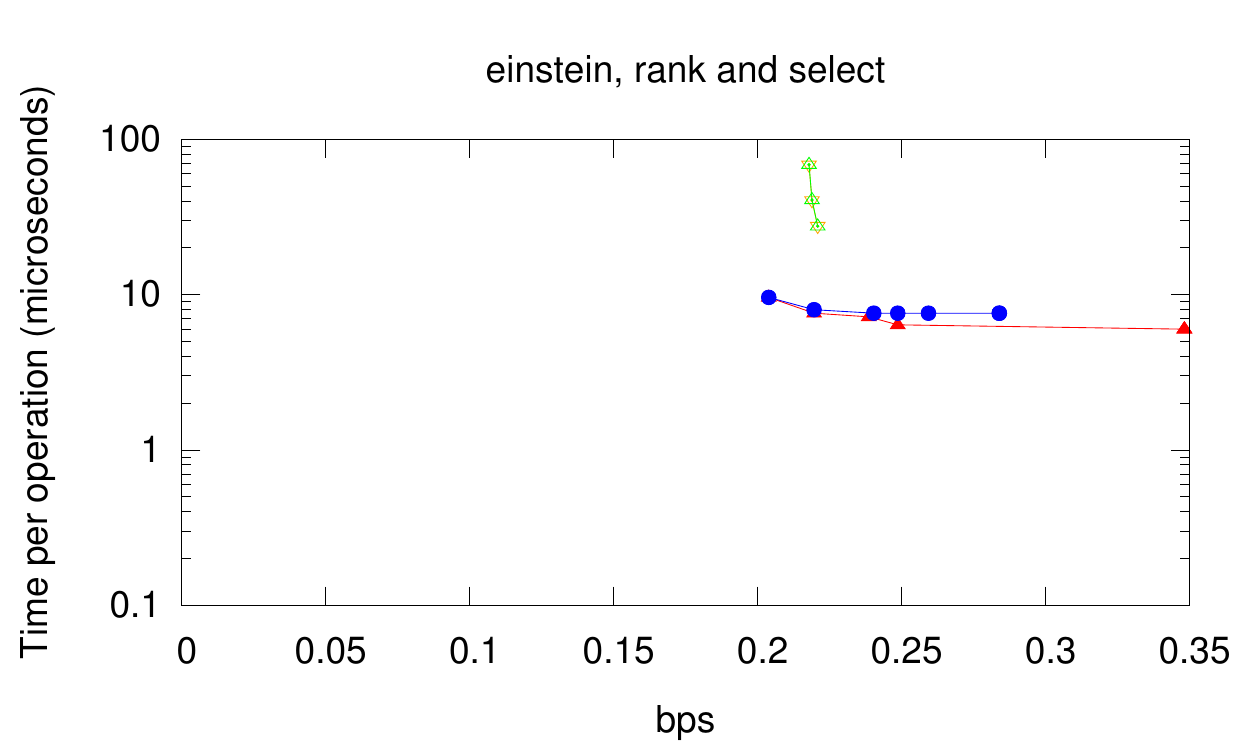}
 \caption{Comparison of $\rank$ and $\select$ performance of $\gccn$ and $\gccc$.}
 \label{fig:cmpsnsc.1}
 \end{figure}

In Section~\ref{sec:gcc} we proposed two sampling approaches for $\gcc$:
$\gccn$ is regular in $S$ and $\gccc$ is regular in $C$. We anticipated that
$\gccc$ should use less space on more repetitive sequences, but it could be
slower. Now we compare both sampling methods on the repetitive sequences with
smaller alphabets described in Table~\ref{table:datasets}.
Figure~\ref{fig:cmpsnsc.1} shows the results for $\rank$ and $\select$ 
($\access$ is equivalent to $\rank$ in our algorithms). 

\no{
 \begin{figure}[]
 \centering

 \caption{Comparison of $\select$ performance of $\gccn$ and $\gccc$.}
 \label{fig:cmpsnsc.2}
 \end{figure}
}

While, as said, $\gccc$ might use less space than $\gccn$ when the sequence is
more repetitive, this occurs in practice only slightly on $\dnaaaa$, and spaces
become closer as repetitiveness decreases on synthetic datasets ($\dnaaa$ to
$\dna$). Still, the differences are very slight, and instead $\gccn$ is much
faster than $\gccc$ for the same space usage. The same occurs in the real
sequences, where $\gccc$ uses less space than $\gccn$ only in $\fitags$. For 
the remaining experiments, we will use only $\gccn$.

\subsection{Performance on small alphabets}\label{exp:small}

We compare our $\gccn$ with $\wt.\M$, $\wth.\M$, and $\wm.\M$. We
also include in the comparison two statistically compressed representations that
are the best for small and moderate alphabets: $\wth.\cm$ and $\wth.\rrr$.

\begin{figure}[p]
 \centering

 \includegraphics[width=0.49\textwidth]{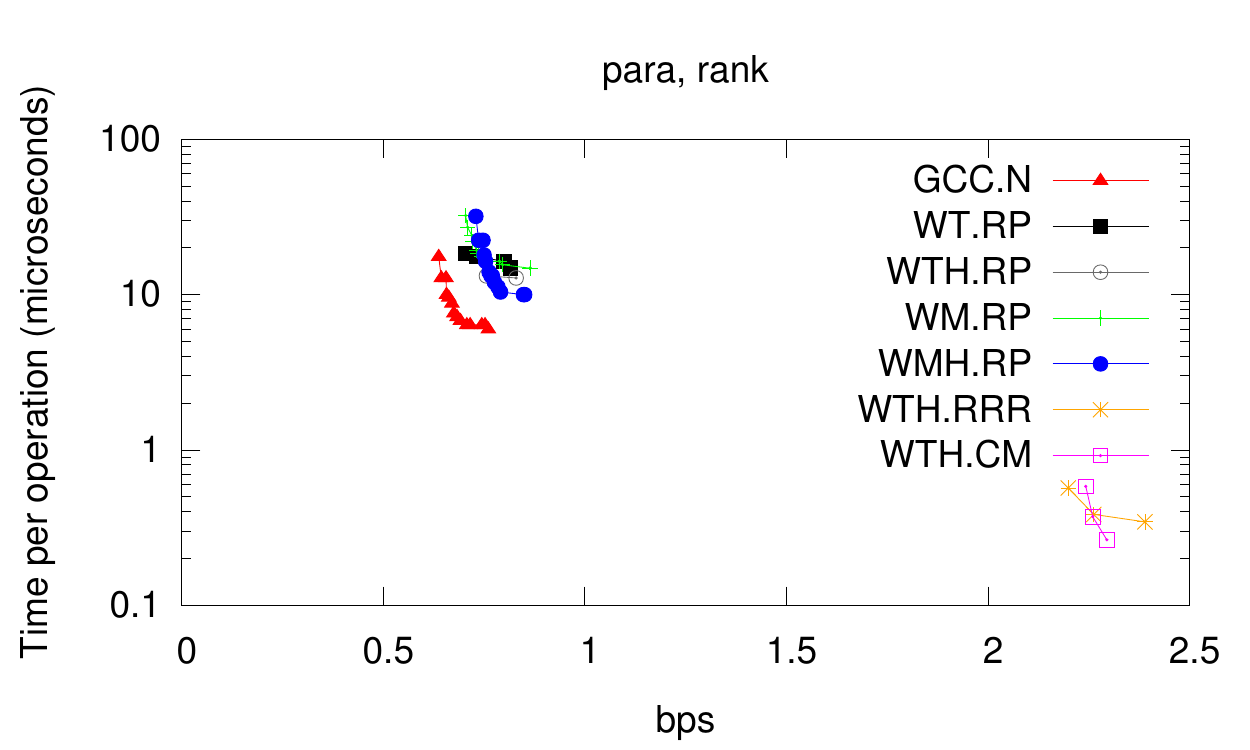}
 \includegraphics[width=0.49\textwidth]{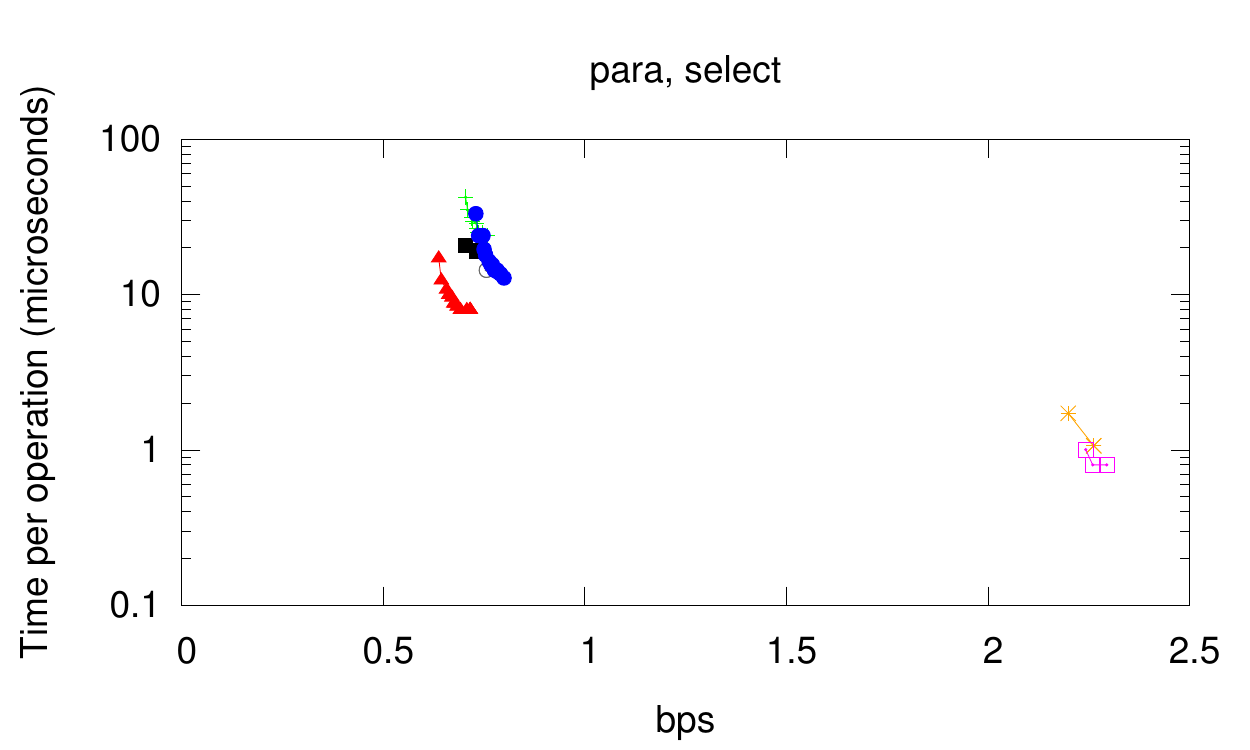}
 
 \includegraphics[width=0.49\textwidth]{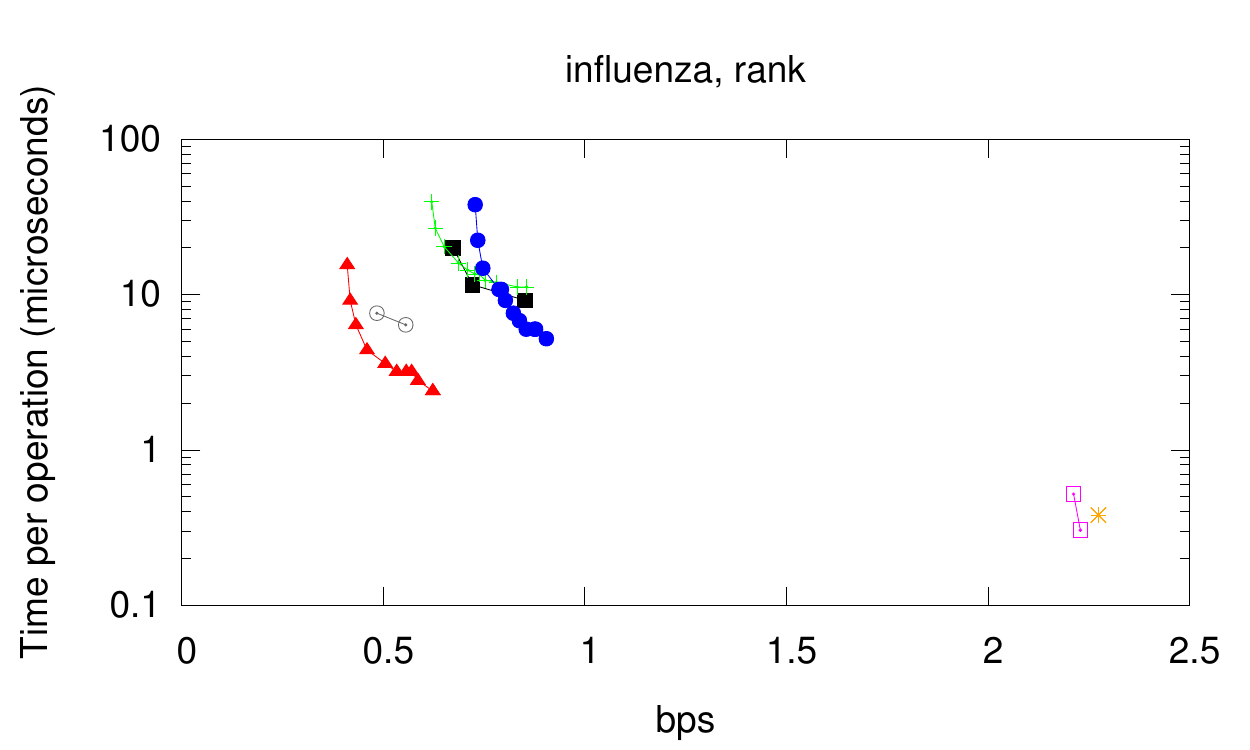}
 \includegraphics[width=0.49\textwidth]{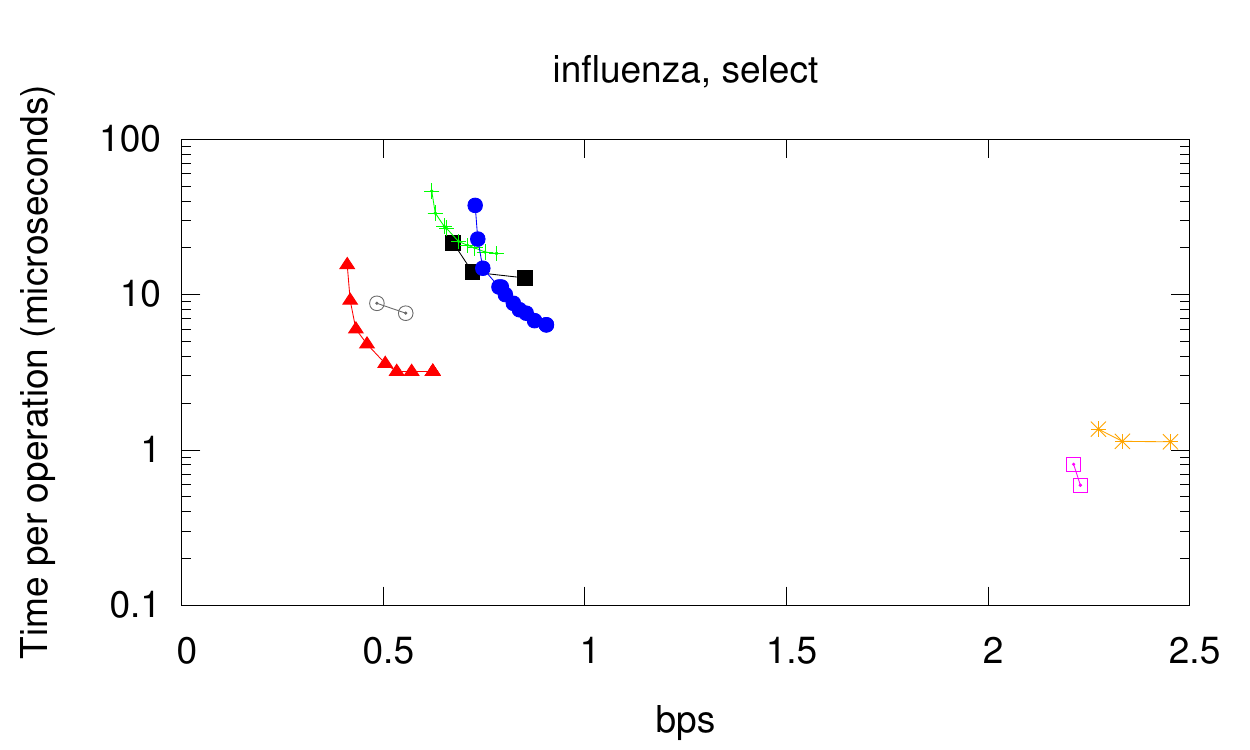}
 
  \includegraphics[width=0.49\textwidth]{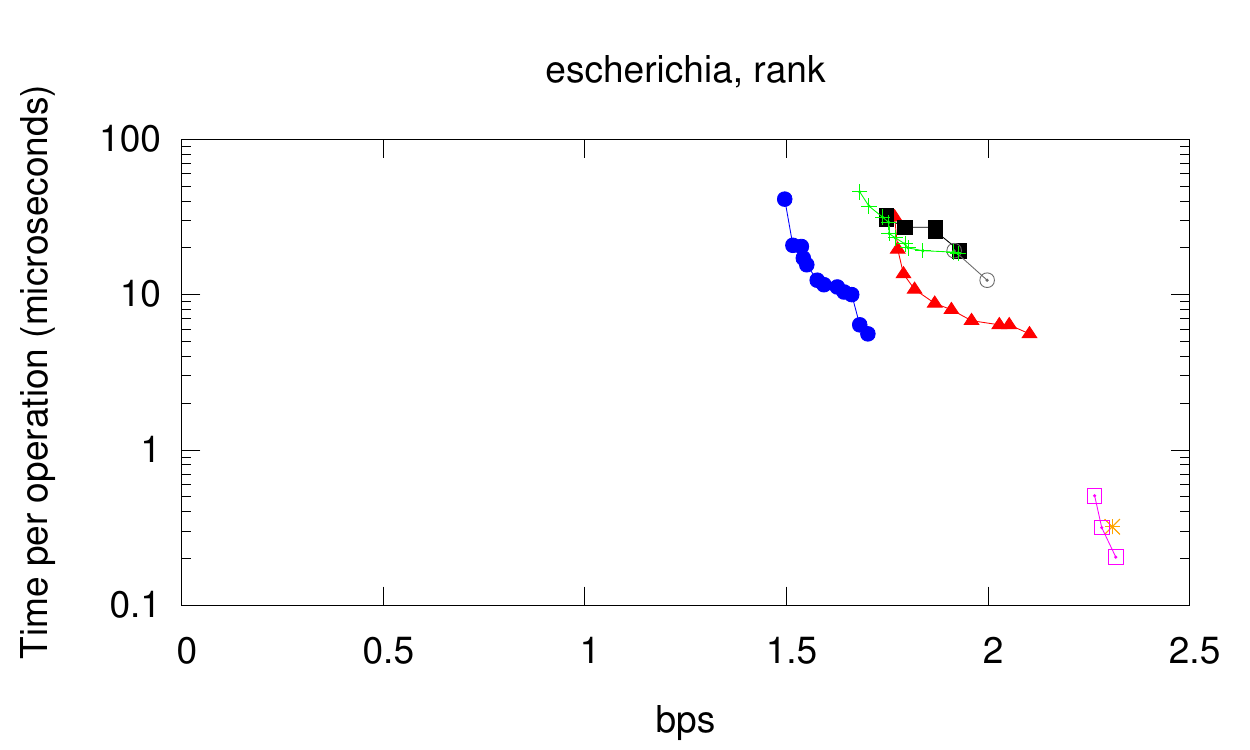}
  \includegraphics[width=0.49\textwidth]{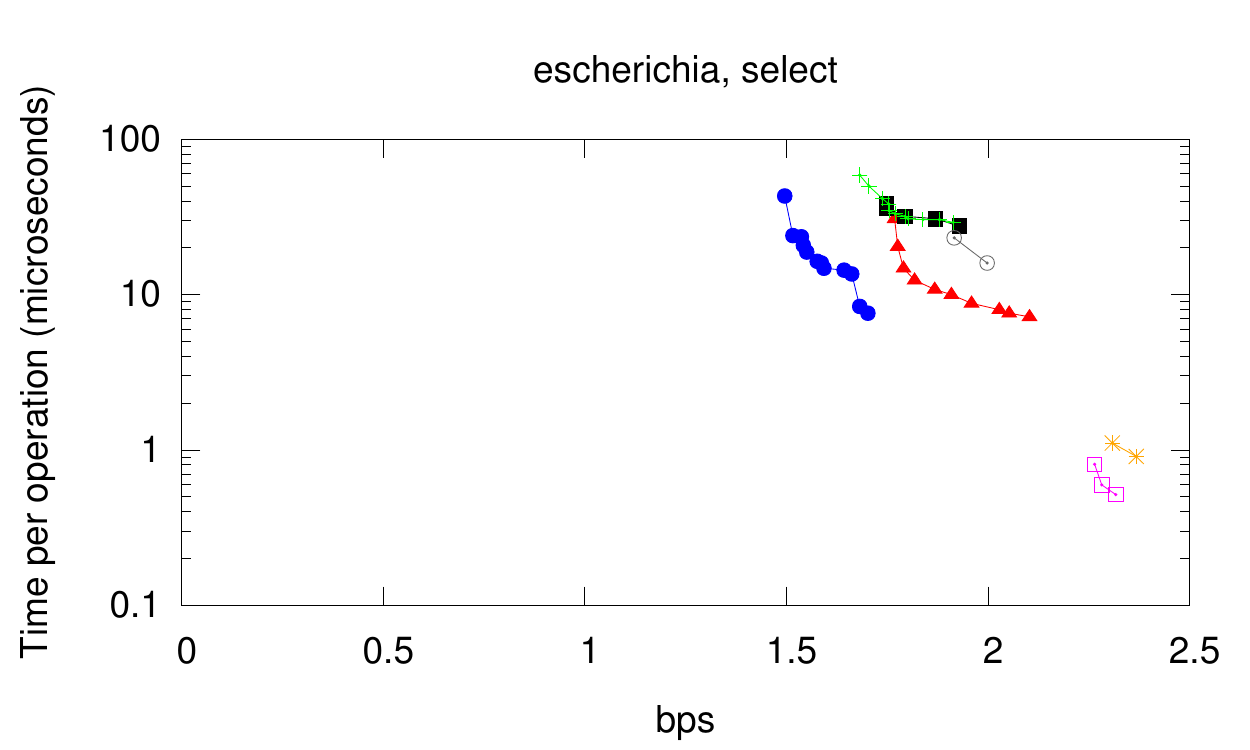}

 \includegraphics[width=0.49\textwidth]{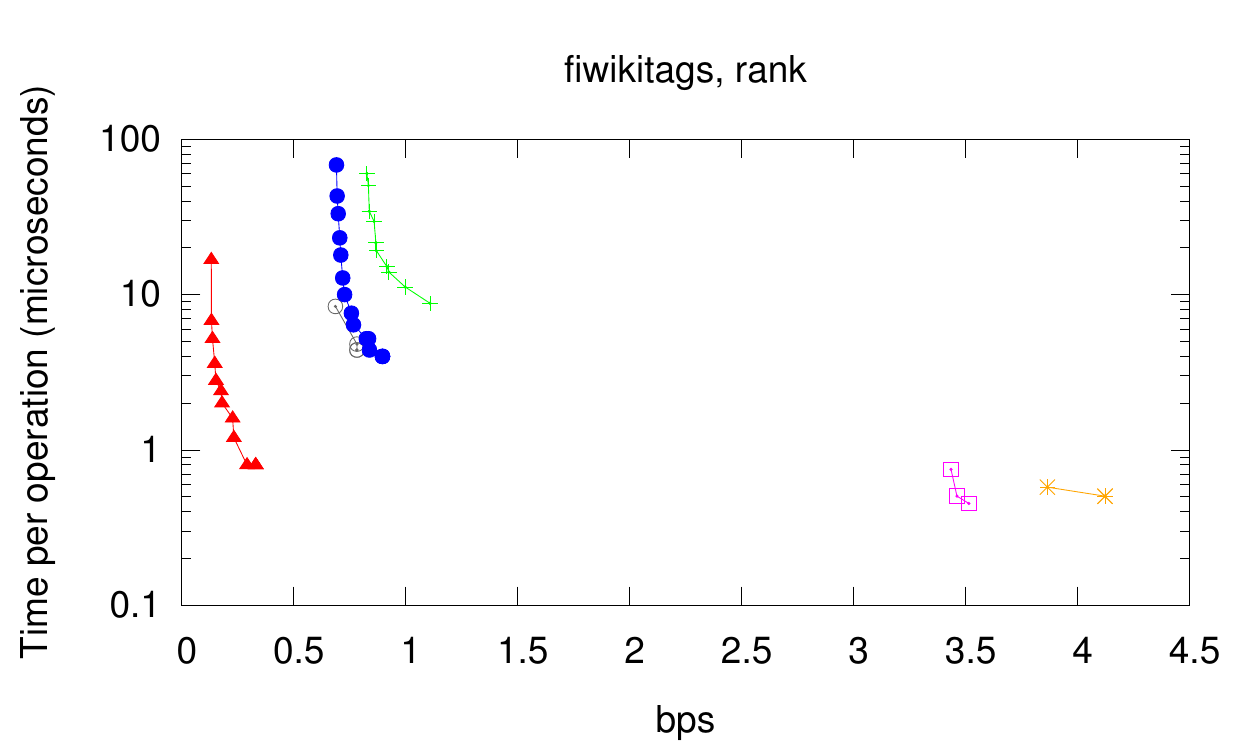}
 \includegraphics[width=0.49\textwidth]{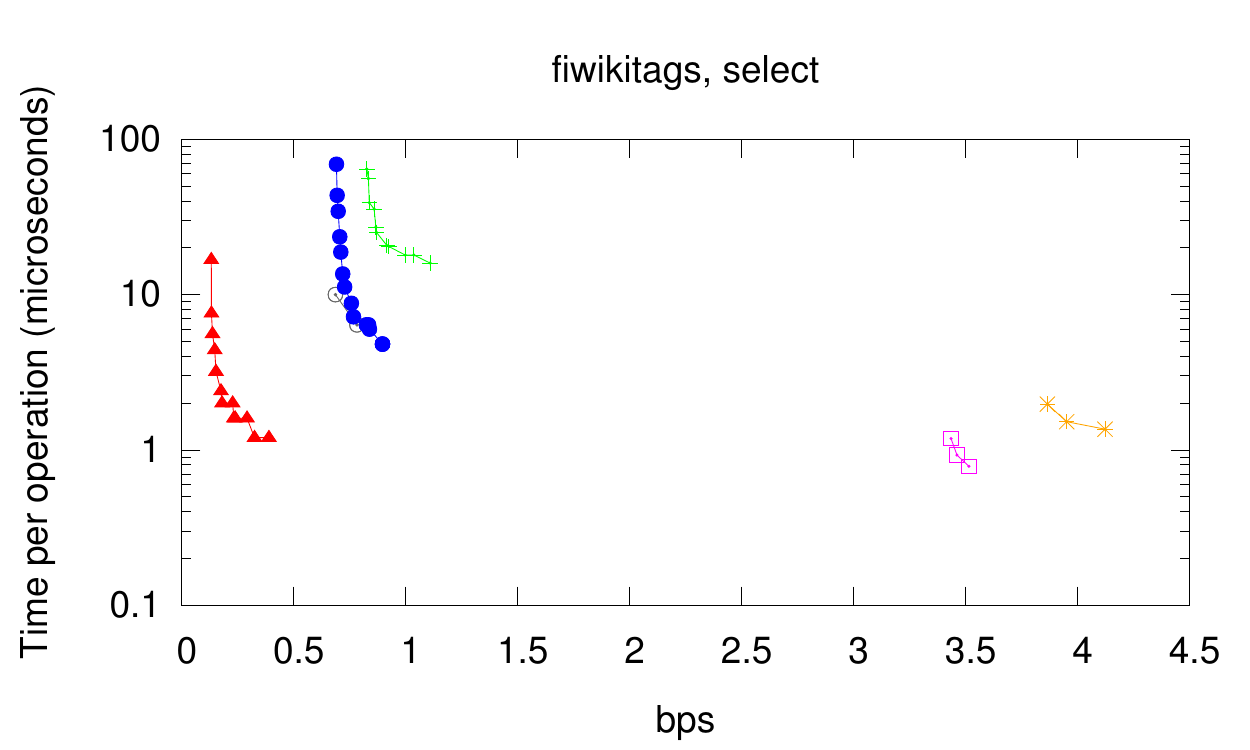}

 \includegraphics[width=0.49\textwidth]{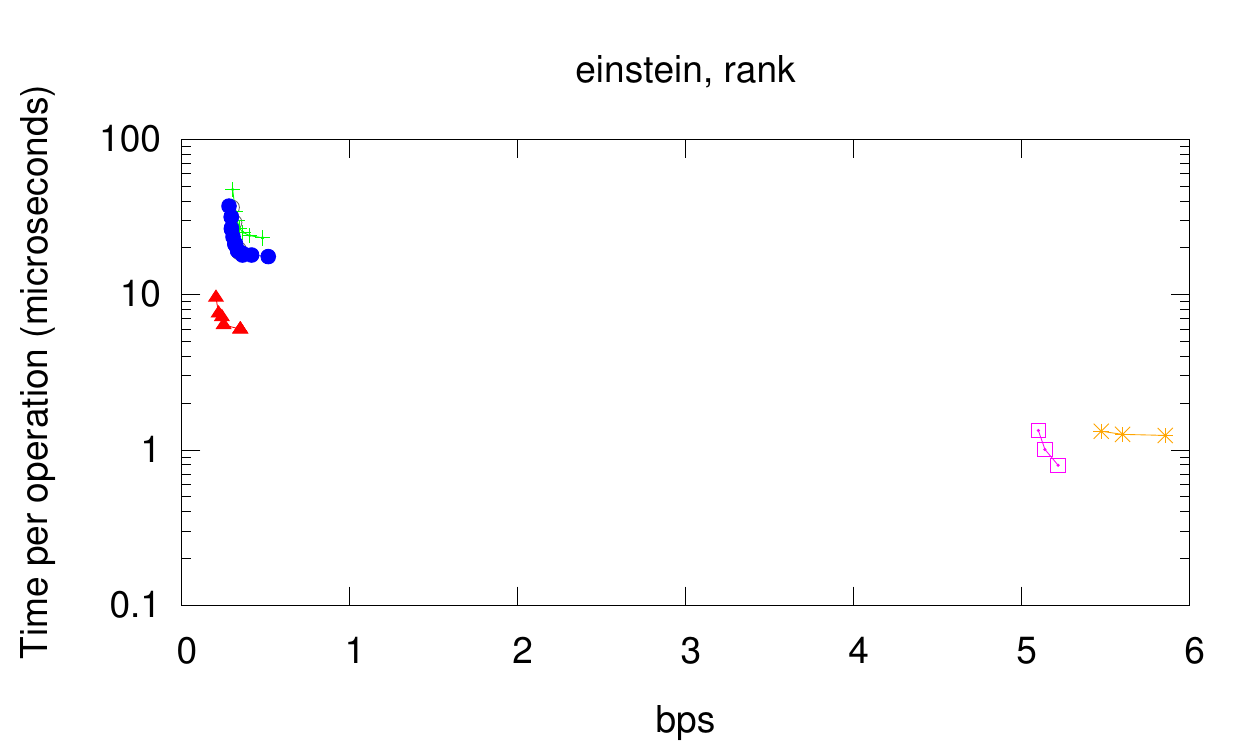}
 \includegraphics[width=0.49\textwidth]{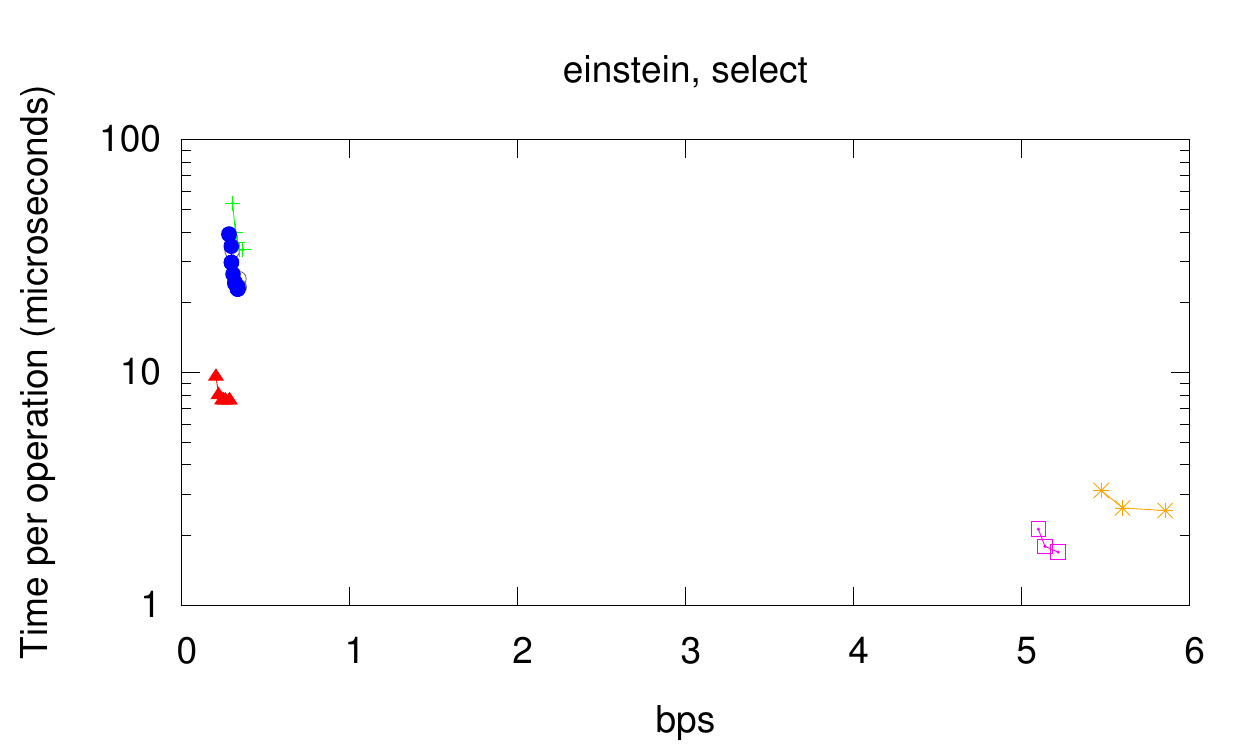}
 
 \caption{Space-time tradeoffs for $\rank$ and $\select$ queries over small
alphabets (time in logscale). }
 \label{fig:small.1}
 \end{figure}

Figure~\ref{fig:small.1} shows the results for $\rank$ and $\select$ on the 
real collections that have small and moderate alphabets (again, the results 
for $\access$ are very similar to those for $\rank$).
It can be seen that $\wth.\M$ generally performs better than $\wt.\M$ in space
and time, as expected. The variant $\wm.\M$ performs slightly better than 
$\wt.\M$ in space, as it represents only one grammar per level and not per
node (the difference would be higher on larger alphabets). In exchange,
$\wm.\M$ is slightly slower than $\wt.\M$ because it performs more
$\rank$/$\select$ operations on the bitmaps represented with $\gcc$.
Finally, $\wmh.\M$ uses less space than $\wm.\M$ only in some cases, but it
generally outperforms it for the same space. It performs particularly well on
{\tt escherichia}, the least repetitive of the datasets.

\no{
 \begin{figure}[t]
 \centering
 
 \includegraphics[width=0.49\textwidth]{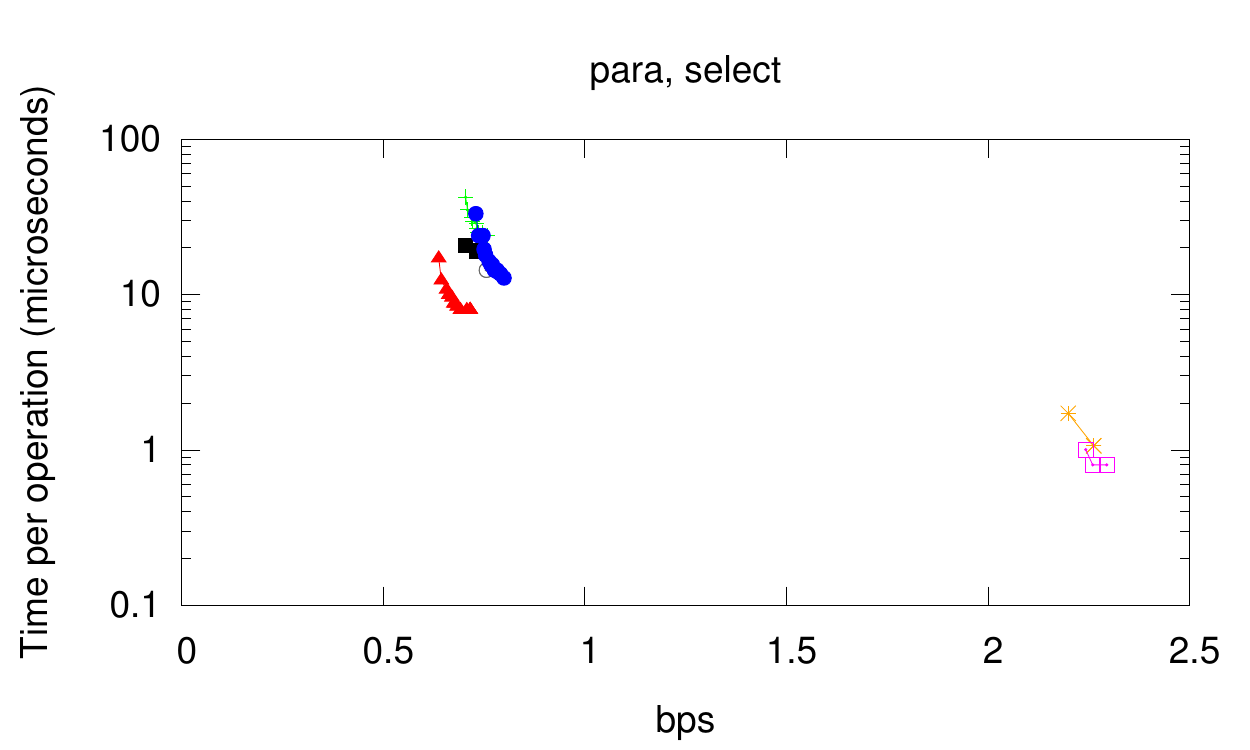}
 \includegraphics[width=0.49\textwidth]{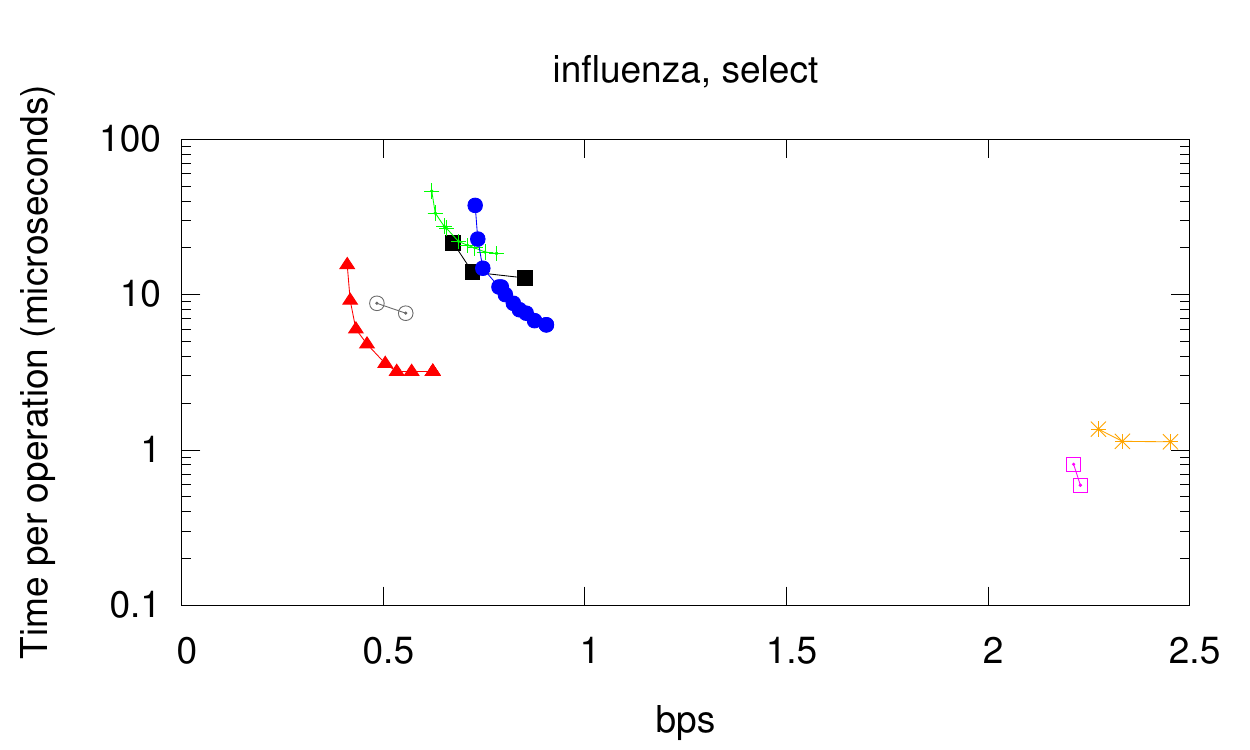}
 \includegraphics[width=0.49\textwidth]{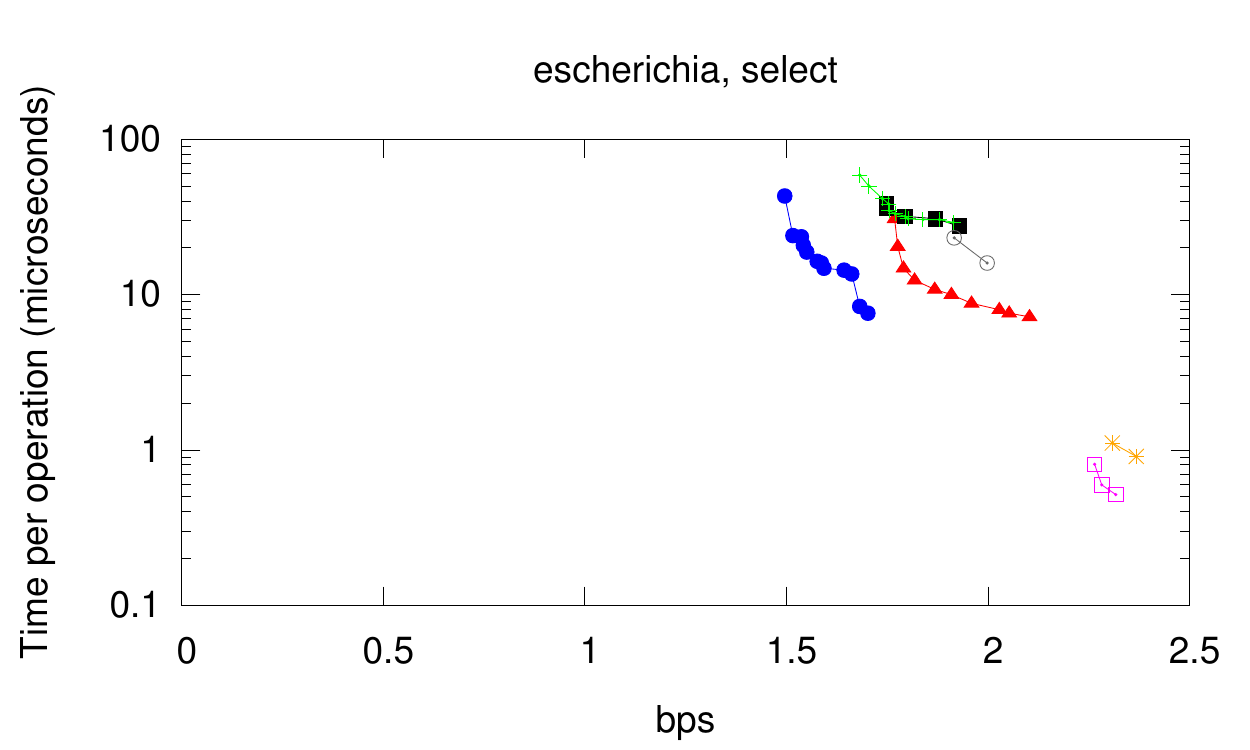}
 \includegraphics[width=0.49\textwidth]{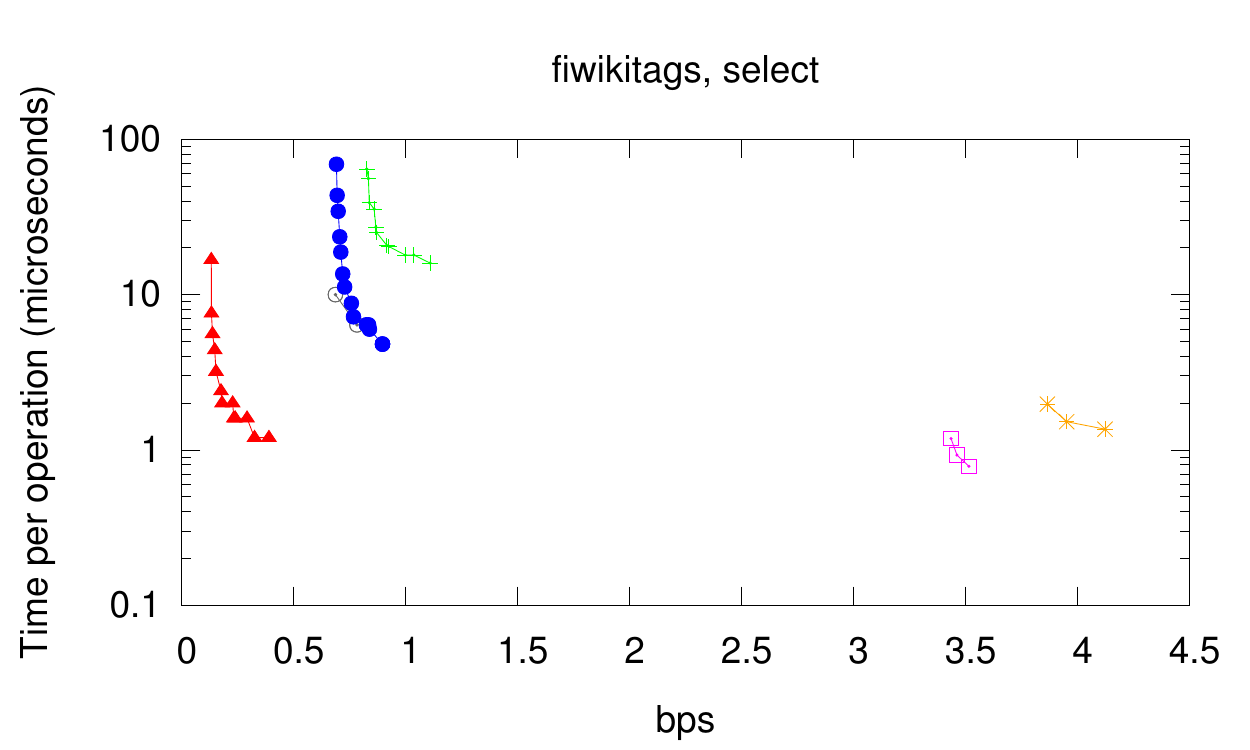}
 \includegraphics[width=0.49\textwidth]{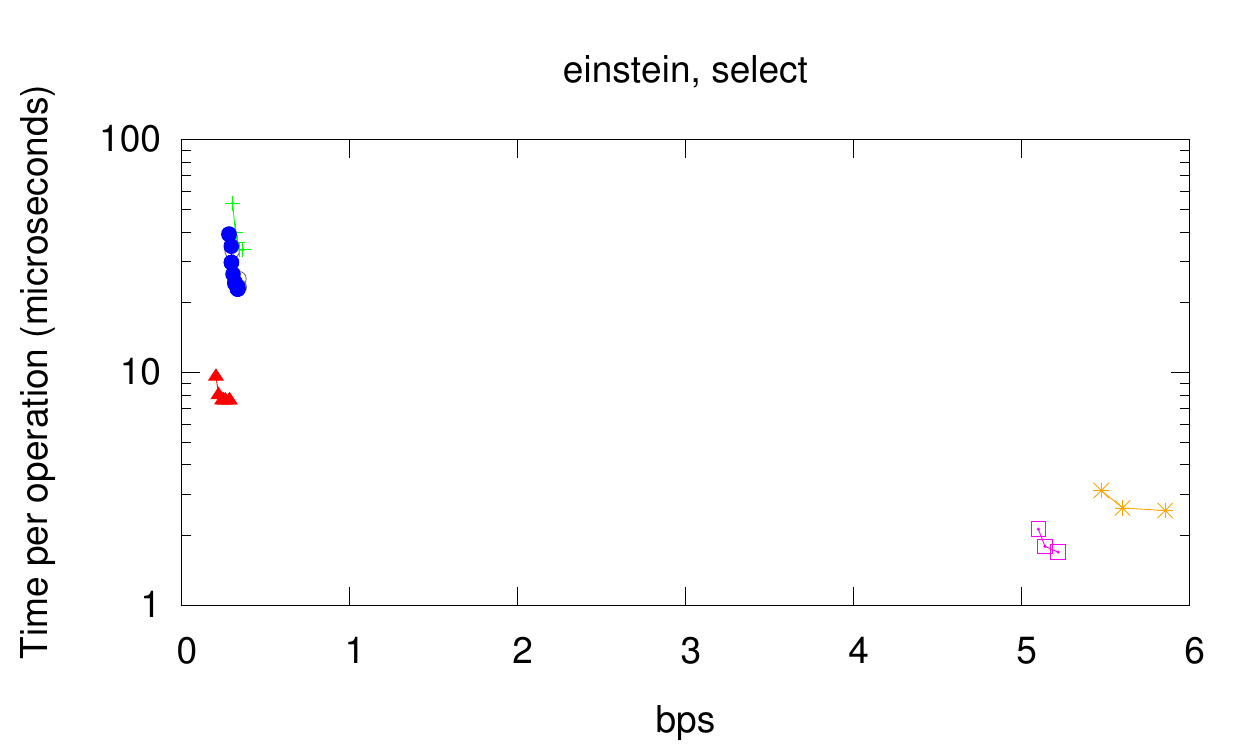}
 
 \caption{Space-time tradeoffs for $\select$ queries over small and moderate
alphabets (time in logscale).}
 \label{fig:small.2}
 \end{figure}
}

%  \begin{figure}[t]
%  \centering
 
%  \includegraphics[width=0.49\textwidth]{charts/esch/rank.pdf}
%  \includegraphics[width=0.49\textwidth]{charts/tags/rank.pdf}
 
%  \includegraphics[width=0.49\textwidth]{charts/esch/select.pdf}
%  \includegraphics[width=0.49\textwidth]{charts/tags/select.pdf}
 
%  \includegraphics[width=0.49\textwidth]{charts/esch/access.pdf}
%  \includegraphics[width=0.49\textwidth]{charts/tags/access.pdf}

%  \caption{Space-time tradeoffs for $\rsa$ queries over small
% alphabets: collections {\tt escherichia} and {\tt fiwikitags} (note logscale in time). }
%  \label{fig:small.2}
%  \end{figure}

Recall that $\wm.\M$ is our improved version of previous work, $\wtrp$ 
\cite{NPVjea13}, and it is now superseded by $\gccn$. The space of $\wm.\M$ is 
in most cases similar to that of $\gccn$, which means that $\wm.\M$ is actually
close to the worst-case space estimation, $O(g\sigma\log n)$. In some cases, 
$\gcc$ is significantly smaller. More importantly, $\gccn$ is 2--15 times 
faster than $\wm.\M$, and also 2--7 times faster than $\wth.\M$, the faster of
the competitors in this family, which also uses more space than $\gccn$.
$\gccn$ handles queries in a few microseconds.

On the other hand, the representations that compress statistically, $\wth.\cm$ and $\wth.\rrr$, are about an order of magnitude faster than $\gccn$, but also
take 5--15 times more space (except on {\tt escherichia}, which is not
repetitive).

\subsection{Performance on large alphabets}\label{exp:large}

Now we use the collections {\tt einstein} (again), {\tt software},
{\tt einstein.words}, {\tt fiwiki}, and {\tt indochina} from 
Table~\ref{table:datasets}, to compare the performance on moderate and large
alphabets. We compare the two versions of our $\aprep$, our $\mwth.\M$, and 
all the statistically compressed or compact schemes for large alphabets: 
$\wm/\wmh$ with $\cm/\rrr$ and $\ap$ (we only exclude $\wm.\cm$, which always
loses to others). In the first two collections, whose alphabet size is
moderate, we also include $\gccn$, to allow comparing its performance with our
variants for large alphabets in these intermediate cases.

\begin{figure}[p]
 \centering
 
\includegraphics[width=0.49\textwidth]{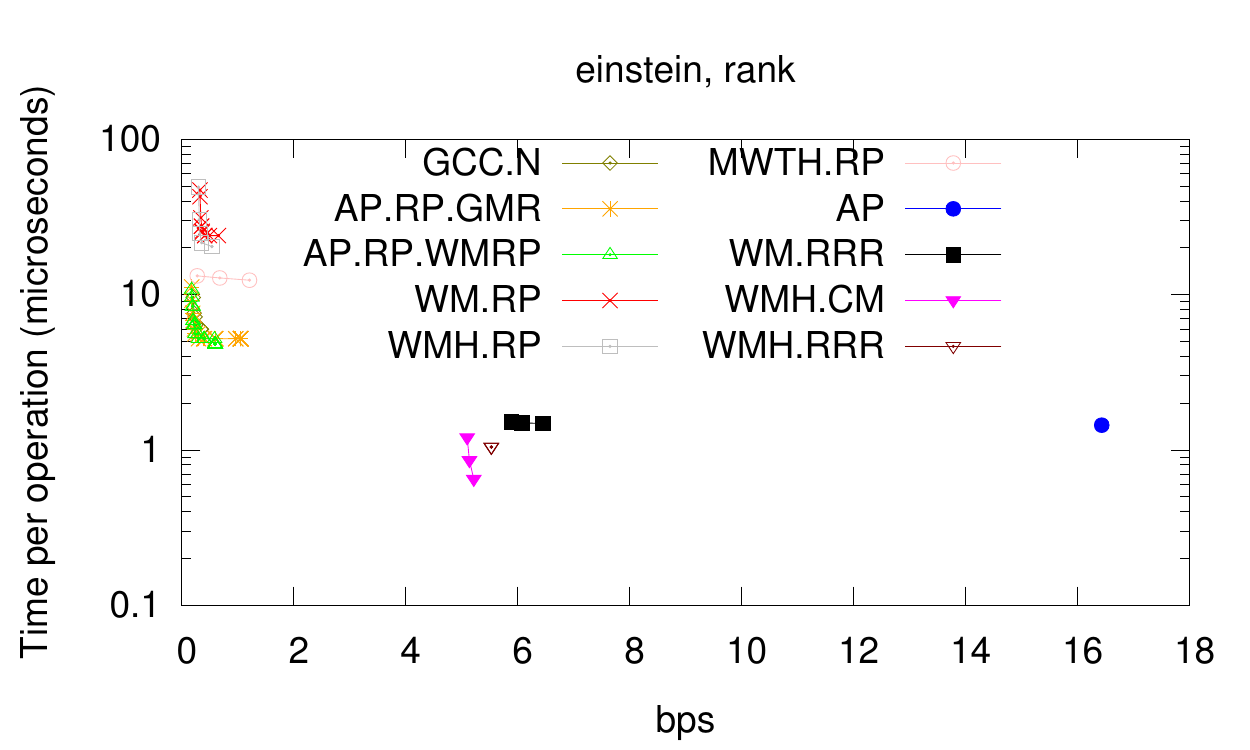}
\includegraphics[width=0.49\textwidth]{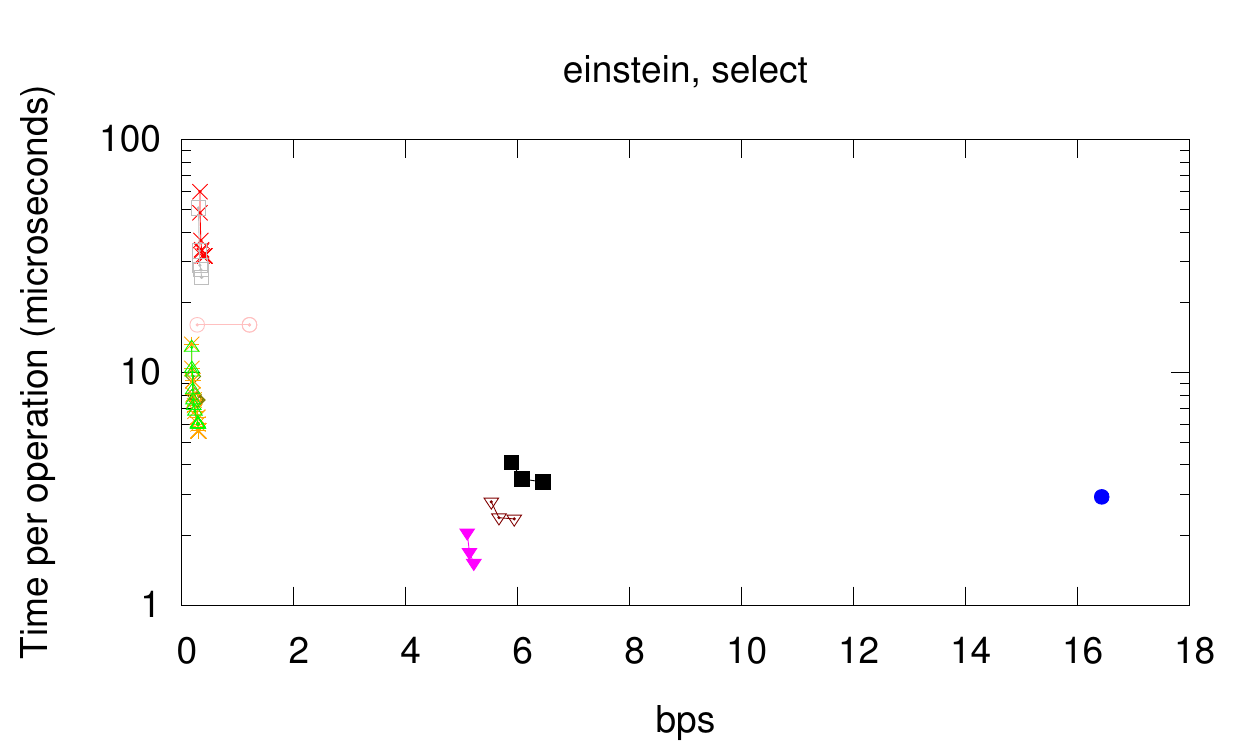}

 \includegraphics[width=0.49\textwidth]{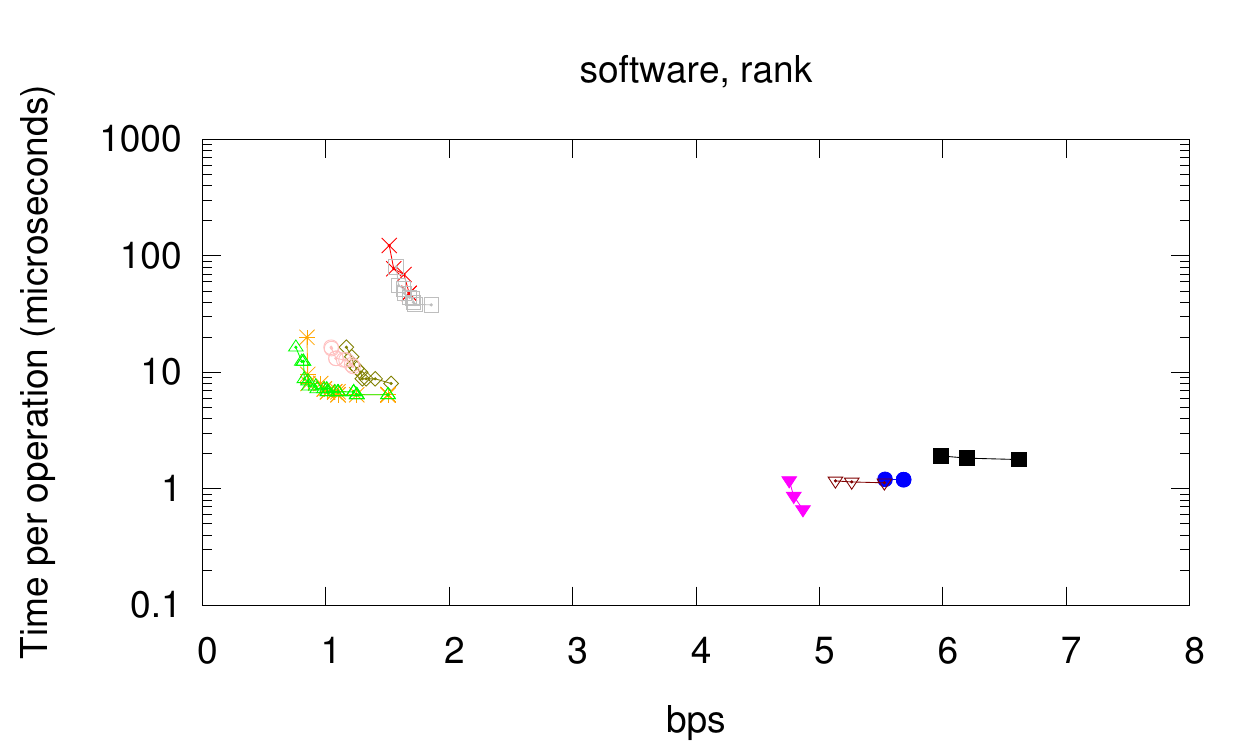}
 \includegraphics[width=0.49\textwidth]{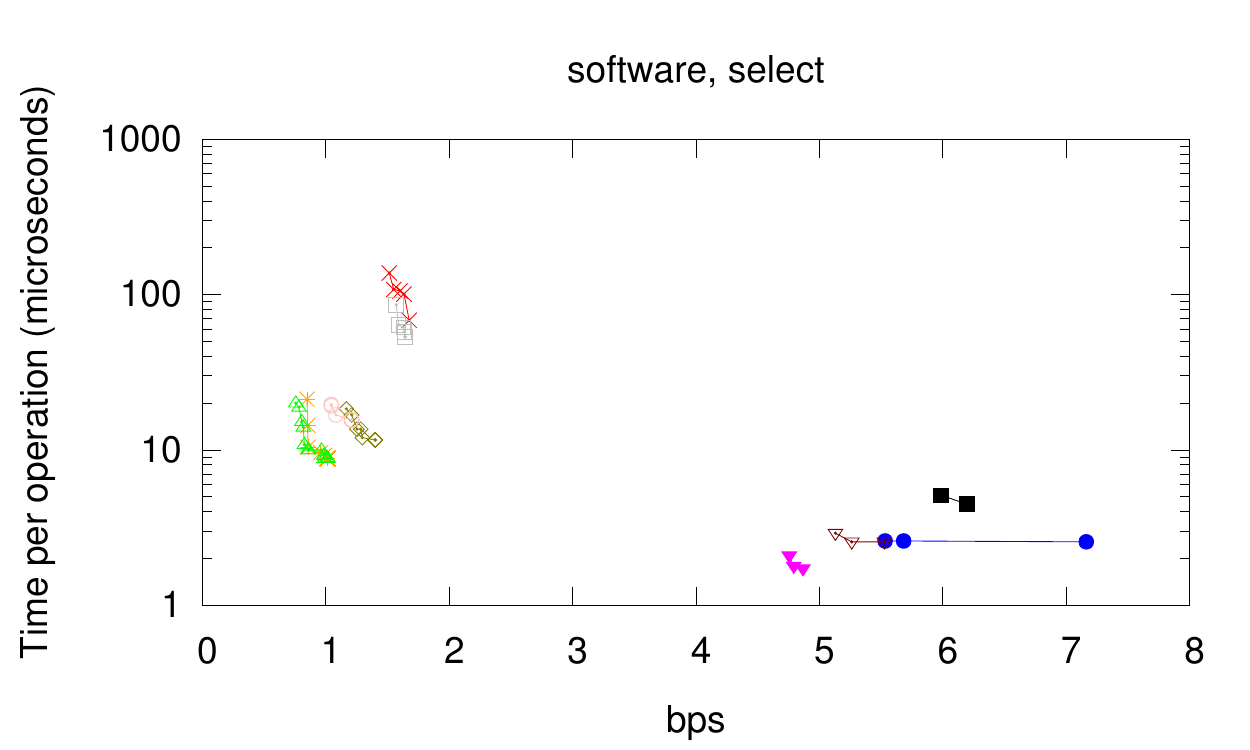}
 
 \includegraphics[width=0.49\textwidth]{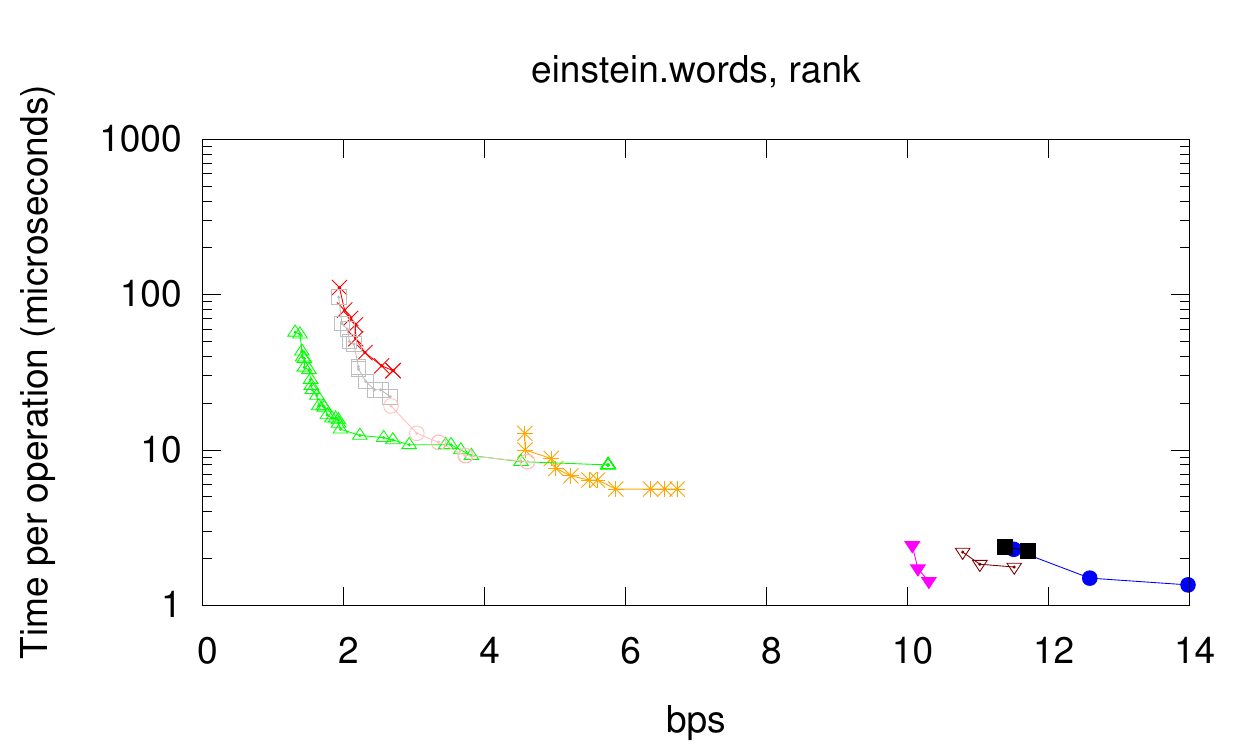}
 \includegraphics[width=0.49\textwidth]{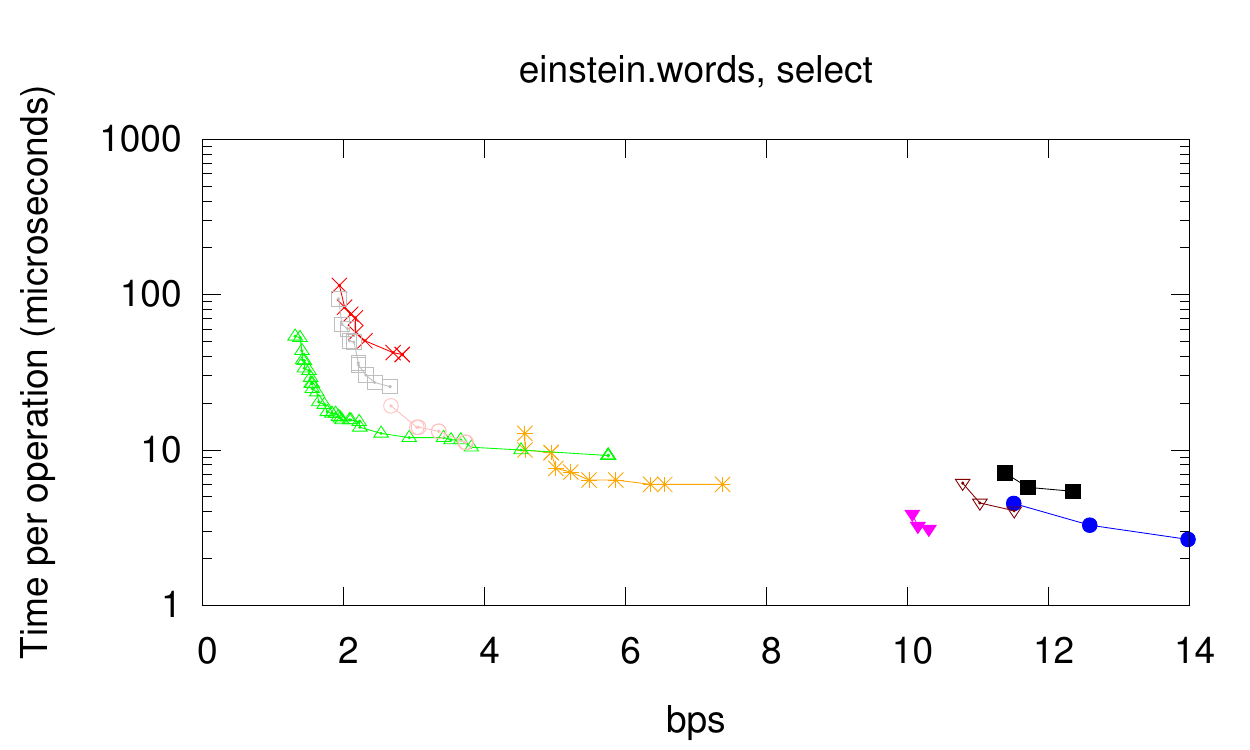}

 \includegraphics[width=0.49\textwidth]{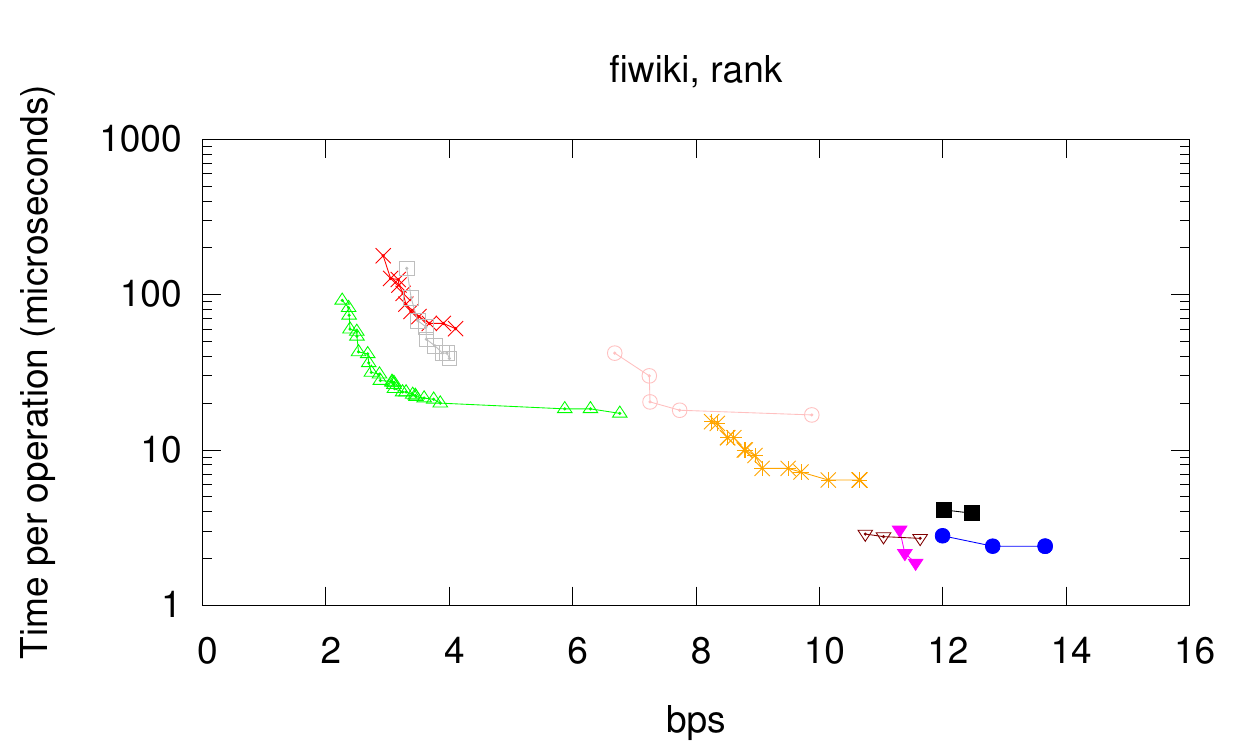}
 \includegraphics[width=0.49\textwidth]{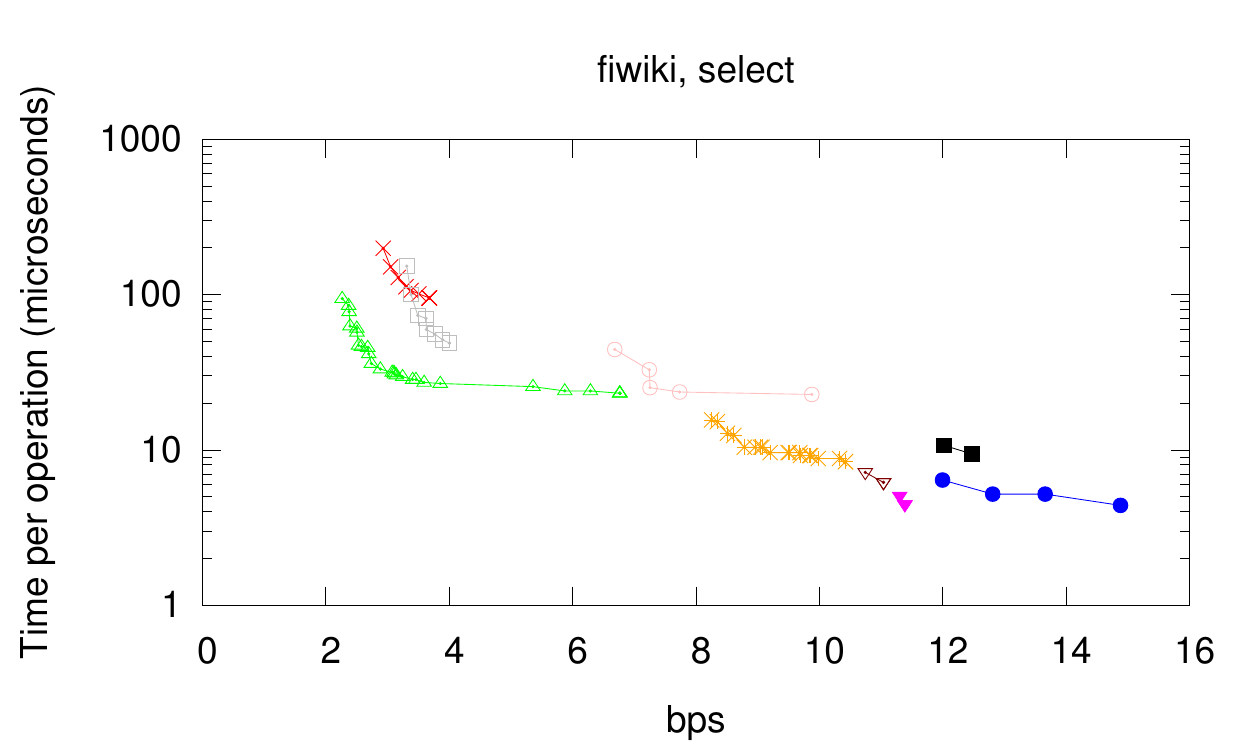}

 \includegraphics[width=0.49\textwidth]{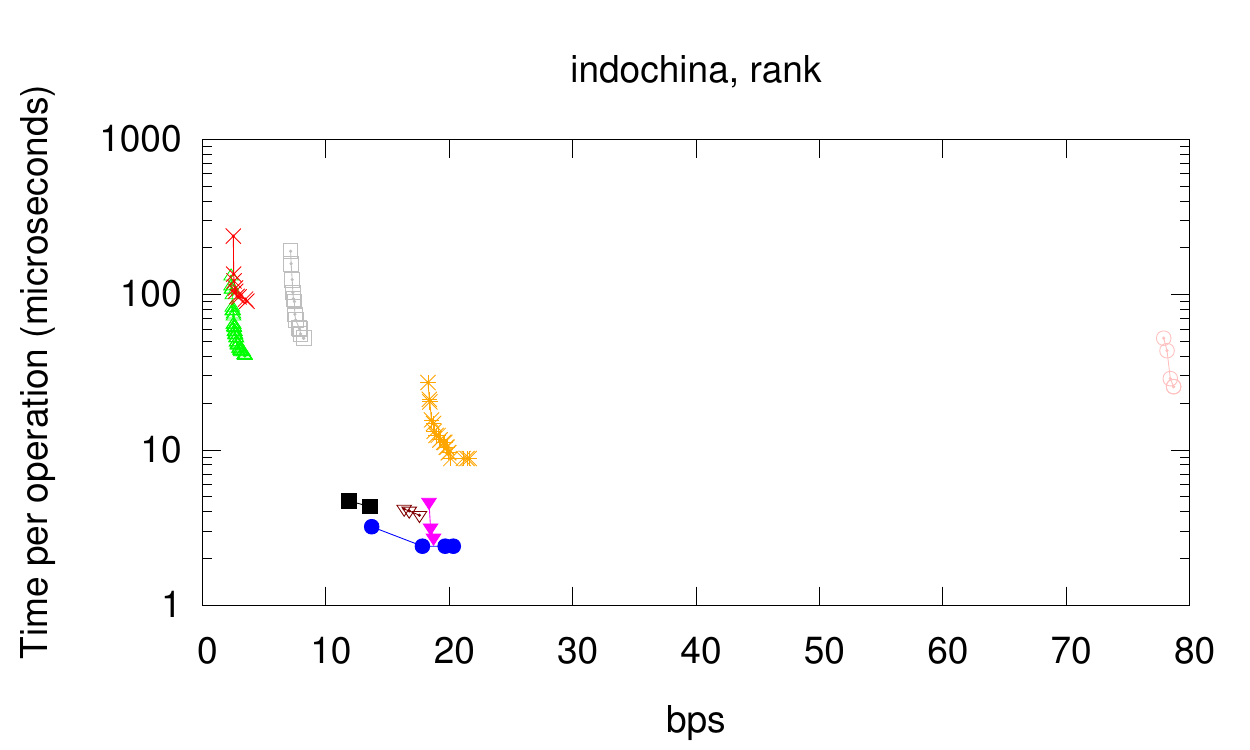}
 \includegraphics[width=0.49\textwidth]{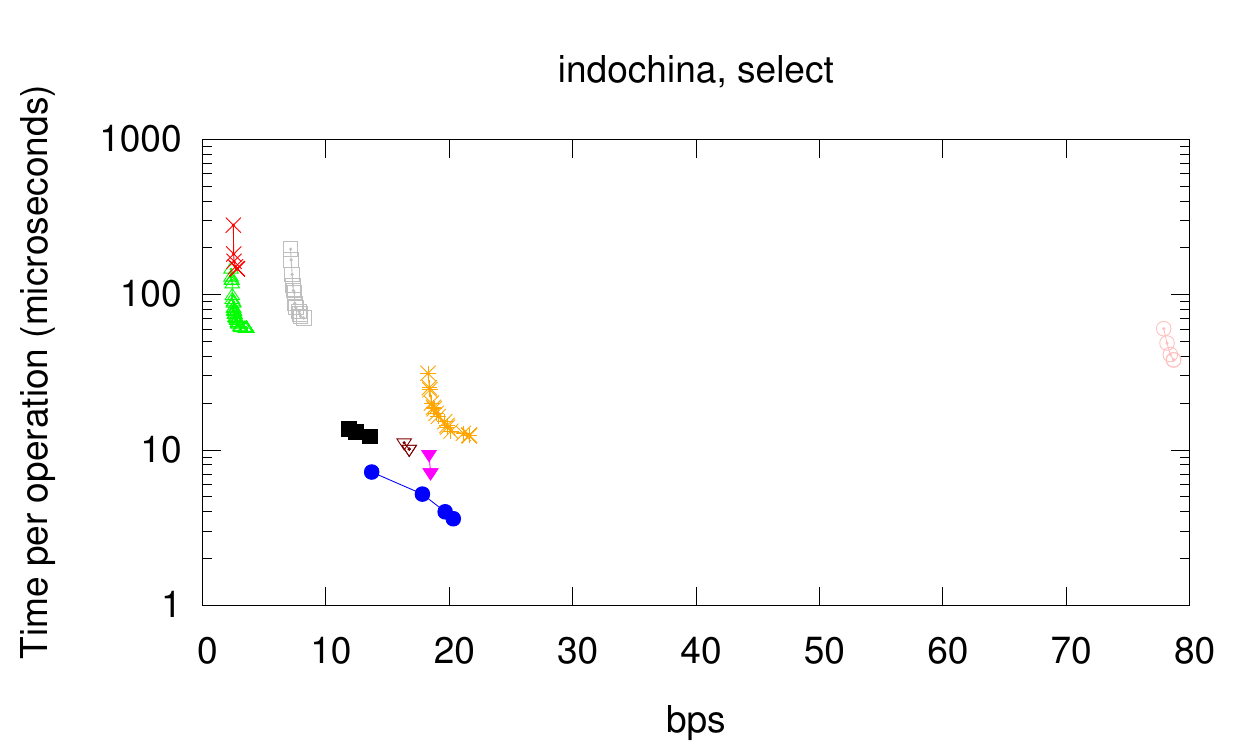}

 \caption{Space-time tradeoffs for $\rank$ and $\select$ queries over moderate
and large alphabets (time in logscale). }
 \label{fig:large.1}
 \end{figure}

Figure~\ref{fig:large.1} shows the results for $\rank$ and $\select$ queries
(once again, $\access$ is omitted for being very similar to the results of
$\rank$).

Recall that $\wm.\M$ is our improvement over the previous work, $\wtrp$
\cite{NPVjea13}. The Huffman-shaped variant, $\wmh.\M$, outperforms it only
slightly in time. Our multi-ary version, $\mwth.\M$, is clearly faster, but
not smaller as one could expect. Indeed, it is larger when $\sigma$
grows, probably due to the use of pointers. What is most interesting, however,
is that all those variants are clearly superseded by our $\ap.\M.\wmrp$, which
dominates them all in time (only reached by $\mwth.\M$ while 
using much more space) and in space (only reached by $\wm.\M$ while using much 
more time). Compared with previous work \cite{NPVjea13}, $\ap.\M.\wmrp$ is
then 2--4 times faster than $\wtrp$, while using the same space or less.
$\ap.\M.\wmrp$ handles queries in a few tens of microseconds.

Note the particularly bad performance of the Huffman-based
versions on {\tt Indochina}. This is because this collection contains inverted
lists, which form long increasing sequences that become runs in the wavelet
tree; the Huffman rearrangement breaks those runs.

Our second variant, $\ap.\M.\gol$, is not so interesting for repetitive
collections. On {\tt einstein} and {\tt software} it performs similarly to
$\ap.\M.\wmrp$. On the others, it is 2--5 times faster, but it uses much more 
space than $\ap.\M.\wmrp$, not so far from that used by statistical 
representations.
Those are, as before, about an order of magnitude faster than $\ap.\M.\wmrp$,
but also use 3--5 times more space. Also, we can see that $\gccn$ is
competitive on {\tt einstein}, which is very repetitive, but not so much on
{\tt software}. Both of our new $\ap.\M$ versions designed for large alphabets
outperform it in space, while they are not slower in time (in some cases they are
even faster).

\no{
 \begin{figure}[t]
 \centering
 
 \includegraphics[width=0.49\textwidth]{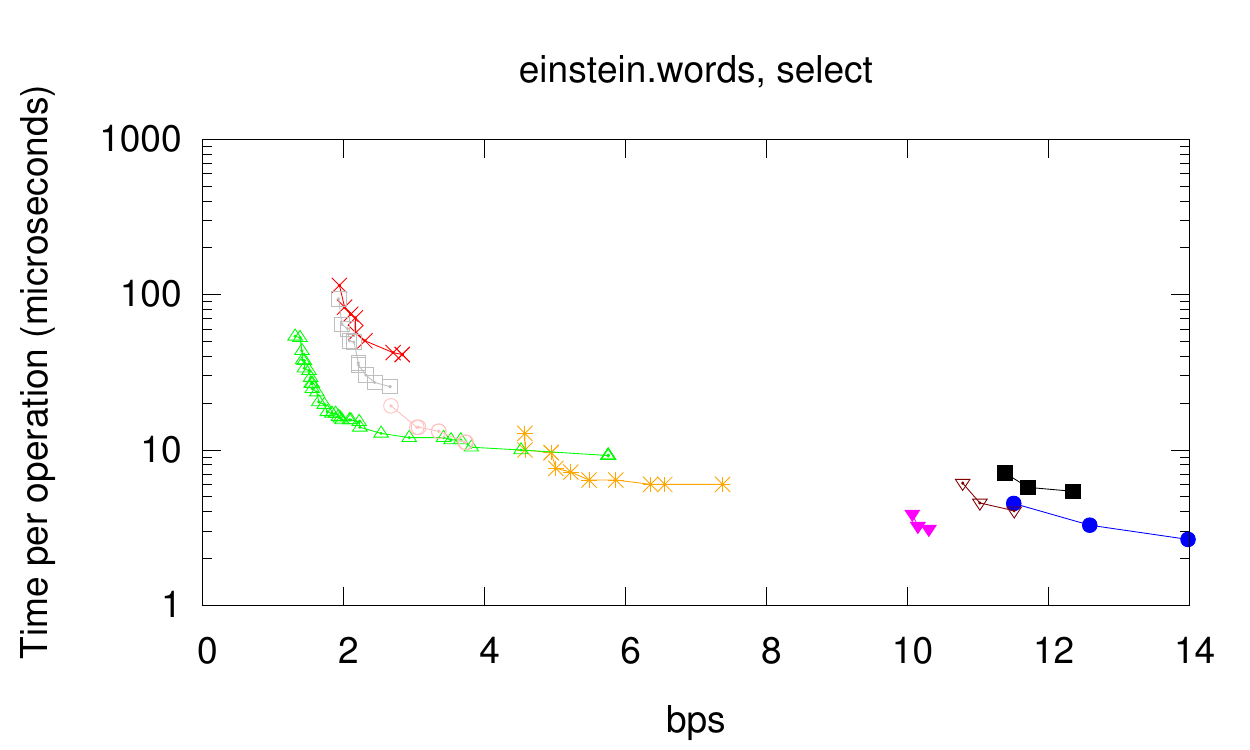}
 \includegraphics[width=0.49\textwidth]{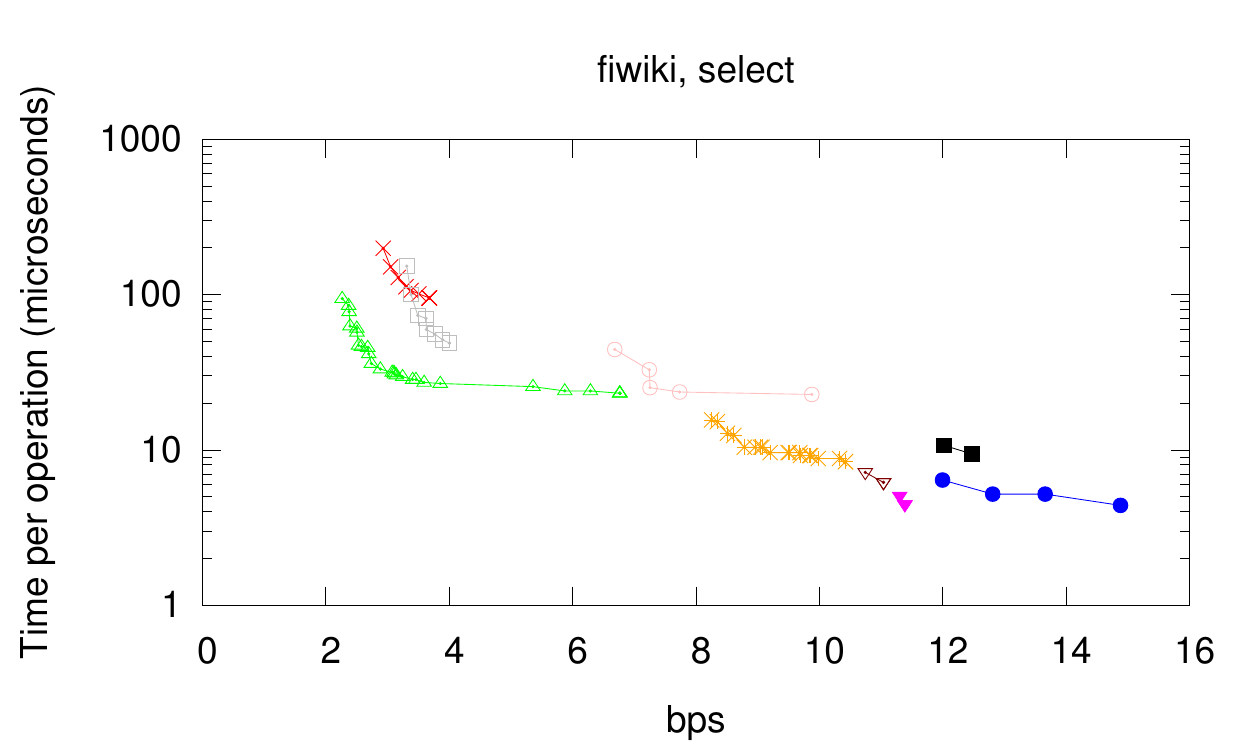}
 \includegraphics[width=0.49\textwidth]{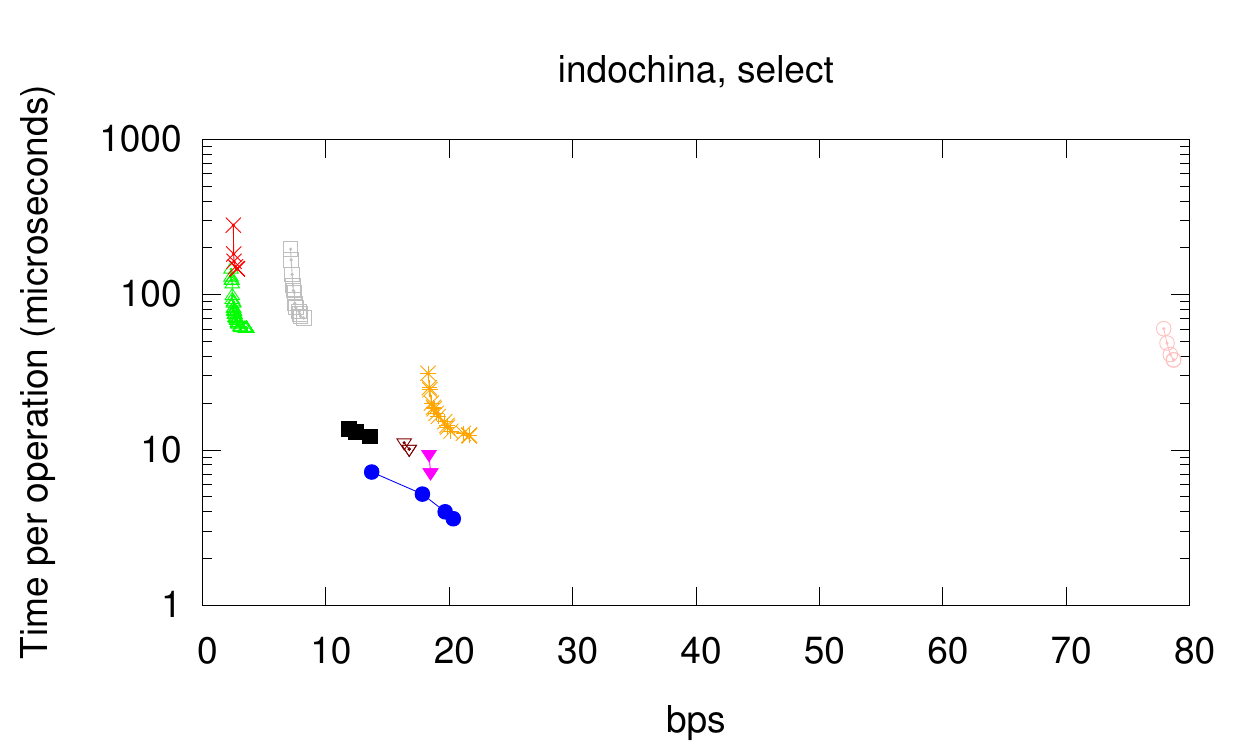}
 
 \caption{Space-time tradeoffs for $\select$ queries over large
alphabets (time in logscale).}
 \label{fig:large.2}
 \end{figure}
}

\section{Applications} \label{sec:app}

We explore now a couple of text indexing applications, where our new
$\rsa$-capable representations can improve the space for repetitive text
collections.

\subsection{Self-indices} \label{sec:app-fm}

Given a string $S[1,n]$ over alphabet $\Sigma = [1,\sigma]$, a
\textit{self-index} is a data structure that represents $S$ and handles operations $\countp(p)$, which returns the number of occurrences of a string pattern $p$ in $S$; $\locate(p)$, which reports the positions of the occurrences of $p$ in $S$; and $\extract(i,j)$, which retrieves $S[i,j]$.

A well-known family of self-indices are the FM-Indices~\cite{FM05}. Modern
FM-Indices \cite{FMMN07} build all their functionality on $\access$ and
$\rank$ queries on the BWT (Burrows Wheeler Transform) \cite{BWT} of $S$,
$S_{bwt}$. Then, operation $\countp$ on $p[1,m]$ takes time $O(m \cdot
\alpha)$, $\alpha$ being the time to answer $\access$ and $\rank$ queries on
$S_{bwt}$. The time to answer $\locate$ and $\extract$ is also proportional to $\alpha$. Therefore, the time of $\rsa$ queries on $S_{bwt}$ directly impacts on the FM-Index performance.

The string $S_{bwt}$ is a reordering of the symbols of $S$, therefore $H_0(S_{bwt})=H_0(S)$. Thus, zero-order-compressed representations of $S$ also obtain zero-order compression of $S_{bwt}$. However, some kinds of zero-order compressors, in particular $\wt.\rrr$ and $\wm.\rrr$, applied on $S_{bwt}$ obtain $nH_k(S)$ bits of space for any $k < \log_\sigma n$ \cite{MNtalg08}. Further, $S_{bwt}$ is typically formed by a few long {\em runs} of equal symbols: the number of runs is at most $nH_k(S)+\sigma^k$ for
any $k$ \cite{MN05}, and the number is much lower on repetitive sequences \cite{MNSV09jcb}. Thus, in a highly repetitive scenario, the runs of $S_{bwt}$ are much longer than $\log_{\sigma}n$ (see Table~\ref{table:datasets}), and typical $k$-order statistical compression of $S_{bwt}$ fails to capture its most important regularities.

Run-Length FM-Indices \cite{MN05,MNSV09jcb} aim to capture these regularities. A Run-Length FM-Index stores in $S^{'}_{bwt}$ the first symbol of each run, marking their positions in a bitmap $R[1,n]$ (they also store a bitmap $R'[1,n]$ with a reordering of the bits in $R$). Compressed Suffix Arrays (CSAs) \cite{GV05,Sad03} (another family of self-indices) have also been adapted to exploit these runs, in a structure called Run-Length CSA \cite{MNSV09jcb}. In general, FM-Indices are preferred over CSAs for sequences over small alphabets, because the cost of $\rsa$ operations increases with $\sigma$, while the equivalent operations on the CSAs do not depend on it.

An alternative to Run-Length FM-Indices is to grammar-compress $S_{bwt}$ with $\gcc$, our $\rsa$ structure for repetitive sequences on small alphabets. To evaluate if grammar compression of $S_{bwt}$ captures more regularities than run-length compression, we compare the following FM-Index implementations:
\begin{itemize}
\item {\tt FMI-$\gcc$}, using the variant $\gccn$ to represent $S_{bwt}$.
\item {\tt FMI-$\ap.\M.\wtrp$}, using the variant $\ap.\M.\wtrp$ to represent 
$S_{bwt}$.
\item {\tt FMI-$\wth.\rrr$}, which uses $\wth.\rrr$ to represent $S_{bwt}$.
\item {\tt FMI-$\wt.\rrr$}, which uses $\wt.\rrr$ to represent $S_{bwt}$.
\item {\tt RLFMI-$\wth$+$\deltae$}, a Run-Length FM-Index \cite{MNSV09jcb} where bitmaps $R$ and $R'$ are compressed with $\deltae$, while $S^{'}_{bwt}$ is represented with $\wth.\rrr$.
\item {\tt RLCSA}, a Run-Length Compressed Suffix Array \cite{MNSV09jcb} setting the sampling rate of its function $\Psi$ to $\{32,64,128\}$.
\end{itemize}

\begin{figure}[t]
 \centering
 
\includegraphics[width=0.49\textwidth]{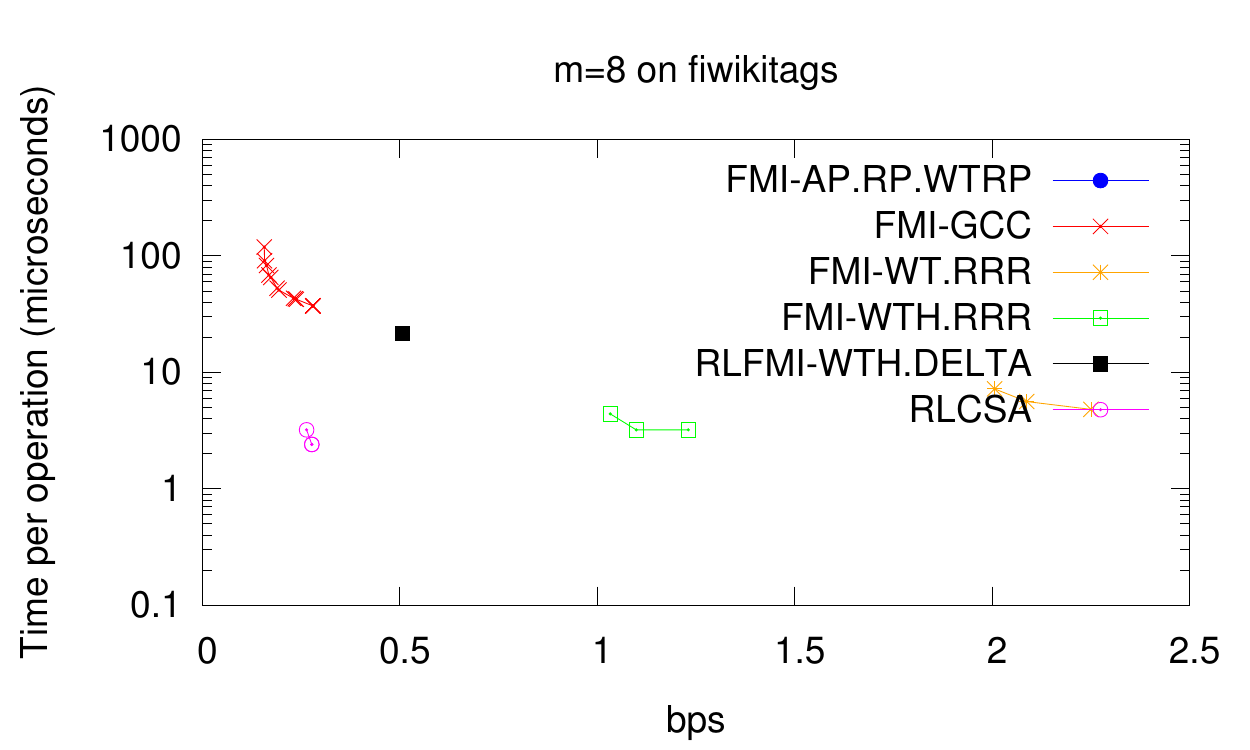}
\includegraphics[width=0.49\textwidth]{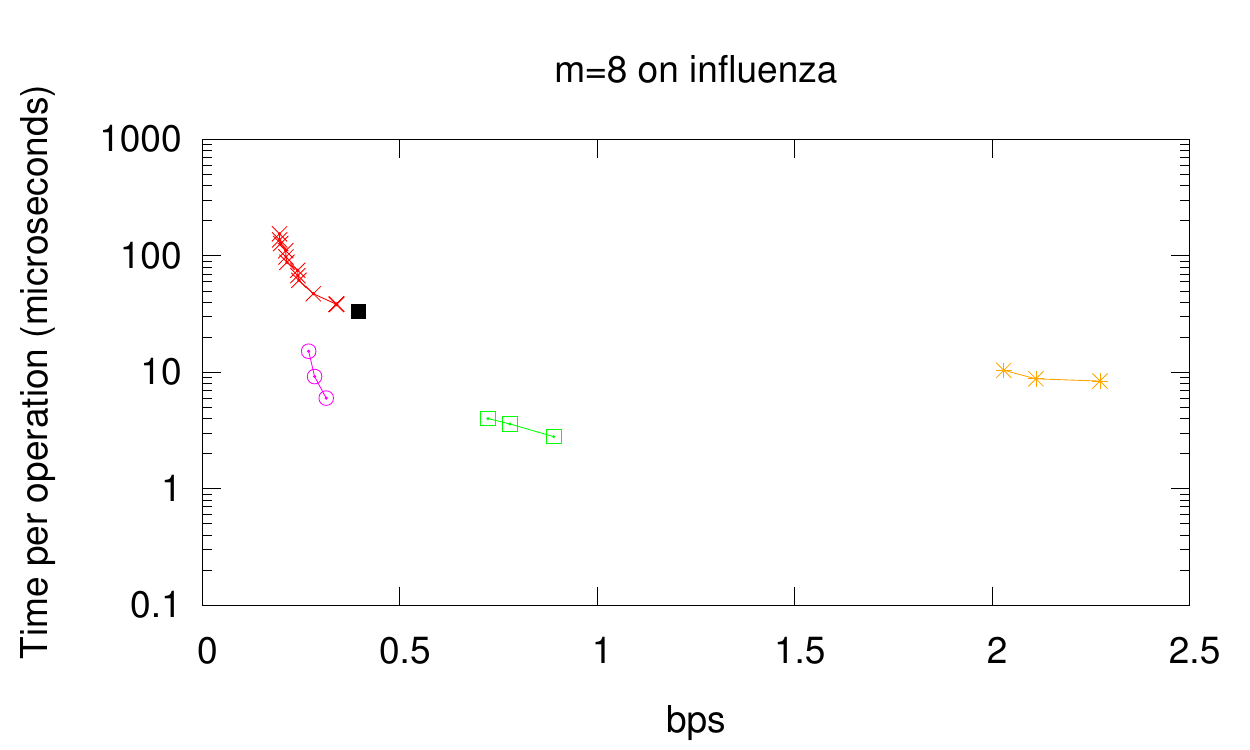}

\includegraphics[width=0.49\textwidth]{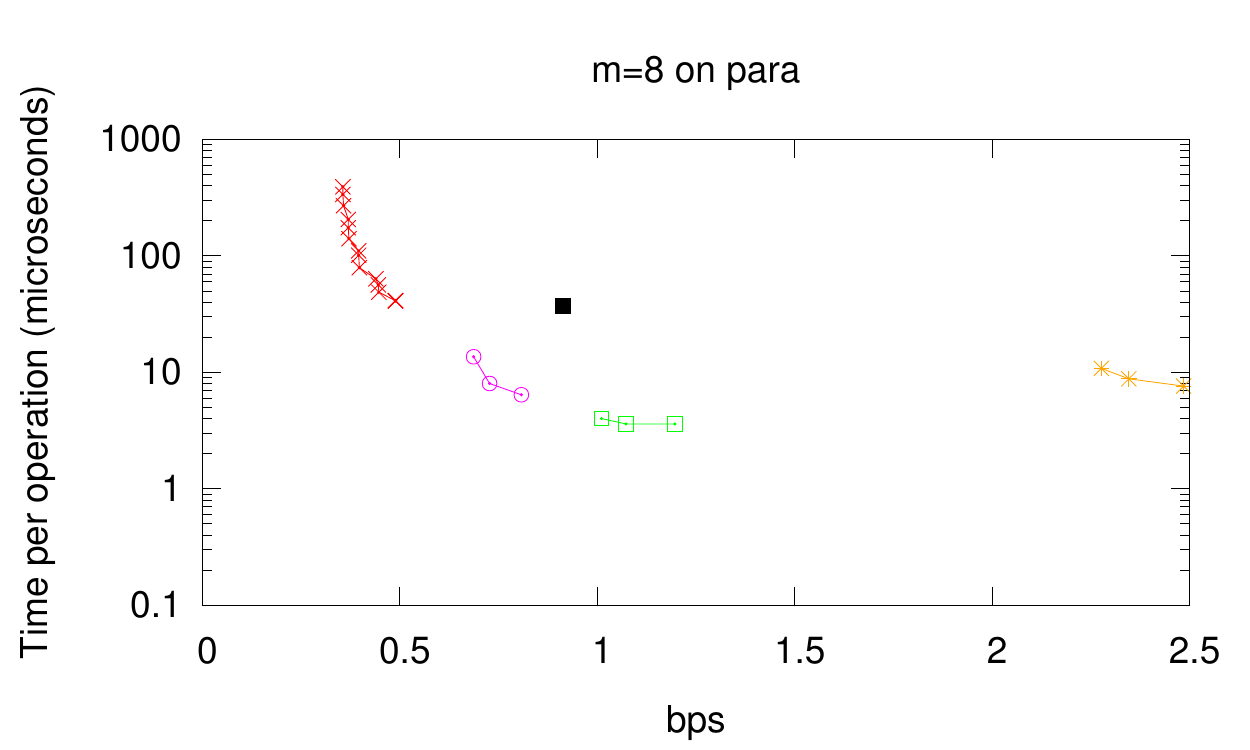}
\includegraphics[width=0.49\textwidth]{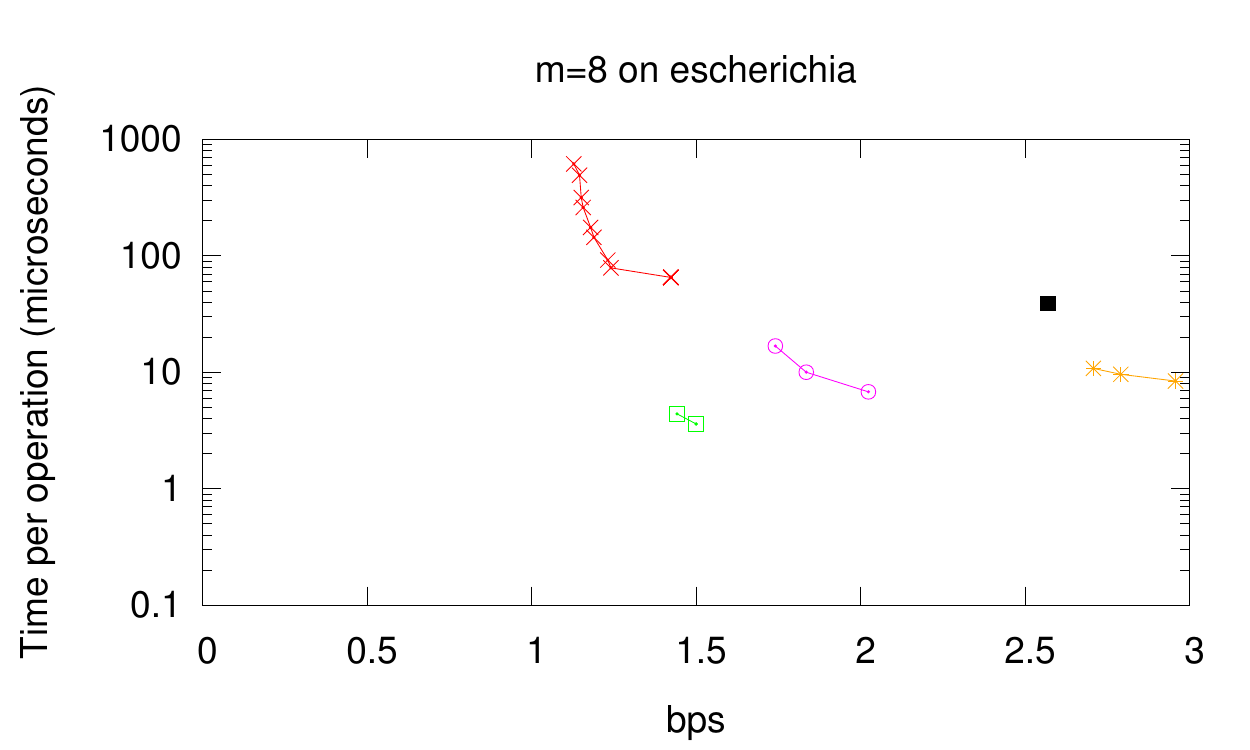}

\includegraphics[width=0.49\textwidth]{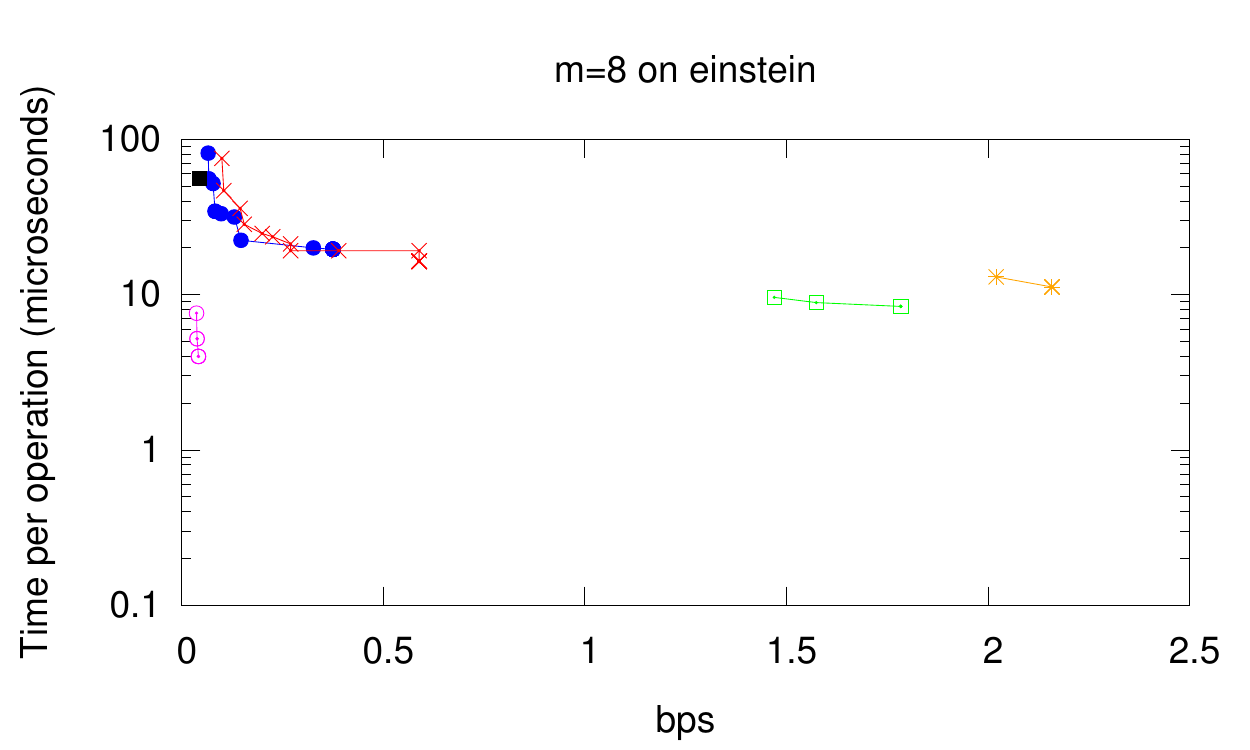}
\includegraphics[width=0.49\textwidth]{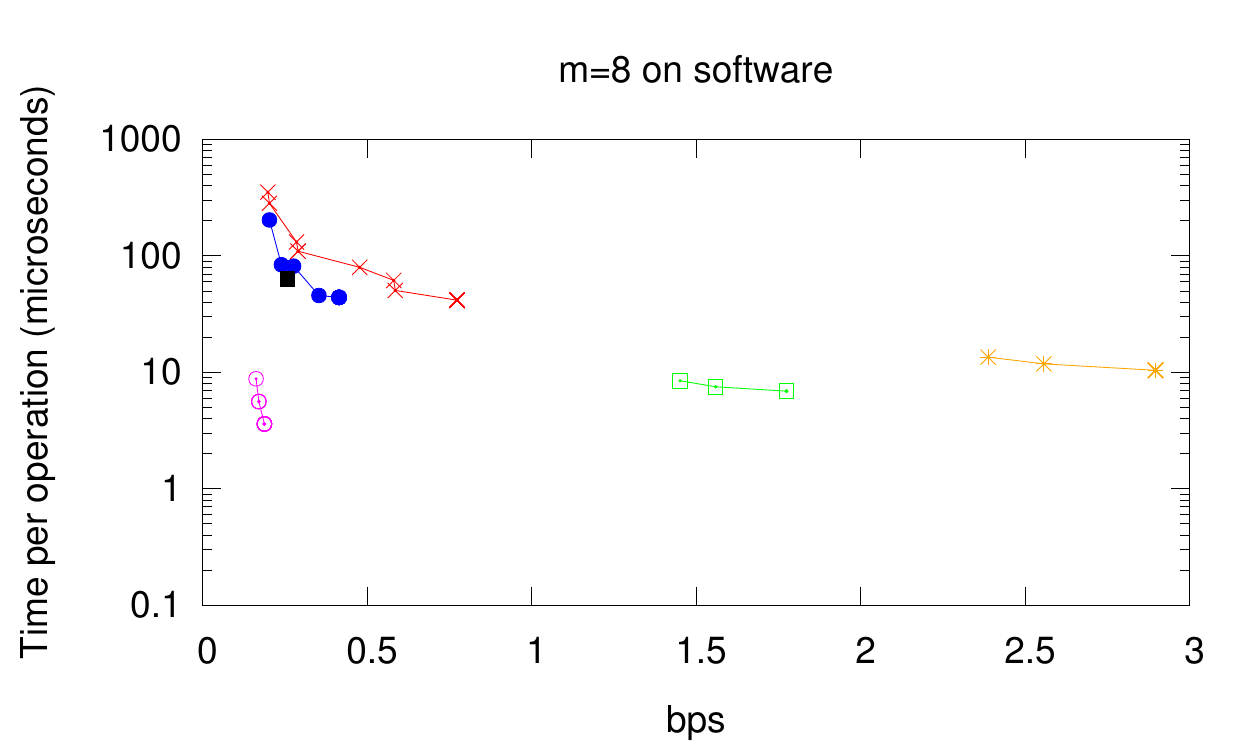}

 \caption{Space-time tradeoffs for operation $\countp$ with $m=8$.}
 \label{fig:count.small}
 \end{figure}

We used the real DNA datasets and {\tt fiwikitags}, as well as {\tt einstein}
and {\tt software} to show the case of larger alphabets. We averaged 
$10{,}000$ queries for patterns picked at random from each dataset. 
%First, we compare the space-time performance of the different FM-Index implementations by evaluating the $LF$-mapping operation, which is the core of all their functionality. The $LF$-mapping is defined as: $$LF(i) = C[S_{bwt}[i]]+\rank_{S_{bwt}[i]}(i)$$
%
%where $C[i]$ stores how many symbols in $S$ are lexicographically smaller than $i$, $i\in \Sigma$. 
%
%
%
%Figures~\ref{fig:lf.small} show the results for this operation (note the {\tt RLCSA}s is not included in this comparison since it is not an FM-Index). As we can see, our implementations, both $\gccc$ and $\gccn$, obtain the best space performance in all the datasets used. It takes around half the space (or even less) than the {\tt RLFMI-$\wth+\deltae$}, which is the next most space efficient solution. Compared with the best statistical approach, the {\tt FMI-$\wth.\rrr$}, differences are even larger: our solutions need only from the 20\% to the 40\% of the space on the most repetitive collections ($\fitags$, $\influ$, and $\para$), but being closer in collection $\esch$, the less- or non-repetitive dataset. In terms of time performance, both $\gccn$ and $\gccc$ runs in the same order of magnitude than the {\tt RLFMI-$\wth+\deltae$} but being slower: from $8\mu s$ in $\influ$ to $1-3\mu s$ in the rest. Compared with {\tt FMI-$\wth.\rrr$}, we run around an order of magnitude slower in all the datasets but paying much less space. 
%
We evaluate the performance of the operation $\countp$ in the indices, for
various pattern lengths. Figure~\ref{fig:count.small} shows the results for
$m=8$, since all the lengths gave similar results.

As it can be seen, the FMI-$\gcc$ obtains the least space on the smaller 
alphabets. The space of the {\tt RLCSA} is close, but still larger than that of the FMI-$\gcc$, in collections $\fitags$ and $\influ$. For $\para$ and $\esch$ the differences are larger, our structure using 60\%--80\% of the {\tt RLCSA} space. Interestingly, grammar compression of $S_{bwt}$ is stronger than the {\tt RLCSA} compression especially when the sequence is not so repetitive. In exchange, the {\tt RLCSA} is about an order of magnitude faster. 

Our index also uses half the space, or less, than the {\tt RLFMI-$\wth$+$\deltae$}, which also adapts to repetitiveness but not as well as grammar compression, and performs badly as soon as repetitiveness starts to decrease. Finally, compared with the best statistical approach, the {\tt FMI-$\wth.\rrr$}, the differences are even larger: our solution needs only 20\%--40\% of the space in the most repetitive collections, only getting closer in $\esch$, which is not so repetitive. 

In terms of time performance, the FMI-$\gcc$ is in the same order of
magnitude of {\tt RLFMI-$\wth$+$\deltae$}, yet it is slower. Compared with
{\tt FMI-$\wth.\rrr$}, our index is about an order of magnitude slower. 

On the larger alphabets, instead, the FMI-$\ap.\M.\wtrp$ outperforms
the FMI-$\gcc$ and uses about the same space as the 
{\tt RLFMI-$\wth$+$\deltae$}, while being faster or equally fast. It is only 
2--4 times slower than the statistical approaches, while using 10\%--20\% of their
space. However, as expected, the {\tt RLCSA} outperforms every FM-index on 
larger alphabets. 
%Yet, in some applications the FM-index 
%cannot be replaced by a {\tt RLCSA}, as specific properties of the BWT are 
%used \cite{Ohl13}.

In the sequel we call {\tt GFMI} to {\tt FMI}-$\gcc$ or {\tt FMI}-$\ap.\M.\wtrp$, whichever is better.

% \begin{figure}[t]
% \centering
% 
%\includegraphics[width=0.49\textwidth]{fiwikitagslf.pdf}
%\includegraphics[width=0.49\textwidth]{influenzalf.pdf}
%
%\includegraphics[width=0.49\textwidth]{paralf.pdf}
%\includegraphics[width=0.49\textwidth]{escherichialf.pdf}
%
% \caption{Space-time tradeoffs for $LF$ operation over 
% small alphabets.}
% \label{fig:lf.small}
% \end{figure}

\subsection{XML and XPath}
\label{sec:appXML}

Now
we show the impact of our new representations in the indexing of repetitive
XML collections. $\sxsi$ \cite{SXSI} is a recent system that represents XML
datasets in compact form and supports XPath queries on them. Its query processing strategy uses a tree automaton that traverses the XML data, using several queries on the content and structure to speed up navigation towards the points of interest. $\sxsi$ represents the XML data using three separate components: (1) a text index that represents and carries out pattern searches over the text nodes (any compressed full-text index \cite{NM07} can be used); (2) a balanced parentheses representation of the XML topology that supports navigation using $2+o(1)$ bits per node (various alternatives exist \cite{ACNS10}); and (3) an $\rsa$-capable representation of the sequence of the XML opening and closing tags.

When the XML collection is repetitive (e.g., versioned collections like
Wiki\-pedia, versioned software repositories, etc.), one can use the {\tt
RLCSA} \cite{MNSV09jcb} as the text index for (1), but now we also consider
using our new {\tt GFMI}. Components (2) and (3), which are usually less relevant in terms of space, may become dominant if they are represented without exploiting repetitiveness. For (2), we consider $\gct$, a tree representation aimed at repetitive topologies \cite{NOsea14.1}, and a classical representation ({\tt FF} \cite{ACNS10}). For (3), we will use our new repetition-aware sequence representations, comparing them with the alternative proposed in $\sxsi$ ({\tt MATRIX}, using one compressed bitmap per tag) and a $\wth$ representation. 

We use a repetitive data-centric XML collection of 200MB from a real software repository. Its sequence of XML tags, called {\tt software}, is described in Table~\ref{table:datasets}. As a proof of concept, we run two XPath queries that make intensive use of the sequence of tags and the tree topology: {\tt XQ1=//class[//methods]}, and {\tt XQ2=//class[methods]}. 

Table \ref{table:results.sxsi} shows the space in bpe (bits per element) of
components (2) and (3). An element is an opening or a closing tag, so there
are two elements per XML tree node. The space of the {\tt RLCSA} without
sampling is always 0.18 bits per character of the XML document, whereas our
new {\tt GFMI} uses 0.15 if combined with $\ap.\M.\wmrp$. The table also shows 
the impact of each component in the total size of the index, considering this
last space. On the rightmost columns, it shows the time to answer both queries.

The original $\sxsi$ ({\tt MATRIX+FF}) is very fast but needs almost 14 bpe,
which amounts to 98\% of the index space in this repetitive scenario (in
non-repetitive text-centric XML, this space is negligible). By replacing the
{\tt MATRIX} by a $\wth$, the space drops significantly, to slightly over 4
bpe, yet times degrade by a factor of 3--6. By using our $\gcc$ for the tags,
a new significant space reduction is obtained, to 2.65 bpe, and the times
increase by a factor of 2, becoming 6--12 times slower than the original
{\tt SXSI}. Finally, changing {\tt FF} by $\gct$ \cite{NOsea14.1}, we can
reach as low as 0.56 bpe, 24 times less than the original {\tt SXSI}, and
using around 60\% of the total space. Once again, the price is the time, which
becomes 50--90 times slower than the basic {\tt SXSI}. The price of using the slower $\gct$ is more noticeable on XQ2, which uses more operations on the tree.

While the time penalty is 1--2 orders of magnitude, we note that the gain in space can make the difference between running the index in memory or on disk; in the latter case we can expect it to be up to 6 orders of magnitude slower. 
%On the other hand, the time differences will blur on queries that do not only access the tags and the tree, but also involve the text, as these cost the same in all the representations. Finally, we note that the {\tt RLCSA} becomes the space bottleneck in {\tt $\gcc$+$\gct$}. It is worthwhile to consider even more compressed text representations, for example based on grammars \cite{CN12} or on LZ77 \cite{KN12}. 

\begin{table}[t]
\centering
\begin{tabular}{l@{~}|@{~}r@{~}r@{~}|@{~}r@{~}r@{~}r@{~}|@{~}r@{~}r}
\textbf{dataset}&\textbf{tags}&\textbf{tree}&\textbf{\%tags}&\textbf{\%tree}&\textbf{\%text}&\textbf{XQ1}&\textbf{XQ2}\\ 
\midrule
% calc con 2.3 bpc para rlcsa
%    {\tt MATRIX+FF}  	&  12.40 & 1.27  & 69.00 &  7.19 & 23.90 & 16 & 35  \\ 
%    {\tt $\wth$+FF}  	&   2.88 & 1.27  & 34.07 & 15.09 & 50.84 & 92 & 113    \\ 
%    {\tt $\gcc$+FF}  	&   0.37 & 1.27  & 6.26 & 21.45 & 72.29 & 442 & 462 \\
%    {\tt $\gcc$+$\gct$} &  	0.37 & 0.19  & 7.66 &  3.93 & 88.42 & 1,032 & 3,302			\\ 
% calc con 0.15 bpc para gfmi
    {\tt MATRIX+FF}  	&  12.40 & 1.27  & 88.89 &  9.12 & 1.99 & 16 & 35  \\ 
    {\tt $\wth$+FF}  	&   2.88 & 1.27  & 65.00 & 28.68 & 6.32 & 92 & 113    \\ 
    {\tt $\gcc$+FF}  	&   0.37 & 1.27  & 19.29 & 66.17 & 14.54 & 184 & 226 \\
    {\tt $\gcc$+$\gct$} &   0.37 & 0.19  & 44.13 & 22.65 & 39.74 & 774 & 3,066			\\ 
\midrule
  \end{tabular}
  \vspace{2mm}
  \caption{Results on XML. Columns \textbf{tags} and \textbf{tree} are in bpe. Columns \textbf{XQ1} and 
  			\textbf{XQ2} show query time in microseconds.}
  \label{table:results.sxsi}
\end{table}

\section{Conclusions}\label{sec:concl}

We have introduced new sequence representations that take advantage of the
repetitiveness of the sequence, by enhancing the output of a grammar
compressor with extra information to support efficient direct access, as well
as $\rank$ and $\select$ operation on the sequence. The only previous
grammar-compressed representation \cite{NPVjea13} is 2--15 times slower and
uses the same or more space than our new representations. Our structures answer
queries in a few tens of microseconds, which is about an order of magnitude 
slower than the times of statistically compressed representations. However, on
repetitive collections, our structures use 2--15 times less space.
We have also explored two applications where repetitiveness is a sharp source 
of compressibility, and have shown how our structures allow one to further
exploit that repetitiveness to obtain significantly less space.

An aspect where our structures could possibly be improved is in the clustering
of the alphabet symbols used when partitioning the alphabet, both in the
simple case of alphabet partitioning and in the hierarchical case of wavelet
trees and matrices. In the first case, we obtained a significant space
improvement by sorting the symbols by frequency, whereas in the second case
none of our attempts performed noticeably better than the original alphabet
ordering. While unsuccessful for now, we believe that some clever clustering
scheme that avoids separating symbols that appear together in repetitive parts
of the sequence could considerably improve the space on large alphabets.

Another future goal is to find ways to improve the time of these grammar
compressed representations. We believe this is possible, even if known lower 
bounds suggest that there must
be a price of at least an order of magnitude compared with statistically
compressed representations. A more far-fetched goal is to build on Lempel-Ziv 
compressed representations. Lempel-Ziv is more powerful than grammar 
compression, but supporting the desired operations on it is thought to be more
difficult.

\section*{References}

\bibliographystyle{plain}
\bibliography{paper}

\begin{thebibliography}{10}

\bibitem{ACNS10}
D.~Arroyuelo, R.~C{\'a}novas, G.~Navarro, and K.~Sadakane.
\newblock Succinct trees in practice.
\newblock In {\em Proc. 12th Workshop on Algorithm Engineering and Experiments
  (ALENEX)}, pages 84--97, 2010.

\bibitem{SXSI}
D.~Arroyuelo, F.~Claude, S.~Maneth, V.~M{\"a}kinen, G.~Navarro,
  K.~Nguy{$\tilde{\hat{\textrm{e}}}$}n, J.~Sir{\'e}n, and N.~V{\"a}lim{\"a}ki.
\newblock Fast in-memory xpath search using compressed indexes.
\newblock {\em Software Practice and Experience}, 45(3):399--434, 2015.

\bibitem{AGCGMO12}
D.~Arroyuelo, V.~Gil-Costa, S.~Gonz{\'a}lez, M.~Mar\'{\i}n, and M.~Oyarz{\'u}n.
\newblock Distributed search based on self-indexed compressed text.
\newblock {\em Information Processing and Management}, 48(5):819--827, 2012.

\bibitem{AGMOS12}
D.~Arroyuelo, S.~Gonz{\'a}lez, M.~Mar\'{\i}n, M.~Oyarz{\'u}n, and T.~Suel.
\newblock To index or not to index: time-space trade-offs in search engines
  with positional ranking functions.
\newblock In {\em Proc. 35th International ACM Conference on Research and
  Development in Information Retrieval (SIGIR)}, pages 255--264, 2012.

\bibitem{AGO10}
D.~Arroyuelo, S.~Gonz{\'a}lez, and M.~Oyarz{\'u}n.
\newblock Compressed self-indices supporting conjunctive queries on document
  collections.
\newblock In {\em Proc. 17th International Symposium on String Processing and
  Information Retrieval (SPIRE)}, LNCS 6393, pages 43--54, 2010.

\bibitem{baeza1999modern}
R.~Baeza-Yates and B.~Ribeiro-Neto.
\newblock {\em Modern Information Retrieval}.
\newblock Addison-Wesley, 2nd edition, 2011.

\bibitem{BCGNNalgor13}
J.~Barbay, F.~Claude, T.~Gagie, G.~Navarro, and Y.~Nekrich.
\newblock Efficient fully-compressed sequence representations.
\newblock {\em Algorithmica}, 69(1):232--268, 2014.

\bibitem{BCN13}
J.~Barbay, F.~Claude, and G.~Navarro.
\newblock Compact binary relation representations with rich functionality.
\newblock {\em Information and Computation}, 232:19--37, 2013.

\bibitem{BHMR11}
J.~Barbay, M.~He, J.~I. Munro, and S.~S. Rao.
\newblock Succinct indexes for strings, binary relations and multilabeled
  trees.
\newblock {\em ACM Transactions on Algorithms}, 7(4):article 52, 2011.

\bibitem{BN13}
J.~Barbay and G.~Navarro.
\newblock On compressing permutations and adaptive sorting.
\newblock {\em Theoretical Computer Science}, 513:109--123, 2013.

\bibitem{BPTesa15}
D.~Belazzougui, P.~H. Cording, S.~J. Puglisi, and Y.~Tabei.
\newblock Access, rank, and select in grammar-compressed strings.
\newblock In {\em Proc. 23th Annual European Symposium on Algorithms (ESA)},
  LNCS 9294, pages 142--154, 2015.

\bibitem{BGGKOPT15}
D.~Belazzougui, T.~Gagie, P.~Gawrychowski, J.~K{\"{a}}rkk{\"{a}}inen,
  A.~Ord{\'{o}}{\~{n}}ez, S.~J. Puglisi, and Y.~Tabei.
\newblock Queries on {LZ}-bounded encodings.
\newblock In {\em Proc. 25th Data Compression Conference (DCC)}, pages 83--92,
  2015.

\bibitem{BNesa12}
D.~Belazzougui and G.~Navarro.
\newblock Optimal lower and upper bounds for representing sequences.
\newblock {\em ACM Transactions on Algorithms}, 11(4):article 31, 2015.

\bibitem{BSODA11}
P.~Bille, G.~M. Landau, R.~Raman, K.~Sadakane, S.~S. Rao, and O.~Weimann.
\newblock Random access to grammar-compressed strings and trees.
\newblock {\em SIAM Journal on Computing}, 44(3):513--539, 2015.

\bibitem{BLN12}
N.~Brisaboa, S.~Ladra, and G.~Navarro.
\newblock {DACs}: Bringing direct access to variable-length codes.
\newblock {\em Information Processing and Management}, 49(1):392--404, 2013.

\bibitem{BWT}
M.~Burrows and D.~Wheeler.
\newblock A block sorting lossless data compression algorithm.
\newblock Technical Report 124, Digital Equipment Corporation, 1994.

\bibitem{CLLPPSS05}
M.~Charikar, E.~Lehman, D.~Liu, R.~Panigrahy, M.~Prabhakaran, A.~Sahai, and
  A.~Shelat.
\newblock The smallest grammar problem.
\newblock {\em IEEE Transactions on Information Theory}, 51(7):2554--2576,
  2005.

\bibitem{CPhd98}
D.~Clark.
\newblock {\em Compact {PAT} Trees}.
\newblock PhD thesis, University of Waterloo, Canada, 1996.

\bibitem{CN08}
F.~Claude and G.~Navarro.
\newblock Practical rank/select queries over arbitrary sequences.
\newblock In {\em Proc. 15th International Symposium on String Processing and
  Information Retrieval (SPIRE)}, LNCS 5280, pages 176--187, 2008.

\bibitem{CNtweb10}
F.~Claude and G.~Navarro.
\newblock Fast and compact {W}eb graph representations.
\newblock {\em ACM Transactions on the Web (TWEB)}, 4(4):article 16, 2010.

\bibitem{CNOis14}
F.~Claude, G.~Navarro, and A.~Ord{\'o}{\~n}ez.
\newblock The wavelet matrix: An efficient wavelet tree for large alphabets.
\newblock {\em Information Systems}, 47:15--32, 2015.

\bibitem{FLMM09}
P.~Ferragina, F.~Luccio, G.~Manzini, and S.~Muthukrishnan.
\newblock Compressing and indexing labeled trees, with applications.
\newblock {\em Journal of the ACM}, 57(1):article 4, 2009.

\bibitem{FM05}
P.~Ferragina and G.~Manzini.
\newblock Indexing compressed texts.
\newblock {\em Journal of the ACM}, 52(4):552--581, 2005.

\bibitem{FMMN07}
P.~Ferragina, G.~Manzini, V.~M{\"a}kinen, and G.~Navarro.
\newblock Compressed representations of sequences and full-text indexes.
\newblock {\em ACM Transactions on Algorithms}, 3(2):article 20, 2007.

\bibitem{GGKNP14}
T.~Gagie, P.~Gawrychowski, J.~K{\"a}rkk{\"a}inen, Y.~Nekrich, and S.~J.
  Puglisi.
\newblock {LZ77}-based self-indexing with faster pattern matching.
\newblock In {\em {Proc. 11th Latin American Symposium on Theoretical
  Informatics (LATIN)}}, LNCS 8392, pages 731--742, 2014.

\bibitem{GMR06}
A.~Golynski, I.~Munro, and S.~Rao.
\newblock Rank/select operations on large alphabets: a tool for text indexing.
\newblock In {\em Proc. 17th Annual ACM-SIAM Symposium on Discrete Algorithms
  (SODA)}, pages 368--373, 2006.

\bibitem{GGMN05}
R.~Gonz\'alez, Sz. Grabowski, V.~M{\"akinen}, and G.~Navarro.
\newblock Practical implementation of rank and select queries.
\newblock In {\em Poster Proc. Volume of 4th Workshop on Efficient and
  Experimental Algorithms (WEA)}, pages 27--38, 2005.

\bibitem{GGV03}
R.~Grossi, A.~Gupta, and J.~Vitter.
\newblock High-order entropy-compressed text indexes.
\newblock In {\em Proc. 14th Annual ACM-SIAM Symposium on Discrete Algorithms
  (SODA)}, pages 841--850, 2003.

\bibitem{GOR10}
R.~Grossi, A.~Orlandi, and R.~Raman.
\newblock Optimal trade-offs for succinct string indexes.
\newblock In {\em Proc. 37th International Colloquium on Algorithms, Languages
  and Programming (ICALP)}, LNCS 6199, pages 678--689, 2010.

\bibitem{GV05}
R.~Grossi and J.~Vitter.
\newblock Compressed suffix arrays and suffix trees with applications to text
  indexing and string matching.
\newblock {\em SIAM Journal on Computing}, 35(2):378--407, 2006.

\bibitem{Huffman:1952}
D.~A. Huffman.
\newblock A method for the construction of minimum-redundancy codes.
\newblock {\em Proceedings of the I.R.E.}, 40(9):1098--1101, 1952.

\bibitem{KY00}
J.~C. Kieffer and E.-H. Yang.
\newblock Grammar-based codes: {A} new class of universal lossless source
  codes.
\newblock {\em IEEE Transactions on Information Theory}, 46(3):737--754, 2000.

\bibitem{KN12}
S.~Kreft and G.~Navarro.
\newblock On compressing and indexing repetitive sequences.
\newblock {\em Theoretical Computer Science}, 483:115--133, 2013.

\bibitem{LM00}
J.~Larsson and A.~Moffat.
\newblock Off-line dictionary-based compression.
\newblock {\em Proceedings of the IEEE}, 88(11):1722--1732, 2000.

\bibitem{LZ76}
A.~Lempel and J.~Ziv.
\newblock On the complexity of finite sequences.
\newblock {\em IEEE Transactions on Information Theory}, 22(1):75--81, 1976.

\bibitem{MN05}
V.~M{\"a}kinen and G.~Navarro.
\newblock Succinct suffix arrays based on run-length encoding.
\newblock {\em Nordic Journal of Computing}, 12(1):40--66, 2005.

\bibitem{MNlatin06}
V.~M{\"a}kinen and G.~Navarro.
\newblock Position-restricted substring searching.
\newblock In {\em {Proc. 7th Latin American Symposium on Theoretical
  Informatics (LATIN)}}, LNCS 3887, pages 703--714, 2006.

\bibitem{MNtalg08}
V.~M{\"a}kinen and G.~Navarro.
\newblock Dynamic entropy-compressed sequences and full-text indexes.
\newblock {\em ACM Transactions on Algorithms (TALG)}, 4(3):article 32, 2008.

\bibitem{MNSV09jcb}
V.~M{\"a}kinen, G.~Navarro, J.~Sir{\'e}n, and N.~V{\"a}lim{\"a}ki.
\newblock Storage and retrieval of highly repetitive sequence collections.
\newblock {\em Journal of Computational Biology}, 17(3):281--308, 2010.

\bibitem{Mun96}
J.~I. Munro.
\newblock Tables.
\newblock In {\em Proc. 16th Conference on Foundations of Software Technology
  and Theoretical Computer Science (FSTTCS)}, LNCS 1180, pages 37--42, 1996.

\bibitem{Nav12}
G.~Navarro.
\newblock Indexing highly repetitive collections.
\newblock In {\em Proc. 23rd International Workshop on Combinatorial Algorithms
  (IWOCA)}, LNCS 7643, pages 274--279, 2012.

\bibitem{NAVacmcs14}
G.~Navarro.
\newblock Spaces, trees and colors: The algorithmic landscape of document
  retrieval on sequences.
\newblock {\em ACM Computing Surveys}, 46(4):article 52, 2014.

\bibitem{Navjda13}
G.~Navarro.
\newblock Wavelet trees for all.
\newblock {\em Journal of Discrete Algorithms}, 25:2--20, 2014.

\bibitem{NM07}
G.~Navarro and V.~M{\"a}kinen.
\newblock Compressed full-text indexes.
\newblock {\em ACM Computing Surveys}, 39(1):article 2, 2007.

\bibitem{NOsea14.1}
G.~Navarro and A.~Ord{\'o\~n}ez.
\newblock Faster compressed suffix trees for repetitive text collections.
\newblock In {\em Proc. 13th International Symposium on Experimental Algorithms
  (SEA)}, LNCS 8504, pages 424--435, 2014.

\bibitem{NOspire14}
G.~Navarro and A.~Ord{\'o\~n}ez.
\newblock Grammar compressed sequences with rank/select support.
\newblock In {\em Proc. 21st International Symposium on String Processing and
  Information Retrieval (SPIRE)}, LNCS 8799, pages 31--44, 2014.

\bibitem{NPVjea13}
G.~Navarro, S.~J. Puglisi, and D.~Valenzuela.
\newblock General document retrieval in compact space.
\newblock {\em ACM Journal of Experimental Algorithmics}, 19(2):article 3,
  2014.

\bibitem{RRR02}
R.~Raman, V.~Raman, and S.~Srinivasa Rao.
\newblock Succinct indexable dictionaries with applications to encoding {\it
  k}-ary trees, prefix sums and multisets.
\newblock {\em ACM Transactions on Algorithms}, 3(4):article 43, 2007.

\bibitem{Sad03}
K.~Sadakane.
\newblock New text indexing functionalities of the compressed suffix arrays.
\newblock {\em Journal of Algorithms}, 48(2):294--313, 2003.

\bibitem{Sak05}
H.~Sakamoto.
\newblock A fully linear-time approximation algorithm for grammar-based
  compression.
\newblock {\em Journal of Discrete Algorithms}, 3(2-4):416--430, 2005.

\bibitem{TTS13}
Y.~Tabei, Y.~Takabatake, and H.~Sakamoto.
\newblock A succinct grammar compression.
\newblock In {\em Proc. 24th Annual Symposium on Combinatorial Pattern Matching
  (CPM)}, LNCS 7922, pages 235--246, 2013.

\bibitem{VY13}
E.~Verbin and W.~Yu.
\newblock Data structure lower bounds on random access to grammar-compressed
  strings.
\newblock In {\em Proc. 24th Annual Symposium on Combinatorial Pattern Matching
  (CPM)}, LNCS 7922, pages 247--258, 2013.

\bibitem{vbyte}
H.~E. Williams and J.~Zobel.
\newblock Compressing integers for fast file access.
\newblock {\em The Computer Journal}, 42(3):193--201, 1999.

\bibitem{witten1999managing}
I.~H. Witten, A.~Moffat, and T.~C. Bell.
\newblock {\em Managing Gigabytes: Compressing and Indexing Documents and
  Images}.
\newblock Morgan Kaufmann, 1999.

\bibitem{ZL77}
J.~Ziv and A.~Lempel.
\newblock A universal algorithm for sequential data compression.
\newblock {\em IEEE Transactions on Information Theory}, 23(3):337--343, 1977.

\end{thebibliography}

\end{document}